 \documentclass[journal,final]{IEEEtran}
\usepackage[T1]{fontenc}

\usepackage[noadjust]{cite}
\usepackage{hyperref}

\ifCLASSINFOpdf
  \usepackage[pdftex]{graphicx}
  \graphicspath{{./Figures/}}
\else
\fi
\usepackage{amsmath}
\usepackage{algorithmic}

\usepackage{array}

\ifCLASSOPTIONcompsoc
 \usepackage[caption=false,font=normalsize,labelfont=sf,textfont=sf]{subfig}
\else
 \usepackage[caption=false,font=footnotesize]{subfig}
\fi
\usepackage{dblfloatfix}
\usepackage{url}

\usepackage{svg}
\usepackage{mathrsfs}
\usepackage{mathtools}
\usepackage{amsfonts} 
\usepackage{epstopdf}
\usepackage{epsf,epsfig,verbatim,amssymb,array,multicol,multirow}
\usepackage{amsmath}

\DeclareMathOperator*{\argmin}{arg\,min}
\usepackage{psfrag,bm,xspace}
\usepackage{hhline}
\usepackage{mathabx}
\usepackage{physics}
\usepackage{bbm}
\usepackage[normalem]{ulem}
\usepackage{enumitem}
\usepackage{xcolor}
\usepackage{ulem}
\usepackage{multicol}
\usepackage{algorithm}
\newtheorem{thm}{Theorem}
\newtheorem{prop}{Proposition}
\newtheorem{dfn}{Definition}

\newtheorem{lem}[thm]{Lemma}
\newtheorem{cor}{Corollary}[thm]

\newtheorem{conjecture}{Conjecture}
\allowdisplaybreaks 
 \global\long\def\11{\mathbbm{1}}
\allowdisplaybreaks 

\newcommand\numberthis{\addtocounter{equation}{1}\tag{\theequation}}
\newcommand{\ra}{\rightarrow}
\newcommand{\la}{\leftarrow}
\newcommand{\mcl}{\mathcal}
\newcommand{\mb}{\mathbb}

\newcommand{\eps}{\epsilon}
\newcommand{\ov}{\overline}
\newcommand{\udl}{\underline}
\newcommand{\bsl}{\boldsymbol}
\def \defeq{\coloneqq}
\def \dg{\dagger}

\newcommand{\wh}{\widehat}
\newcommand{\wt}{\widetilde}

\def \rwc{ R^{c}_W}
\def \rw{ R_W }
\def \kw{ K_W }
\def \ws{(W,S)}
\def \vt{(V,T)}
\newcommand{\totalmismatch}[1]{M_{#1} }
\newcommand{\indunchoke}[1]{\11{\{Y_{opt}=#1\}} }

\newcommand{\nmbr}[2]{ c^{(#1)}_{#2} }
\newcommand{\nmbrn}[1]{ \ov{c}_{#1} }
\newcommand{\nmbrr}[1]{ \udl{c}_{#1} }
\newcommand{\dnmbr}[2]{ d^{(#1)}_{#2} }
\newcommand{\freq}[2]{ \pi^{(#1)}_{#2} }
\newcommand{\freqn}[1]{ \ov{\pi}_{#1} }
\newcommand{\freqr}[1]{ \udl{\pi}_{#1} }

\newcommand{\mismatchr}[1]{ \ov{m}_{#1} }

\newcommand{\setrares}[1]{ \underline{R}_{(#1,W,S)} }
\def \statex{\mathbf{x}}
\def \statey{\mathbf{y}}
\newcommand{\cw}[1]{ C^{(#1)}_{W} }
\def \c1{C^{(1)}}

\def \Nw{ N_W }
\def \Nu{ N_U }
\newcommand{\nwj}[1]{ N^{(#1)}_{W} }

\newcommand{\Gammaexact}[2]{\Gamma_{#1}^{(#2)}}
\newcommand{\Gammalower}[2]{\udl{\Gamma}_{#1}^{(#2)}}
\newcommand{\Gammaupper}[2]{\ov{\Gamma}_{#1}^{(#2)}}

\newcommand{\indrares}[1] { \11{ \{ r \in \setrares{#1} \} } }

\newcommand{\regrw}[2]{\mcl{R}_{#1}^{#2}}
\def \x{ |\statex| }
\def \xw{ |\statex|_W}
\def \xws{ x_W^{(S)} }
\def \xvt{ x_V^{(T)} }
\def \mw{ C_W^{(3)} }
\def \pw{ P_W }
\def \bw{ \beta_W }
\def \aw{ \alpha_W }
\def \slmdas{| \bsl{\lambda} |}
\newcommand{\Psiwr}[2]{\Psi_{#1}^{(#2)} }
\def \setaw{ \mcl{A}_{W} }
\def \setawc{ \mcl{A}^c_W }

\newcommand{\sw}[1]{ \mcl{S}^{#1}_{W} }
\def \rs{ \rw \setminus S }
\def \trs{ (T \cap \rw) \setminus S }

\def \sumskrnew{ \sum \limits_{ \substack{ S:|S \cap W|<\kw-1, \\ r:r \in \rw \setminus S} }  }

\def \sumskrarestnew{ \sum \limits_{ \substack{S:|S \cap W|<\kw-1, \\ r:r \in \setrares{\mcl{F}} } }  }

\def \sumrsvt{ \sum \limits_{ \substack{ r \in \rw,\\ S: W \setminus S \supsetneq \{r\},\\V: W\in\mcl{W}_V, T: r \in T } } }

\def \sumvtr{ \sum \limits_{  \substack{ V: W\in\mcl{W}_V,\\ T: r \in T }  } }

\usepackage{cuted} 

\newcounter{relctr} 
\everydisplay\expandafter{\the\everydisplay\setcounter{relctr}{0}} 

\newcommand\labelrel[2]{%
  \begingroup
    \refstepcounter{relctr}%
    \stackrel{\textnormal{(\alph{relctr})}}{\mathstrut{#1}}%
    \originallabel{#2}%
  \endgroup
}
\AtBeginDocument{\let\originallabel\label} 
\usepackage{longtable}
\begin{document}


%
\title{Rarest-First with Probabilistic-Mode-Suppression (RFwPMS)}
%
%
%

\author{Nouman~Khan,~\IEEEmembership{Member,~IEEE,}
        Mehrdad~Moharrami,
        ~and~Vijay~Subramanian,~\IEEEmembership{Senior~Member,~IEEE}
\thanks{A short version of this work appeared in \textit{IEEE INFOCOM 2020 - IEEE Conference on Computer Communications}, 2020, pp. 1153-1162.}
}

\maketitle

\begin{abstract}
Recent studies suggested that the BitTorrent’s rarest-first protocol, owing to its work-conserving nature, can become unstable in the presence of non-persistent users. Consequently, for any provably stable protocol, many peers, at some point, would have to be endogenously forced to hold off their file-download activity. In this work, we propose a tunable piece-selection policy that minimizes this (undesirable) requisite by combining the (work-conserving but not stabilizing) rarest-first protocol with only an appropriate share of the (non-work conserving and stabilizing) mode-suppression protocol. We refer to this policy as ``Rarest-First with Probabilistic Mode-Suppression'' or simply RFwPMS. We study RFwPMS using a stochastic abstraction of the BitTorrent network that is general enough to capture 
a multiple swarm setting of non-persistent users -- each swarm having its own altruistic preferences that may or may not overlap with those of other swarms. 
Using Lyapunov drift analysis, we show that for all kinds of inter-swarm behaviors and all arrival-rate configurations, 
RFwPMS is stable. Then, using the Kingman's moment bound technique, we further show that the expected steady-state sojourn time of RFwPMS is independent of the arrival-rate in the single-swarm case (under a mild additional assumption). Finally, 
our simulation-based performance evaluation confirms our theoretical findings and shows that the steady-state expected sojourn time\footnote{The time a peer remains in the system collecting all the pieces of the file.} is linear in the file-size (compared to our loose estimate of a polynomial with degree 6). Overall, an improved performance is observed in comparison to previously proposed stabilizing schemes like mode-suppression (MS).

\end{abstract}

\begin{IEEEkeywords}
P2P File-Sharing, BitTorrent, Rarest-First, Mode-Suppression, Foster-Lyapunov Theorem, Kingman's bound
\end{IEEEkeywords}


%
\IEEEpeerreviewmaketitle

\section{Introduction}\label{sec:introduction}
\IEEEPARstart{C}{onsider} the task of distributing a large file to peers in an unstructured peer-to-peer (P2P) network. The file is initially available with a distinguished peer (usually termed as the seed) and each peer can initiate a transfer connection with any other peer \cite{tcfss}.

One efficient method to perform the above  task is to chop the file into a large number of small and roughly equally-sized pieces/chunks, and to allow peers to share the pieces with each other. Chopping the file allows peers to distribute parts of it before possessing it completely -- this is the key idea behind the popular ``BitTorrent protocol'' \cite{Cohen03incentivesbuild}. Such an upload-while-download scheme reduces the average file-download time and more importantly, enables the network to scale its throughput with the number of peers. As a result, the BitTorrent protocol has gained large popularity over the years. Even today, despite the growth of streaming services like Netflix, Hulu, and Youtube, BitTorrent sharing remains a significant source of internet traffic \cite{gipreport2020}. In the research literature also, the protocol has gained extensive interest. For instance, on the theoretical side, various mathematical models have been recently studied \cite{tcfss,crs,mps,bh,optus,ssus,va,gs,ms,ms:arxiv,ms:journal,srikant,ourinfocom,bundling_menasche,de_Souza_e_Silva_2019}, 
with each model providing a high-level abstraction of the detailed workings of the actual protocol.

Two key features of BitTorrent-like networks are the  \emph{tit-for-tat} and \emph{optimistic-unchoke} interactions among the users. Tit-for-tat, as the name indicates, are interactions where both peers share the file-contents with each other based on mutual benefit; on the other hand, opportunistic-unchoking interactions are those in which a peer altruistically offers pieces to other peers. Both these interactions are important for BitTorrent's performance; opportunistic-unchoking helps the incoming empty-peers to get some pieces of the file whereas the tit-for-tat interactions help ward off \emph{free-riders} (users who download the file but do not contribute any of it back). 

A common occurrence in BitTorrent-like networks is that a peer usually spends relatively more time downloading the last few pieces of the file. This phenomenon, known as the \textit{delay-in-endgame-mode}, is because the last few pieces are often the rare ones in the network. Inspired by this, Hajek and Zhu \cite{mps} studied a stochastic abstraction and showed that an unstructured\footnote{Each peer contacts another peer uniformly at random.} BitTorrent-like network that employs any work-conserving\footnote{A work-conserving piece-selection policy is one in which a piece transfer always happens if the uploading peer has a piece that the downloading peer needs.} piece-selection policy (e.g., Random-Novel (RN), Rarest-First (RF)) becomes unstable if the arrival rate of peers exceeds the fixed seed's upload rate and if each peer departs immediately upon completing their own file-download. 

The cause behind instability is a phenomenon called the \textit{Missing-Piece Syndrome} or \textit{Last-Piece Syndrome (LPS)} \cite{mps}, wherein the network converges to, and then cannot escape from, the \textit{one-club scenario}. In this scenario, there are a large number of peers who possess all the pieces of the file except one (such peers are called \textit{one-club peers}), a very small number of peers who possess that one piece (such peers are called \textit{infected peers}), and a very small number of peers that are in neither of these two groups (they are called \textit{young peers}). Since a vast majority of the peers are one-club peers, owing to random peer contacts and the work-conserving piece-selection policy, all the infected peers leave the network quickly and almost all young peers (including the new-comers) join the one-club peers. Inevitably, the fixed seed is tasked with uploading that one rare piece to the entire one-club whose growth rate is larger than its upload capacity. The network thus remains trapped in this configuration and the size of the one-club grows to infinity, causing instability. This result was later substantiated with experimental checks performed by Mendes, de Souza e Silva, Menasché, Leão and Towsley \cite{expcheck}, where it was observed that when seeds have a small effective service capacity, or when seeds are intermittent, the throughput saturates as the population size grows. Importantly, the authors also demonstrated that the last piece syndrome unfolds in a closed BitTorrent network in which each departure causes an arrival of an empty peer.

Another typical phenomenon in BitTorrent-like networks is the \emph{low availability of chunks in the start-up phase} when only one user (the seed) has the complete contents of the file and the rest are empty.\footnote{Flash-crowd is a start-up phase when a large number of users enter the network and there is only a few existing users with the complete file.} Reference \cite{expcheck} calls this scenario the \emph{first block problem}.\footnote{Indeed, \cite{expcheck} argues that the problem of last-piece-syndrome is overestimated whereas the first-block-problem is much more critical in practice.} Understandably, the duration of this phase, for any choice of piece-selection algorithm, depends on the upload capacity of the seed. However, if the incoming peers choose to leave the network immediately upon download completion, then the choice of piece-selection policy is important -- to the effect that it should enforce some form of lingering instead of a blind replication to all pieces.

Finally, multiple file downloads occurring simultaneously is commonly observed in practice. This is the multi-swarm setting~\cite{optus,ssus} for which there isn't as yet a piece-selection policy that guarantees stability for all arrival rates, particularly when the swarms interact with each other.

In the above discussion, we have highlighted multiple design criteria for BitTorrent-like file-sharing networks. These are:
\begin{itemize}
\item Stability guarantees for all arrival-rates in both single-swarm and multi-swarm settings.
\item Improved steady-state file-delivery times.
\item Improved transient responses (e.g. low flush-out times in flash-crowds).
\end{itemize}
Our goal in this paper is to propose a flexible piece-selection policy that can attain the aforementioned design criteria. To this end, in Section \ref{sec:relatedwork}, we will first go through the existing work and then in \ref{sec:contributions}, list the contributions of this manuscript.
\subsection{Related Work}\label{sec:relatedwork}
A series of recent works have appeared of which the relevant papers are \cite{bh,crs,mps,tcfss,va,gs,ms,ms:arxiv,ms:journal,optus,ssus}.

\begin{itemize}
\item
Zhu and Hajek \cite{bh}, in a follow-up to \cite{mps}, showed that if, after completing their file-download, each peer remains in the system long enough to upload one additional piece, then the network is stable under any positive seed uploading capacity and any peer arrival rate. This demands persistence of peers, which might not hold, especially with wireless users who are sensitive to their energy and bandwidth usages. Thus, recent works have also covered piece-selection policies that ensure stability when peers are strictly non-persistent.

\item
Massoulié and Vojnović \cite{crs} considered a BitTorrent-like stochastic system where before entering the system, users obtain one piece (referred to as \textit{coupon} in \cite{crs}) from a central bootstrap server. \textit{Under this assumption}, they showed that the performance of the network does not depend critically on either persistence of users or on load balancing piece-selection policies like RF.

\item
Norros, Reittu and Eirola \cite{tcfss} proposed the \textit{Enforced Friedman algorithm} in  which a peer makes three contacts simultaneously (with replacement) and if there are `minority pieces' (pieces possessed by exactly one of the three peers), then the peer downloads one of them uniformly at random. If there are no minority pieces, then the peer waits for the next triple contact. The stability of the protocol was shown for a two-chunk file-sharing system whereas for the case of a multi-chunk system, it was left as a conjecture, with numerical simulations providing good evidence of stability. In a follow-up work, Oguz, Anantharam and Norros \cite{va} proved the stability of the Enforcement Friedman protocol for multi-chunk systems and also proposed a provably stable improvement under the name of \textit{Common Chunk protocol}. In this protocol, only new peers who arrive with no pieces follow the rules in \cite{tcfss}, and the peers who lack only one piece contact three peers and are allowed to download the last piece only if every piece they have is present with at least two of the three contacted peers.

\item
Bilgen and Wagner \cite{gs, bilgen20} proposed the \textit{Group Suppression Protocol} in which peers who share the same piece profile are defined as a \textit{group} and the group with the largest population is defined as the \textit{largest club}. The protocol states that a peer belonging to the largest club uploads only to those peers who hold greater number of pieces than it does, and refuses the upload to all other peers if it contacts them. The protocol was proved to be stable for a two-chunk file-sharing system and its stability for multi-chunk systems was left as a conjecture.

\item
Reddyvari, Parag and Shakkottai \cite{ms} took a chunk-level viewpoint and proposed the \textit{Mode-Suppression} (MS) protocol. Here, the transfer of pieces in the mode (present with the most number of peers) is prohibited, except when all pieces are in the mode, and a random-novel piece not in the mode (if any) is sent. In a follow-up \cite{ms:arxiv,ms:journal}, Reddyvari, Parag and Shakkottai prove the stability and scalability of \textit{Threshold Mode-Suppression} (TMS) wherein pieces in the mode are prohibited from transfer when the largest-mismatch (difference between chunk-counts of the most-abundant and the rarest chunks, see \ref{def:mismatch}) crosses a certain fixed threshold. This line of work showed stability and scalability for all arrival rates in the single-swarm setting.

\item
Previous works have also considered bundling different swarms together in the same network and then allowing content-sharing across them. This was proposed by Zhou, Ioannidis and Massoulié \cite{optus} with the claim that such ``universal swarms'' can increase the stability region of a BitTorrent network. Then Zhu, Ioannidis, Hegde and Massoulié \cite{ssus} formally characterized the stability region of such networks when a work-conserving piece-selection policy is used. The stability region is indeed larger than in the single-swarm setting, however, yet again, it doesn't hold for all arrival rate configurations. 
\end{itemize}

\subsection{Contributions}\label{sec:contributions}
The contributions of this paper are listed below:
\begin{enumerate}
\item[a)]
Developing a tractable model (that extends the stochastic model from \cite{mps} and \cite{bh}) for the analysis of file-sharing P2P networks that employ tit-for-tat and optimistic-unchoke mechanisms; 
\item[b)] 
Demonstrating that instability occurs when a hard tit-for-tat rule is used without optimistic-unchoking. Here, instability occurs due to a \emph{first-piece syndrome (FPS)} where newly arriving peers have to rely solely on the seed to get their first piece (Proposition \ref{prop:fps});
\item[c)] 
Showing that soft tit-for-tat or optimistic-unchoking mechanisms cannot ward off the \emph{last-piece-syndrome} in the presence of a work-conserving piece-selection policy (Proposition \ref{prop:lps});
\item[d)] 
In a general multi-swarm setting, proposing the swarm-based RFwPMS piece-selection policy and proving its stability using a novel Lyapunov function (Theorem \ref{thm:stability}).
\item[e)] 
Demonstrating the scalability of swarm-based RFwPMS in the single-swarm case, i.e., the expected steady-state sojourn time is upper-bounded by a constant which is independent of the arrival-rate of the incoming peers.\footnote{Establishing the scalability in the general multi-swarm case is left as an open problem.} (Theorem \ref{thm:scalability}).
\item[f)] 
Showing improved performance of swarm-based RFwPMS in steady-state and flash-crowds via numerical simulations.
\end{enumerate}

A preliminary version of this work was presented in \cite{ourinfocom} with a partial theoretical analysis and a sub-case of the general stochastic model studied here. The current work adds to the contributions of \cite{ourinfocom} -- items a) through f) above -- and provides the complete details associated with the stability analysis of swarm-based RFwPMS.

\subsection{Organization}
Since the single-swarm model is a special case of the multi-swarm model, henceforth we only consider the multi-swarm model, except when we discuss our scalability results. The remainder of the paper is organized as follows. In Section \ref{sec:model}, we introduce the multi-swarm model which is built upon the model in \cite{ssus}. Section \ref{sec:analysis} presents the main stability theorem along with the preliminary setup needed for the detailed proof (the detailed proof is provided in Appendix \ref{sec:appendix:stability}). Section \ref{sec:discussion} discusses the working of RFwPMS as well as a few types of swarm behaviors that can be relevant in specific P2P environments and follow naturally as a result of the general nature of our assumption on inter-swarm behaviors. Section \ref{sec:simulations} presents few important snapshots of numerical results on stability, scalability and performance of RFwPMS. Finally, in Section \ref{sec:conclusion}, we give concluding remarks and few future-work directions. 

\subsection{Notation}
To aid the reading experience, we encourage the reader to refer to the relevant notation and definitions detailed in Appendix \ref{sec:appendix:notationtable}.
	
\section{System Model}\label{sec:model}
In this section, we introduce the stochastic model and describe our proposed piece-selection policy.
\subsection{Key Model Assumptions}
A \textbf{master-file}, denoted by $\mcl{F}$, is chopped into at least two equally-sized pieces, i.e., $\mcl{F}=[K]$\footnote{For $c, d\in \mathbb{Z}=\{\dotsc,-1,0,1,\dotsc\}$ and $c <d$, we use the standard notation $[c,d]:=\{c+1,\dotsc,d\}$ and $[d]$ for $[0,d]$ when $d\in \mb{N}=\{1,2,\dotsc\}$.} where $K\ge 2$. 
There is a distinguished peer, the \textbf{seed}, which holds this master-file $\mcl{F}$ and stays in the network indefinitely. The existence of the seed for indefinite period of time ensures that every piece is available in the network at all times, thus allowing us to study the long-term behavior of the network.
We define \textbf{file-$W$} as any non-empty subset of the master-file, thus, $ W \ne \emptyset$ and $ W \subseteq \mcl{F}$. The number of pieces in file-$W$ is denoted by $\kw$, i.e., $K_W = |W|$. With this definition of file, \textbf{swarm}-$W$ is defined as the set of peers who are primarily interested in downloading (pieces of) file-$W$. We note that peers entering the network can be interested in any file, i.e., the files need not be disjoint subsets of $\mcl{F}$. Besides their primary interest in file-$W$, swarm-$W$ peers may also have a secondary preference for some other pieces of the master-file. Thus, the set of pieces that a swarm-$W$ peer can download during its stay in the network is given by $\mcl{F}_{W}$ where $ W \subseteq \mcl{F}_{W} \subseteq \mcl{F}$. It is assumed that peers enter the network according to independent Poisson processes, i.e., a swarm-$W$ peer enters the network according to a Poisson process of rate $\lambda_W > 0$, independent of other swarms. Let $\mcl{W}$ denote the set of all swarms that join the network and let $\bsl{\lambda} = (\lambda_{W} : W \in \mcl{W})$ denote the vector of their arrival-rates. We now list three important assumptions of the model.
\begin{enumerate}
\item[a)] \textit{Empty Cache upon Arrival}:
Each peer maintains a \textbf{cache} to store the pieces it downloads. The cache is empty upon arrival and has a capacity of $|\mcl{F}_{W}|$ pieces. The part of the cache that is devoted for the pieces of secondary interest is called the \textbf{excess-cache} (where pieces from the set $\mcl{F}_{W} \setminus W$ are stored). In the context of \cite{crs}, empty cache assumption aligns with the case when the central bootstrap server is bottlenecked, for example, in the case of a high peer arrival-rate or during a flash-crowd\footnote{A situation in which the network suddenly encounters a very large number of empty peers; this is commonly seen with torrents of popular files.}.
\item [b)]
\textit{Ally Swarms}: While peers interested in the same file (i.e., belonging to the same swarm) exchange pieces with each other, they may or may not prefer to collaborate with peers who are interested in other files (i.e. peers belonging to a different swarm). Thus, each swarm has an associated set of ally-swarms. Formally, the \textbf{ally-set} of swarm-$W$, denoted by $\mcl{W}_W$, is a non-empty subset of $\mcl{W}$ that consists of swarm-$W$ as well as any other swarms to which its peers can upload pieces ($ W \in \mcl{W}_{W} \subseteq \mcl{W}$); and 
\item[c)]
\textit{Strictly Non-persistent Peers}: Once a peer finishes downloading their pieces of primary interest, they leave the network immediately.
\end{enumerate}

\subsection{State Description}
The notation used in this paper is a combination of related notations in \cite{ssus} and \cite{ms}. We classify peers into types according to the swarm they belong to, and the set of pieces in their cache. Hence, a peer in swarm $W$ holding $S \subseteq \mcl{F}_{W}$ is said to be of type $\ws$.
We denote the number of $\ws$-type peers at time $t \ge 0$ by $\xws(t) \in \mathbb{Z}_{+} \defeq \{0,1,\dots \}$. Then the \textbf{state} of the network at time $t$ is given by the vector 
\begin{align*}\label{eq:state}
\statex(t) \defeq  (\xws(t) : W \in \mcl{W}, S \subseteq \mcl{F}_{W}, \text{ and } W \setminus S \ne \emptyset ). \numberthis
\end{align*}
Note that as a result of aforementioned assumptions, the cache-profile $S$ of a swarm-$W$ peer always satisfies the conditions $S \subseteq \mcl{F}_{W}$ and $W \setminus S \ne \emptyset$, with the latter condition capturing strictly non-persistent peers. For the sake of brevity, from hereon, we will omit writing these two conditions. Also, since the fixed seed is always present in the network, we do not include it in $\statex(t)$.
The \textbf{population} of swarm-$W$, i.e., the number of swarm-$W$ peers at time $t$ is given by
\begin{align*}\label{eq:xw}
|\mathbf{x}(t)|_{W} \defeq \sum_S \xws(t). \numberthis
\end{align*}
Similarly, the total number of peers at time $t$ is given by
\begin{align*}\label{eq:x}
|\mathbf{x}(t)| \defeq \sum_{W\in\mcl{W}} |\statex(t)|_W. \numberthis 
\end{align*}
From hereon, for brevity, we will write $\statex (t)$ as $\statex$ since the dependence on time $t$ will be clear from the context. 

\subsection{Peer-Contact Policy}\label{subsec:contactpolicy}
Consistent with the stochastic models of \cite{tcfss,crs,mps,bh,gs,ms,ms:arxiv,ssus,optus,va}, we assume that the network employs random peer-contacts. Specifically, it is assumed that each peer has a fixed number of contact-links, denoted by $L\ge 1$. In normal peers, the first $\left(L-\indunchoke{1}\right)$ of these links are reserved for tit-for-tat based piece exchanges whereas the $L^{th}$ link is used for optimistic-unchoking if and only if $Y_{opt}=1$; otherwise it is also used for tit-for-tat based piece exchanges. Here, $Y_{opt}$ is a binary parameter that is set to 1 if optimistic-unchoking is desired as part of the peer-contact policy. In contrast to normal peers, all the $L$ links of the fixed seed are used for optimistic-unchokes.

For normal peers, we assume that each tit-for-tat link is activated according to an independent Poisson process of rate $\mu>0$ and the optimistic-unchoke link is activated according to another independent Poisson process of rate $\hat{\mu}>0$. Upon activation of the link, the peer contacts, uniformly at random, some other (normal) peer from the network.\footnote{The motivation behind introducing a separate rate for the optimistic-unchoke link comes from how BitTorrent operates. In practice, by default, the number of links $L$ is 5, and which peer is optimistically-unchoked is rotated roughly every third tit-for-tat period (see \cite{Cohen03incentivesbuild}). By introducing $\mu$ and $\hat{\mu}$, this can be captured by our model by setting $\hat{\mu}=\frac{\mu}{3}$. For more details, see \cite{Cohen03incentivesbuild}.} For the fixed seed, each of its (optimistic-unchoke) links is activated according to a Poisson process of rate $U>0$. We assume that the transfer of piece/s occurs instantaneously with the contact (in reality, this will take more time than initiating a contact).

To define the tit-for-tat and optimistic-unchoking mechanisms, let us imagine a contact in which a peer, say peer-(1), 
which is of type $(W_1,S_1)$, has contacted another peer, say peer-(2), 
which is of type $(W_2,S_2)$. The interaction between the two peers depends on the type of contact that peer-(1) 
has made with peer-(2). 
Thus, there are two cases.
\begin{enumerate}
\item[a)] \textit{Tit-for-Tat Contact}: In this case, both peers first check the swarm-identity of each other following which they reveal their cache-profiles (based on how they view each other's swarm). Specifically, for each $k=1,2$, the following events happen sequentially.
\begin{enumerate}
\item[i)] Peer-$(k)$ shares its swarm-identity $W_{k}$ with peer-($-k$).\footnote{Here, $-k$ denotes the element in $\{1,2\}\setminus\{k\}$.}
\item[ii)] If $W_{-k} \notin \mcl{W}_{W_k}$, Peer-$(k)$ reveals an empty cache-profile, $\widehat{S}_k=\emptyset$\footnote{This ensures that no piece is transferred to a non-ally peer.}, otherwise it shows its true cache-profile, $\widehat{S}_k=S_k$. We call $\widehat{S}_k(\cdot)$ the \textbf{revealed cache-profile} of peer-$(k)$.
\item[iii)] Peer-$(k)$ checks if the contact is potentially useful, i.e.,  if $(\widehat{S}_{-k} \cap W_k)\setminus S_k$ is non-empty.
\footnote{Checking $(\widehat{S}_{-k} \cap W_k) \setminus S_k$ instead of $(\widehat{S}_{-k} \cap \mcl{F}_{W_k})\setminus S_k$ matches with our assumption that extra pieces (in $\mcl{F}_{W}\setminus W$) are given secondary preference.} If $(\widehat{S}_{-k} \cap W_k)\setminus S_k$ is non-empty, peer-$(k)$ commits to transfer some piece to peer-$(-k)$ from $\widehat{S}_{k}$. If $(\widehat{S}_{-k} \cap W_k)\setminus S_k$ is empty, the network forces peer-$(k)$ to conduct a Bernoulli($p$) trial, only upon the success of which, it must commit to transfer some piece to peer-$(-k)$ from $\widehat{S}_{k}$. Once peer-$(k)$ has committed to transfer a piece to peer-$(-k)$ from $\widehat{S}_k$, we say that ``peer-$(k)$ has \textbf{push-contacted} peer-$(-k)$ under revealed cache-profile $\widehat{S}_k$
'' 
or equivalently ``peer-$(-k)$ has pull-contacted peer-$(k)$ with revealed cache-profile $\widehat{S}_{k}$
''. Here, importantly, we note that the probability of this push-contact is at least $p$. 
\end{enumerate}
\item[b)] \textit{Optimistic Unchoke}: In this case, peer-$(1)$ checks the swarm-identity $W_{2}$ and then push-contacts peer-$(2)$ with revealed cache-profile $\widehat{S}_1(W_{2})$.
\end{enumerate}

From a) and b), it is clear that no piece is transferred from peer-$(k)$ to peer-$(-k)$ if $W_{-k} \notin \mcl{W}_{W_k}$ ($k=1,2$). Hence, we need not consider push-contacts from a peer to its non-ally peer. Once a push-contact has been made, which exact piece is chosen for transfer and whether the transfer is successful or not, is determined by the network's piece-selection policy. This is described next.

\subsection{Piece-Selection Policy (Swarm-based RFwPMS)}
Suppose that at time $t \ge 0$, a $\vt$-peer has push-contacted its ally-peer, say a $\ws$-peer ($W \in \mcl{W}_V$), with revealed cache-profile $\wh{T}$. To describe Swarm-based RFwPMS, some definitions (\textit{a'la} \cite{ms,ms:arxiv,ms:journal}) are needed.
\begin{dfn}
	The \textbf{frequency} of piece $i$ in swarm-$W$'s population is denoted by 
	$ \freq{i}{W}(\statex) $ and is defined as follows:
	\begin{align*}\label{eq:freqwi}
	\freq{i}{W}(\statex) \defeq
	\begin{cases}
	\frac{1}{|\mathbf{x}|_{W}} \sum\limits_{ S: i \in S } \xws(t) & \text{if $ |\statex|_{W} > 0$}, \\
	0 & \text{if $|\statex|_{W}=0 $}. \numberthis
	\end{cases}
	\end{align*}
	The \textbf{chunk-count} of piece $i$ in swarm-$W$ is then given by $\nmbr{i}{W}(\statex)=\freq{i}{W}(\statex)\xw$. The maximum and minimum chunk frequencies in swarm-$W$ are denoted by 
	$\ov{\pi}_W(\statex)$ and
	$\udl{\pi}_W(\statex)$ respectively,
	\begin{align*}
	\freqn{W}(\statex) \defeq \max \limits_{i \in W} \pi_W^{(i)} (\statex) \hspace{5pt} \text{ and } \hspace{5pt} \freqr{W}(\statex) \defeq \min \limits_{i \in W} \pi_W^{(i)} (\statex). \numberthis \label{eq:maxminfreqw} 
	\end{align*}
	Similar definitions hold for maximum and minimum chunk-counts of swarm-$W$, denoted respectively by $\nmbrn{W}(\statex)$ and $\nmbrr{W}(\statex)$, that is,
	\begin{align*}
	\nmbrn{W}(\statex) \defeq \freqn{W} (\statex)\xw \hspace{5pt} \text{ and } \hspace{5pt} \nmbrr{W}(\statex) \defeq \freqr{W} (\statex)\xw. \numberthis \label{eq:maxminnmbrw} 
	\end{align*}
	Importantly, both are computed over the file of primary interest, $W$, instead of the complete set of downloadable pieces, $\mcl{F}_{W}$.
\end{dfn}
\begin{dfn}
	The \textbf{total chunk-count} in swarm-$W$ is given by
	\begin{align*}\label{eq:chunkcountw}
	\pw(\statex) \defeq \sum \limits_{i \in W} \nmbr{i}{W}(\statex). \numberthis
	\end{align*}
\end{dfn}
\begin{dfn}\label{def:mismatch}
The \textbf{mismatch} of piece $i$ in swarm-$W$ is given by
\begin{align*}\label{eq:mwi}
m^{(i)}_{W}(\statex)\coloneqq \nmbrn{W}(\statex)-\nmbr{i}{W}(\statex).\numberthis
\end{align*} Importantly, we shall be interested in the \textbf{largest-mismatch} 
\begin{align*}\label{eq:largestmwi}
\ov{m}_{W}(\statex)\coloneqq (\nmbrn{W}-\nmbrr{W})(\statex),\numberthis
\end{align*}
and the \textbf{total-mismatch}
\begin{align*}\label{eq:totalmismatch}
\totalmismatch{W}(\statex)\defeq\sum_{i\in W}m_W^{(i)}(\statex).\numberthis
\end{align*}
Here, we note that 
\begin{align*}\label{eq:largestvstotalmismatch}
\ov{m}_W(\statex)\le M_W(\statex)\le (\kw-1)\ov{m}_W(\statex).\numberthis
\end{align*}
\end{dfn}
\begin{dfn}
For a swarm-$W$, the \textbf{complementary chunk-count} of piece $i$, denoted by $\dnmbr{i}{W}(\statex)$, is its number of copies in the ally-swarms of swarm-$W$. That is, 
\begin{align*}\label{eq:diw}
\dnmbr{i}{W}(\statex) \defeq \sum\limits_{\substack{V\ne W\\ W\in \mcl{W}_V}} \nmbr{i}{V}(\statex).\numberthis
\end{align*}
\end{dfn}

\begin{dfn}
	The set of \textbf{rare pieces} in swarm-$W$, denoted by 
	$\rw(\statex)$, is defined as follows:
	\begin{align*}
	\rw(\statex) \defeq
	\begin{cases}
	\{  i \in W : \nmbr{i}{W}(\statex) < \nmbrn{W}(\statex)  \} & \text{if $ \nmbrn{W}(\statex) \ne \nmbrr{W}(\statex)$}, \\
	W & \text{if $\nmbrn{W}(\statex) = \nmbrr{W}(\statex) $}.
	\end{cases} \label{eq:rw} \numberthis
	\end{align*}
\end{dfn}

\begin{dfn}
	The set of \textbf{non-rare pieces} in swarm-$W$ is given by ${R}^c_{W} (\statex) \coloneqq  W \setminus \rw(\statex)$.
\end{dfn}

\begin{dfn}
	The set of \textbf{extra pieces} for swarm-$W$ is given by $ \mcl{F}_{W} \setminus W$.
\end{dfn}

Unless otherwise noted, we will refer to a rare-piece by $r$, a non-rare piece by $n$, and importantly many times, the rarest piece (one with the lowest chunk-count) by $\udl{r}$. We now list the rules of swarm-based RFwPMS for a possible piece transfer from the  $\vt$-peer to (its ally) $\ws$-peer, when $(V,T)$-peer reveals cache-profile $\wh{T}$:
\begin{enumerate}
\item [a)]
\textit{Download of a Rare Piece}: If a rare-piece is available for transfer, i.e., $ \left( \wh{T} \cap R_{W}(\statex)  \right) \setminus S \ne \emptyset$, then the $\vt$-peer uploads the rarest piece it can offer, i.e., a piece chosen uniformly at random from the set
\begin{align*}\label{eq:setrares}
\setrares{\wh{T}}( \statex ) \defeq \argmin \limits_{j \in \left( \wh{T} \cap \rw (\statex) \right) \setminus S} \nmbr{j}{W}(\statex).\numberthis
\end{align*}
\item[b)]
\textit{Download of a Non-rare Piece}: In the case that no rare piece is available for transfer but a non-rare piece is, i.e., 
$ \left( \wh{T} \cap R_{W}(\statex)  \right) \setminus S = \emptyset$ and $(\wh{T} \cap W) \setminus S \ne \emptyset $, 
then the $\vt$-peer chooses some non-rare piece $n \in (\wh{T} \cap W) \setminus S$ uniformly at random, and uploads it only if the result of a Bernoulli$\left(\zeta_W^{(n)}(\statex)\right)$ trial is a success. We refer to $\zeta_W^{(n)}(\statex)$ as the \textit{non-rares' sharing factor} and it is defined as follows.
\begin{align*}
\zeta_W^{(n)}( \statex )  \defeq
\begin{cases}
\exp\left(-\frac{\left( \mismatchr{W}(\statex) + \left( \dnmbr{n}{W}(\statex)\right)^{\alpha_W} \right)}{\bw\kw}   \right) &\text{if $\bw > 0$}, \\
0 & \text{if $\bw = 0 $}.
\end{cases} \numberthis \label{eq:zetaw}
\end{align*}
Here, $\bw \ge 0, \alpha_W \in (0,1]$ are tuning parameters. (As an aside, note that complete mode-suppression is covered as we allow $\beta_W=0$). 

\textit{Intuition}: Let us develop some intuition about our specific choice of $\zeta_W^{(n)}(\statex)$. As indicated earlier in Section \ref{sec:introduction}, the mode-suppression (MS) protocol \cite{ms:arxiv} strictly forbids the replication of non-rare pieces. This does a good job of maintaining a uniform chunk-distribution throughout the network's evolution. However, there is an accompanying undesirable effect -- no piece is transferred in all those push-contacts wherein only the non-rare pieces are available for transfer. As shown in Section \ref{sec:simulations}, this can incur a high penalty on the file-delivery time during a flash-crowd.\footnote{A situation in which the network suddenly encounters a very large number of empty peers; this is commonly seen with torrents of popular files.} Besides flash-crowds, even under normal operating conditions, completely suppressing non-rare pieces is unnecessary and, as indicated in \cite{ms:arxiv}, a trade-off exists between their suppression and sharing. Swarm-based RFwPMS allows for tuning this trade-off via $\zeta_W^{(n)}$. 

Even though different swarms are coupled together in the same system, one can optimistically expect that if each swarm tries to uniformly maintain its own chunk-distribution, the stochastic system should hopefully converge to some form of equilibrium (after possibly some temporary transient behavior due to the effects of other swarms). Keeping this in mind, the first term of $\zeta_W^{(n)}$, i.e., 
$$\exp\left(-\frac{\mismatchr{W}( \statex )}{\bw\kw}  \right), $$ 
is intended to increase the probabilistic-suppression of non-rare pieces as the total-mismatch becomes larger. (We use $\mismatchr{W}$ as a simple proxy for $\totalmismatch{W}$). 

Ideally, we would have liked $\zeta_W^{(n)}$ to consist of the first term only. But showing stability of our multi-swarm model in that case is hard. The key technical difficulty behind this is the form of the Lyapunov function that we have engineered (see \eqref{eq:stab:v}, \eqref{eq:stab:vwdef}), and our analysis of the unit-transition drift (where we are effectively trying to decouple the inter-swarm effects). There, if we consider states with large $\mismatchr{W}$ and a large complementary chunk-count $\dnmbr{n}{W}$, then the suppression through $\exp\left(- \frac{\mismatchr{W}( \statex )}{\bw\kw}  \right) $ turns out to be insufficient to  satisfy the unit-transition drift conditions. (The term $\exp\left(- \frac{\mismatchr{W}( \statex )}{\bw\kw}  \right) $ cannot shrink a polynomial term in $\dnmbr{n}{W}$). All is not lost however, as we circumvent this technical issue, by introducing the second suppression term 
$$\exp\left(- \frac{  \left(\dnmbr{n}{W}( \statex ) \right)^{\alpha_W}   }{\bw\kw}  \right), $$ 
where by choosing $\aw\in (0,1]$ sufficiently small, $ \zeta_W^n(\statex)$ effectively becomes a function of the ratio of swarm-$W$'s total-mismatch to its file-size $\kw$. The higher this ratio, the lesser the likelihood that the (non-rare) piece $n$ gets replicated. The choice of the ratio instead of just the total-mismatch matches the intuition that a file with large number of pieces should allow more sharing of non-rare pieces as opposed to a file with smaller number of pieces.
\item[c)]
\textit{Download of an Extra Piece}: If, from rules (b) and (c), no piece of file-$W$ could be uploaded to the $\ws$-peer, then the $\vt$-peer uploads an extra piece chosen uniformly at random from the set $(\wh{T} \cap \mcl{F}_{W}) \setminus (S \cup W) $.
\end{enumerate}

With the above three rules, the \textbf{transferable-set} $A(\statex,\wh{T}, W, S)$ for swarm-based RFwPMS can be computed according to Algorithm \ref{alg:transferable_set}. By transferable set, we mean the set of pieces chosen for transfer by the piece-selection policy -- in our case, the transferable set will either be empty or a singleton set.

\begin{algorithm}[!hbt]
    \caption{Transferable-set of swarm-based RFwPMS.}
    \label{alg:transferable_set}
    \begin{algorithmic}[1]
    \renewcommand{\algorithmicrequire}{\textbf{Input:}}
    \renewcommand{\algorithmicensure}{\textbf{Output:}}
    \REQUIRE $\statex, W, S, \wh{T}$.
    \ENSURE Transferable-set, $A(\statex, \wh{T}, W, S)$.
    \STATE Set $A \leftarrow \emptyset$.
    \STATE Set $H_1 \la \left( \wh{T} \cap R_{W}(\statex) \right) \setminus S$.
    \STATE Set $H_2 \la \left( \wh{T} \cap W \right) \setminus S$.
    \STATE Set $H_3 \la \left(\wh{T} \cap \mcl{F}_W\right) \setminus \left(S \cup W \right)$.
    \IF {$ H_1 \ne \emptyset $}
        \STATE Choose $r$ randomly from $\setrares{\wh{T}}$.
        \STATE Set $A \ra A \cup \{r\}$.
    \ELSIF{$ H_2 \ne \emptyset$}
        \STATE Choose $n$ randomly from $\left( \wh{T} \cap W \right) \setminus S$.
        \STATE Conduct Bernoulli trial with success probability $\zeta_W^{(n)}$.
        \IF{Success}
            \STATE Set $A \leftarrow A \cup \{n\}$.
        \ELSE
            \IF{$ H_3 \ne \emptyset$}
                \STATE Choose $e$ randomly from $H_3$.
                \STATE Set $A \la A \cup \{e\}$.
            \ENDIF
        \ENDIF
    \ELSE
        \IF{$H_3 \ne \emptyset$}
                \STATE Choose $e$ randomly from $H_3$.
                \STATE Set $A \la A \cup \{e\}$.
        \ENDIF 
    \ENDIF
    \RETURN $A$.
    \end{algorithmic}
\end{algorithm}

In order to extend our rules to the seed, we assume that the seed is a $(\dg,\mathcal{F})$-type peer\footnote{For example, $\dg$ can be $\mcl{F}\cup\{K+1\}$.} with the ally-set $\mcl{W}_{\dagger}=\mcl{W}$. Thus, the set of rare pieces when the seed push-contacts a normal peer is given by
\begin{align*}
\setrares{\mcl{F}}(\statex) = \argmin \limits_{j \in \rw (\statex) \setminus S} \nmbr{j}{W}(\statex).
\end{align*}

\textit{Remark}: Note that when there is no rare piece at offer, a non-rare piece $n \in (T \cap W) \setminus S$ is transferred only \textit{probabilistically}, so swarm-based RFwPMS is not a work-conserving scheme. 

\textit{Remark}: With rules a), b) and c), we note that each swarm maintains its own chunk-table and also prioritizes its primary pieces over other pieces of the master-file. For these two reasons, we call our policy, ``\textit{swarm-based}'' RFwPMS.

\subsection{Process Description / Suitable Bounds on Transition Rates}
For the sake of notational simplicity, from hereon, we will write function $f(\cdot)$ as $f$ when the dependence on the argument is clear. At any given current state $\statex$, the next state of the network depends solely on $\statex$ because the piece-selection policy solely depends on $\statex$ and the new arrivals (empty peers) are determined by an independent Poisson process. Hence, the evolution of the network described by the process $\{\statex(t): t \ge 0\}$ is a continuous-time, time-homogeneous, and irreducible (easily shown) Markov chain with state space,
\begin{align*}
\mcl{S} \defeq \mathbb{Z}_{+}^{\sum \limits_{W \in \mcl{W}} \left| \left\{S : S \subseteq \mcl{F}_{W} \text{ and } W \setminus S \ne \emptyset \right\} \right|  }. \numberthis \label{eq:statespace}
\end{align*}
Given a state $\statex$, there are different events that can lead to state-transitions, namely, the arrival of a new empty peer, the download of a single-piece or a (two-sided) piece-exchange (download of two pieces simultaneously as a result of some tit-for-tat contact). Here, we do not list the transition rates of (two-sided) piece-exchanges: later in Section \ref{sec:analysis}, we will see that with our choice of Lyapunov function, such transitions can be viewed as two separate single-piece-download events. The result of each event is given below.
\begin{enumerate}
\item[a)] \textit{Arrival of an Empty-peer}:
Recall that a swarm-$W$ peer with an empty cache enters the network according to a Poisson process of rate $\lambda_W$. This results in a unit increase in the number of swarm-$W$ peers with no pieces. Let us denote this transition by $\{(\emptyset,+)_W\}$ and its corresponding rate by $q_W^{(\emptyset+)}$. Then
\begin{align*}
q_W^{(\emptyset+)} = \lambda_W.\numberthis\label{eq:qwemptysetplus}
\end{align*}
\item[b)] \textit{Download of Piece of Primary Interest}: 
Consider the event that a $\ws$-peer missing piece $i \in W$ downloads piece $i$. The necessary condition for this event is that $\ws$-peer gets push-contacted by a $\vt$-peer such that $W \in \mcl{W}_{V}$ and $i \in T$. There are three distinct ways in which this can happen.
\begin{enumerate}
\item[i)] The $\ws$-peer gets push-contacted by the $\vt$ peer as a result of a tit-for-tat contact initiated by the $\vt$-peer. By the superposition and thinning properties of Poisson processes, the rate at which an ally peer of swarm $W$ with piece $i$ makes a tit-for-tat contact with a $\ws$-peer is
\begin{align*}
\left( \mu \sum_{\substack{V:W\in\mcl{W}_V,\\T:i\in T }}\xvt \right) \cdot (L- \indunchoke{1}) \frac{\xws}{\x}\frac{\x}{\x-1}.
\end{align*}
\item[ii)] The $\ws$-peer gets push-contacted by the $\vt$-peer as a result of an optimistic-unchoke (initiated by the $\vt$-peer). The rate at which a $\ws$-peer gets contacted in an optimistic-unchoke by an ally-peer having piece $i$ is given by
\begin{align*}
&UL\cdot \frac{\xws}{\x} + \left(\indunchoke{1} \hat{\mu}\sum_{\substack{V:W\in\mcl{W}_V,\\T:i\in T }}\xvt \right) \\
&\hspace{120pt} \times \frac{\xws}{\x}\frac{\x}{\x-1}.
\end{align*}
\item[iii)] The $\ws$-peer gets push-contacted by the $\vt$-peer as a result of a tit-for-tat contact initiated by the $\ws$-peer. The rate at which a $\ws$-peer makes a tit-for-tat contact with an ally-peer that has piece $i$ is
\begin{align*}
\left( \mu \xws \right) \cdot (
L-\indunchoke{1}) \frac{\sum_{\substack{V:W\in\mcl{W}_V,\\T:i\in T }}\xvt}{\x}\frac{\x}{\x-1}.
\end{align*}
\end{enumerate}
Let us denote by $\Gamma_{W}^{(i)}=\Gamma_{W}^{(i)}(\statex)$ the aggregate rate at which an ally peer of swarm-$W$ with piece $i$ \textbf{push-contacts} some other peer in the network. The exact form of $\Gamma_{W}^{(i)}$ is complicated, however, from i), ii), iii) and our descriptions of tit-for-tat and optimistic-unchoke, we can bound it both from above and below. Let
\begin{align*}
\udl{\Gamma}_{W}^{(i)}(\statex) &\defeq LU + \Delta_p\xi_1 
\left(\nmbr{i}{W}+\dnmbr{i}{W}\right)\\
&\ge LU + \Delta_p \nmbr{i}{W}. \numberthis\label{eq:Gammalower2}
\end{align*}
and
\begin{align*}
\ov{\Gamma}_{W}^{(i)} (\statex)&\defeq LU + \Delta_1\xi_1 
\left(\nmbr{i}{W}+\dnmbr{i}{W}\right)\\
&\le LU + 2\Delta_1 \left(\nmbr{i}{W}+\dnmbr{i}{W}\right),\numberthis\label{eq:Gammaupper2}
\end{align*}
where, for brevity, we have introduced $\xi_1\defeq \xi_1(\statex) = \frac{\x}{\x-1} \in (1,2]$ (for all $\x\ge 2$), and $\Delta_t\defeq 2(L-\indunchoke{1})\mu t +\indunchoke{1}\hat{\mu}$. It is clear that
\begin{align*}\label{eq:Gammabounds}
\udl{\Gamma}^{(i)}_{W} \le \Gamma^{(i)}_{W} \le \ov{\Gamma}^{(i)}_{W}. \numberthis 
\end{align*}
and
\begin{align*}\label{eq:Gammaratios}
\frac{\Gammaupper{W}{i}}{\Gammalower{W}{i}}&=\frac{LU+\Delta_1 \xi_1\left(\nmbr{i}{W}+\dnmbr{i}{W}\right) }{LU+\Delta_p\xi_1 \left(\nmbr{i}{W}+\dnmbr{i}{W}\right) }\\
&\le\xi_2\defeq
\begin{cases}
1 + \frac{2(L-1)\mu+\hat{\mu}}{2(L-1)\mu p +\hat{\mu}} &\text{if $Y_{opt}=1$,}\\
1 + \frac{1}{p} &\text{if $Y_{opt}=0$.}\\
\end{cases}\numberthis
\end{align*}
Here, $\xi_2=\xi_2(Y_{opt},L,p,\mu,\hat{\mu})$ is a constant that depends on the model parameters.

After the $\ws$-peer has downloaded piece $i$, depending on $S$, it will either remain in the network or leave it immediately. Therefore, we have two cases:
\begin{enumerate}
\item[i)] if $W \setminus S \supsetneq \{i\}$, the peer stays in the system. Let us denote this transition by $\{(S,i+)_W\} $ and the corresponding rate by $q_W^{(S,i+)} $. Based on the description of swarm-based RFwPMS,  $q_W^{(S,i+)} $ depends on whether $i\in R_W$ or $i\in\rwc$. If $i=r$ is some rare piece in $R_W$, then we have
\begin{align*}\label{eq:qwsrplus}
&\frac{\xws}{\x}\udl{\Gamma}_W^{(r)}\xi_2 \ge \frac{\xws}{\x}\ov{\Gamma}_W^{(r)} \ge q_W^{(S,r+)}\ge\dots\\
&\hspace{10pt} \ge \frac{\xws}{\x} \left[ \frac{LU \11{\{r \in \setrares{\mcl{F} }\}}}{|\setrares{\mcl{F}}|} \right. \\
&\hspace{10pt} \left. + \Delta_p\xi_1 \sumvtr \frac{\11{ \{r \in \setrares{T} \}}}{|\setrares{T}|}\xvt \right], \numberthis
\end{align*}
where the two indicator terms ensure that piece $r$ is transferred {\it only if it is the rarest of all the available rare pieces}. Importantly, when the chunk-distribution in swarm-$W$ is uniform i.e., $\nmbrn{W} =\nmbrr{W}$, by definition, $R_W=W$ and the above expression assumes a more tractable form. Therefore, for each swarm-$W$, we partition the state space into two regions, namely  $\sw1$ and  $\sw2$, where $\sw1=\{\statex \in \mcl{S}: R_W(\statex) \subsetneq W\}$, and $\sw2=\{\statex \in \mcl{S}: R_W(\statex) = W\}$.\footnote{When $\statex \in \sw2 $, both the indicator terms in \eqref{eq:qwsrplus} evaluate to 1.} 
If $i=n$ is some non-rare piece in $\rwc$, then
\begin{align*}\label{eq:qwsnplus}
\udl{q}_W^{(S,n+)} 
\le q_W^{(S,n+)} \le \frac{\xws}{\x}\ov{\Gamma}_W^{(n)}\zeta_W^{(n)}\le
\frac{\xws}{\x}\udl{\Gamma}_W^{(n)}\xi_2\zeta_W^{(n)},\numberthis
\end{align*}
where
\begin{align*}\label{eq:qwsnplus2}
\udl{q}_W^{(S,n+)} &\defeq \frac{\xws}{\x}\left[\frac{LU \11{\{R_W\setminus S =\emptyset\}}}{|W\setminus S|} \right.\\
&\hspace{-20pt} \left. +\Delta_p\xi_1\sum_{\substack{V:W\in\mcl{W}_V,\\T:n\in T}} \frac{\xvt \11{[\trs=\emptyset]}}{|(T\cap W)\setminus S|} \right]\zeta_W^{(n)}.\numberthis
\end{align*}
\item[ii)] If $W \setminus S= \{i\}$, the only piece of file-$W$ that is missing with the $\ws$-peer is piece $i$. Consequently, for both $\statex \in \sw1$ and $\statex \in \sw2$, piece $i$ is the rarest piece transferable to the $\ws$-peer, which leaves the network immediately upon downloading it. Let us denote this transition by $\{( S,i-)_W\}$ and its corresponding rate by $q_W^{( S,i-)} $.
Then,
\begin{align*}\label{eq:qwsiminus}
\frac{\xw}{\x}\frac{\xws}{\xw}  \udl{\Gamma}_W^{(i)}a_W^{(i)} \le q_W^{(S,i-)} \le \frac{\xw}{\x}\frac{\xws}{\xw}\udl{\Gamma}_W^{(i)}\xi_2 a_W^{(i)},\numberthis
\end{align*}
where, for brevity, we introduced $a_W^{(i)}=a_W^{(i)}(\statex)\defeq \zeta_W^{(i)}\11{\{i\in\rwc\}}+\11{\{i\in R_W\}}$.
\end{enumerate}

\item[c)] \textit{Download of Extra Piece}: Recall that extra pieces are preferred only when no pieces from the file of primary interest are transferable. We shall see that the drift analysis of our Lyapunov function does not depend on the download of extra pieces. Consequently, we skip listing the associated rates.
\end{enumerate}

\subsection{Last-Piece-Syndrome vs. Tit-for-Tat}
While it is clear that a tit-for-tat mechanism helps combat the \textit{last-piece-syndrome}, the issue of quantifying its efficacy via a suitable model was left as an open problem in \cite{mps}. Our stochastic model described above is partly motivated by this question. In this subsection, we show that introducing tit-for-tat does not prevent instability if a work-conserving piece-selection policy (like RF and RN) is used.
\begin{prop}\label{prop:fps}
Consider the multi-swarm network model as described in Section \ref{sec:model}, with a hard tit-for-tat mechanism ($p=0$) without optimistic-unchoking ($Y_{opt}=0$). If $\sum_{W \in \mcl{W}} \lambda_W > LU $, then the network is unstable under any piece-selection policy.
\end{prop}

\begin{IEEEproof}
We divide the network into two subsystems, where subsystem-1 consists of empty-peers and subsystem-2 consists of the rest of the peers. The two subsystems are connected in tandem as shown in \figurename \ref{fig_lemma_fps}.
\begin{figure}[!t]
	\centering
	\includegraphics[width=\linewidth]{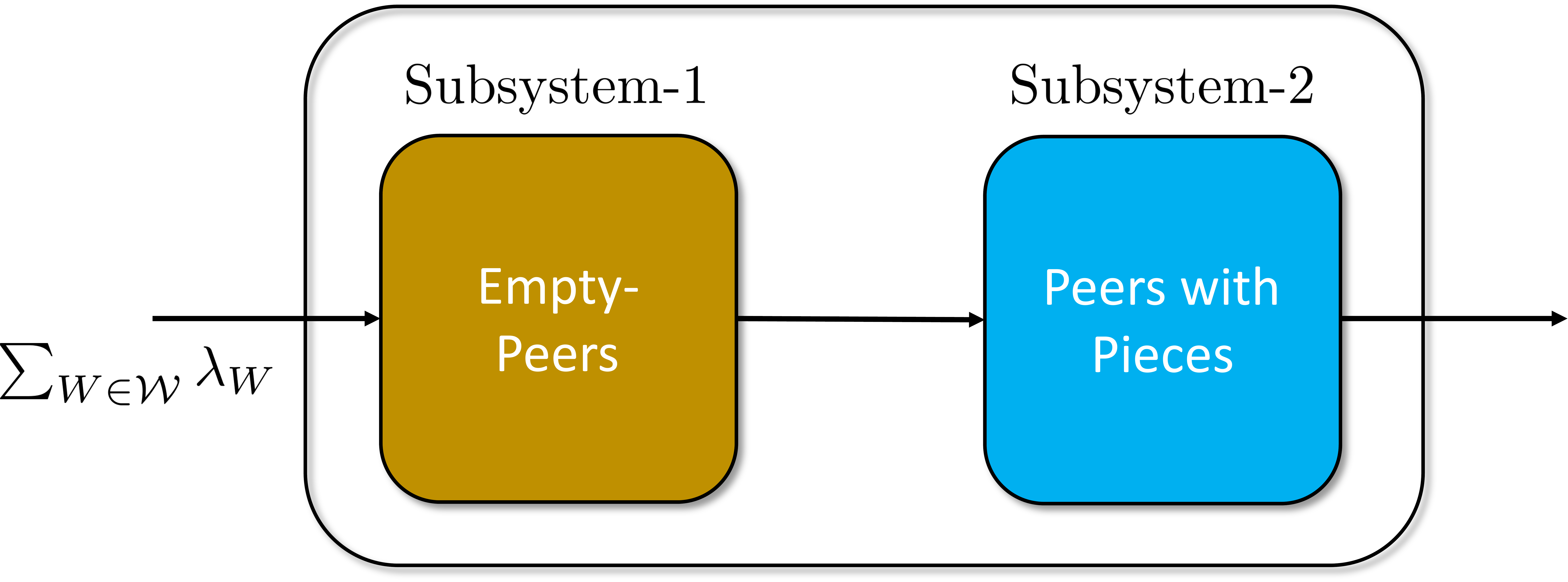}
	\caption{Subsystems 1 and 2 in Tandem.}
	\label{fig_lemma_fps}
\end{figure}
The arrival rate of incoming peers in subsystem 1 is $\sum_{W \in \mcl{W}} \lambda_W$ and the departure rate is upper-bounded by $LU$ (at each tick of the fixed seed's contact-link, at most $1$ empty peer can depart from subsystem-1 and there are $L$ such links). It is well-known that subsystem-1 is unstable if $\sum_{W \in \mcl{W}} \lambda_W > LU$ rendering the network unstable. Since this instability is manifested by the build-up of empty peers, we call this phenomenon, the \textit{first-piece syndrome (FPS)}. \end{IEEEproof}

\begin{figure}[!b]
	\centering
	\includegraphics[width=\linewidth]{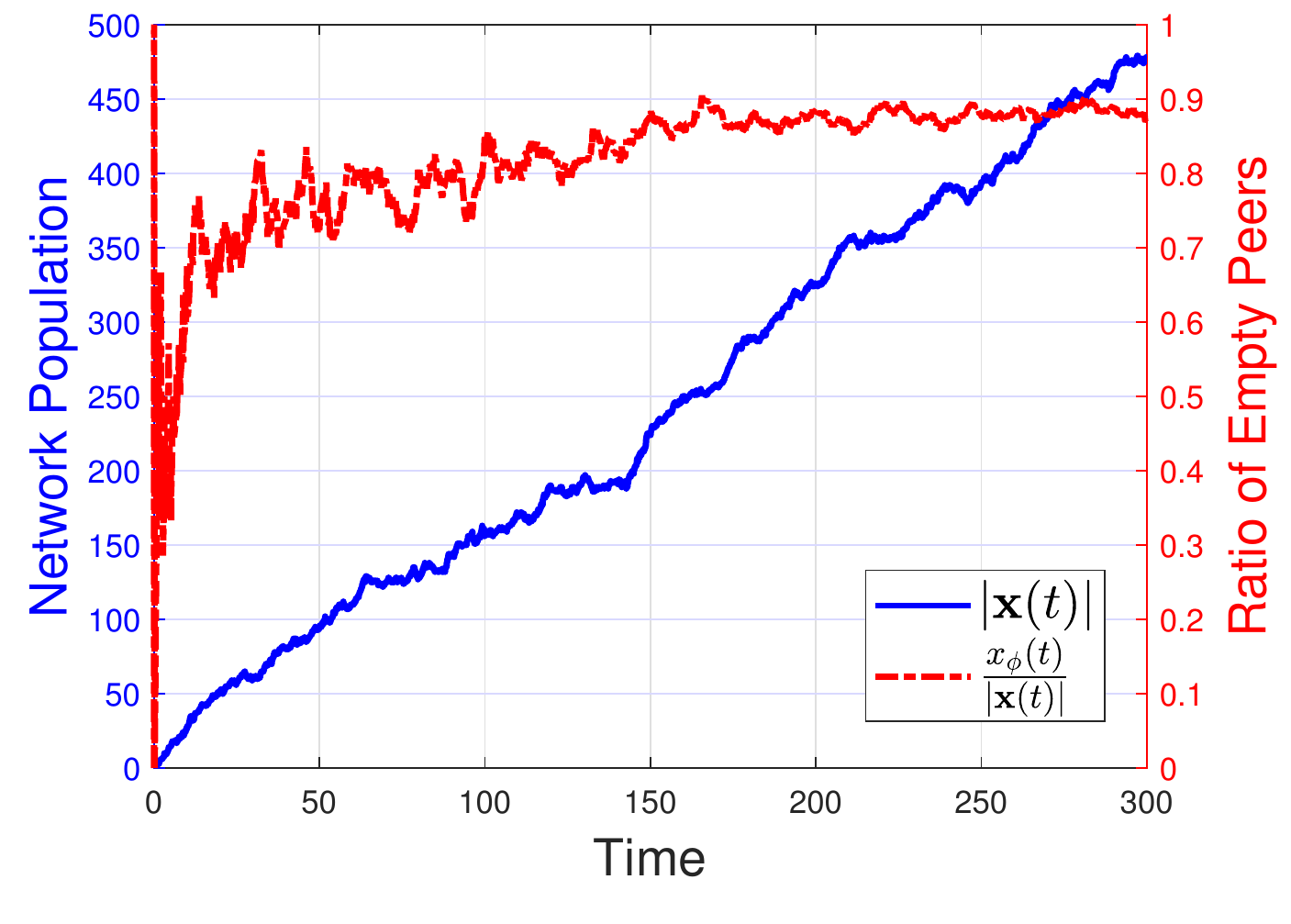}
	\caption{First Piece Syndrome in a Single-Swarm Network with Hard Tit-for-Tat ($p=0$) and no Optimistic-unchoking ($Y_{opt}=0$). Assignment of other model parameters was $W=[10]$, $\lambda_W=4$, $U=\mu=1$, $\wh{\mu}=\frac{1}{3}$, $L=3$, $p=Y_{opt}=0$.} \label{fig_hard_tt}
\end{figure}

\figurename \ref{fig_hard_tt} shows the first-piece-syndrome manifested in a single-swarm network using a hard tit-for-tat mechanism without optimistic-unchoking.

\begin{prop}\label{prop:lps}
Consider the single-swarm network model, as presented in Section \ref{sec:model}, (a) with a soft tit-for-tat mechanism ($p>0$) or, (b) with a hard tit-for-tat mechanism ($p=0$) and optimistic-unchoking ($Y_{opt}=1$). If a work-conserving piece-selection policy is used, then the network is unstable if $\lambda > LU $.
\end{prop}
\begin{IEEEproof}
The method of the proof remains the same as in \cite[Proposition 2.1 (i)]{mps}. In soft tit-for-tat mechanism, a young peer can get a piece from a one-club peer with probability at least $p >0$ whereas in a hard tit-for-tat mechanism with optimistic-unchoking, it can get a piece from a one-club peer via the optimistic-unchoke. In either case, by taking the initial size of the one-club peers suitably large, it can be shown that the last-piece-syndrome event has a positive probability, establishing the transience of the system.
\end{IEEEproof}

Propositions \ref{prop:fps} and \ref{prop:lps} motivate the design of a piece-selection policy that is provably stable in file-sharing networks that employ 
the ``Soft Tit-for-Tat'' and the ``Hard Tit-for-Tat with Optimistic-Unchoking'' mechanisms.
\section{Stability of swarm-based RFwPMS in the Multi-Swarm}\label{sec:analysis}
In this section, we present our main result on the stability of swarm-based RFwPMS. The proof is established using the Foster-Lyapunov theorem \cite{hajekbook,srikantbook}; see Proposition \ref{prop:foster_kingman} in Appendix \ref{sec:appendix:fosterlyapunov}.

\begin{thm}\label{thm:stability}
For the multi-swarm model with non-persistent peers as described in Section \ref{sec:model}, swarm-based RFwPMS is stable over the parameter region
$$\left\{U > 0, \bsl{\lambda} > \mathbf{0}, L\ge 1, \Delta_p>0
, \text{ and } \kw \ge 2 \ \forall\  W \in \mcl{W}\right\}.$$
\end{thm}
\emph{Remark}: Here, $\Delta_p>0$ includes
\begin{itemize}
\item ``Soft Tit-for-Tat'': $p\mu>0$ and ($L=1,Y_{opt}=0$ OR $L>1$),
\item ``Hard Tit-for-Tat with optimistic-unchoking'': $p=0$ AND ($L\ge 1, Y_{opt}=1, \hat{\mu}>0$), and
\item Stability result of our earlier work \cite{ourinfocom}: $Y_{opt}=1, L=1, \hat{\mu}>0$.
\end{itemize}

The proof of Theorem \ref{thm:stability} uses a novel Lyapunov function that is based on the refinement of our earlier intuition in \cite{ourinfocom}. The key ideas of the proof are as follows:
\begin{itemize}
\item 
In the case of large total-mismatch, extract negative drift from the download of rare-pieces, and use the non-rares sharing factor to upper-bound the positive component of the drift from the download of non-rare pieces; (this is proved in Lemma \ref{lem:rfrn} in Appendix \ref{sec:appendix:rfrn} and Lemma \ref{lem:upperboundnnonrares} in Appendix \ref{sec:appendix:upperboundnonrares}); and 
\item 
When the total-mismatch is not too large, extract negative drift from the download of all primary pieces (this is incorporated into the proof of Lemma \ref{lem:getrid} in Appendix \ref{sec:appendix:getrid}).
\end{itemize}

The Lyapunov function we use is given by
\begin{align*}\label{eq:stab:v}
V(\statex) &\defeq \sum_{W\in\mcl{W}} V_W (\statex), \numberthis
\end{align*}
where
\begin{align}\label{eq:stab:vwdef}
\begin{split}
V_W(\statex) &\defeq V_{W,1}(\statex) + V_{W,2}(\statex)+ V_{W,3}(\statex),\\
V_{W,1}(\statex) &\defeq \left(\left( 
\totalmismatch{W}-\eta\nmbrn{W}	
\right)^+\right)^2,\\ 
V_{W,2}(\statex) &\defeq C_W^{(1)} \left(\kw \xw-\pw \right),\\
V_{W,3}(\statex) &\defeq C_W^{(2)} \left( \mw-\pw \right)^+.\footnotemark
\end{split}
\end{align}
Here, $\eta\in(0,1)$ and $\cw1, \cw2, \mw \in \mb{R}_{>0}$ are suitably large constants. Instead of presenting the relations satisfied by $\cw1$, $\cw2$, and $\mw$ in the onset, we will go through the analysis and present them naturally at places where they are needed.\footnotetext{For all $d\in\mb{R}$, we use the standard notation $(d)^+=\max\{d,0\}$.}

\textit{Intuition}: The exact details of the Lyapunov function given by \eqref{eq:stab:v} and \eqref{eq:stab:vwdef} are complicated. However, the following lines of thought will prove to be of substantial value in following the proof of Theorem \ref{thm:stability}.
\begin{itemize}
\item
Given each swarm maintains its own chunk-counts, $V$ being the superposition of $V_W$'s, reflects the intuition that the network shall be stable if and only if each swarm is stable.

\item 
To develop some intuition behind the choice of $V_W$, let us consider a multi-swarm network in which each swarm's ally-set contains that swarm only (later in Section \ref{sec:discussion}, we call this the setting of selfish swarms). In that case, the non-rares sharing factor $\zeta_W^{(n)}$ (assuming $\beta_W>0$) takes the form $\exp\left( -\frac{\ov{m}_W}{\beta_W\kw}   \right)$. 

\begin{itemize}
\item Given that swarm-based RFwPMS would allow the download of non-rare pieces with high probability whenever the largest-mismatch $\ov{m}_W \ge \frac{\totalmismatch{W}}{\kw-1}$ is relatively small, it is reasonable, as a starting point, to guess a term of the form $(\totalmismatch{W})^2$. Unfortunately, as it turns out, with the bounds that we have derived, working with $(\totalmismatch{W})^2$ fails in satisfying the unit-transition drift stability conditions. The primary reason behind this is the fact \textit{the rarity of the rarest piece and the abundance of the non-rare pieces are not related to each other}, that is, knowing that $\nmbrn{W}$ is large does not necessarily imply that the largest-mismatch $\ov{m}_W$ is large as well. We circumvent this issue by using $V_{W,1}$ (see \eqref{eq:stab:vwdef}) as a ``proxy'' for $(\totalmismatch{W})^2$. Specifically, by using $\left(\left(\totalmismatch{W}-\eta\nmbrn{W}\right)^+\right)^2$ instead of $(\totalmismatch{W})^2$, we avoid any drift penalization (i.e., positive drift) whenever $\totalmismatch{W}$ is less than $\eta\nmbrn{W}$. Therefore, in essence, whenever $\totalmismatch{W}<\eta\nmbrn{W}$, as concerns $V_{W,1}$, our Lyapunov function discards the distinction between rare and non-rare pieces. On the other hand, if $\totalmismatch{W}\ge \eta\nmbrn{W}$, this implies that $\mismatchr{W} \ge \frac{\eta}{(\kw-1)}\nmbrn{W}$, and now, the abundance of non-rare pieces, i.e., large $\nmbrn{W}$, also implies a high largest-mismatch $\ov{m}_W$. This will trigger our non-rares sharing factor $\zeta_W^{(n)}$ which will ensure that the positive component of the drift from the non-rare pieces decays to zero with increasing $\nmbrn{W}$.

\item In contrast to $V_{W,1}$, the second term, $V_{W,2}$, promotes the download of every piece of file-$W$. The intuition behind including $V_{W,2}$ is to ensure a negative drift is generated from every piece download. This helps us generate sufficient negative drift in the case when $V_{W,1}$ is inactive, i.e., when $\totalmismatch{W}<\eta\nmbrn{W}$. Furthermore, an important technical purpose that it serves is to ensure that $V$ has finite level-sets.

\item Finally, the last term $V_{W,3}$ handles all the states in which swarm-$W$ is seriously piece-deprived (states in which very few pieces of file-$W$ are present with swarm-$W$).
\end{itemize}
\end{itemize}

Next, we present Lemmas \ref{lem:chunkcounts} and \ref{lem:finitelevelsets} together which establish that $V$ has finite level-sets.

\begin{lem}[Bounds on Chunk-Counts]\label{lem:chunkcounts}
For all $ \statex $, the total chunk-count in swarm-$W$ is upper-bounded by $(\kw-1)\xw$. Consequently, the fraction of swarm-$W$ peers who have the rarest piece, i.e., $ \udl{\pi}_W $, is upper-bounded by $(\kw -1)/\kw$ and the fraction of swarm-$W$ peers who are missing the rarest piece is lower-bounded by $1/\kw$.
\end{lem}
\begin{IEEEproof}
For all $ \statex $, we can lower bound the total chunk-count in swarm-$W$ as follows,
\begin{align*}
\kw \nmbrr{W} &\le \sum_{i\in W} \nmbr{i}{W} = \pw
= \sum_S \xws |S \cap W|
\le \dots\\
&\le (\kw - 1) \sum_S \xws
=(\kw - 1) \xw.
\end{align*}
Hence, $\udl{\pi}_W \le \frac{ \kw -1}{ \kw }$ and $1 - \udl{\pi}_W \ge \frac{1}{ \kw }$. The second inequality is true because $W\setminus S \ne \emptyset$.
\end{IEEEproof}

\begin{lem}[Finite Level-Sets of $V$]\label{lem:finitelevelsets}
$V(\statex) \ra \infty $ as $ \x \ra \infty $.
\end{lem}
\begin{IEEEproof}
We have $\pw\le(\kw-1)\xw$ for every $\statex$. So, $V_W(\statex)\ge V_{W,2}(\statex)\ge \cw1\xw\ra \infty$ as $\xw\ra\infty$. Since $\x \ra \infty$ only if $\xw \ra \infty$ for some $W \in \mcl{W}$, it follows that $V(\statex) \ra \infty$ as $\x \ra \infty$. Consequently, for all $C \in \mb{R}_{+}$, the set $\{\statex : V(\statex) \le C \}$ is finite.
\end{IEEEproof}

\subsection{Upper Bounds on Potential Changes}\label{subsec:vupperbounds}
Having established the finite level-sets of $V$, we evaluate the potential change for each possible transition. Note that, with our choice of $V$, any transition that occurs in swarm-$W$, affects the term $V_W$ only.

\subsubsection{Arrival of an Empty Peer}
The arrival of a $(W,\emptyset)$-peer results in a unit increase in $\xw$ but does not affect any chunk-count. Therefore, $\totalmismatch{W},\nmbrn{W},\pw$ stay the same. The potential change as a result of this transition, denoted by ${\Delta V}_W^{(\emptyset,+)}$, is component-wise given by
\begin{align*}
{\Delta V}_{(W,1)}^{(\emptyset,+)} &=0.\\
{\Delta V}_{W,2}^{(\emptyset,+)} &= \kw\cw1.\\
{\Delta V}_{(W,3)}^{(\emptyset,+)} &=0.
\end{align*}
Overall,
\begin{align*}\label{eq:stab:vwemptysetplus}
{\Delta V}_W^{(\emptyset+)} &=  \kw\cw1. \numberthis
\end{align*}

\subsubsection{Download of a Single Piece $i\in W$ with Peer-Departure}
Assume that a $(W,S)$-peer with $W\setminus S=\{i\}$ downloads piece $i$ and leaves the system. The departure causes a unit-decrease in $\xw$; a unit-decrease in chunk-count of every piece $j\in W\setminus\{i\}$. The chunk-count of piece $i$, i.e., $\nmbr{i}{W}$, stays the same. Therefore, $\pw$ decreases by $\kw-1$. The resulting potential change, denoted by ${\Delta V}_W^{(S,i-)}$, can be component-wise upper-bounded as
\begin{align*}
&{\Delta V}_{W,2}^{(S,i-)} \le -\cw1.\\
&{\Delta V}_{W,3}^{(S,i-)} \\ &=\cw2\left[\left(\mw-\pw+(\kw-1)\right)^+ -\left(\mw-\pw\right)^+ \right]\\
&\le\cw2(\kw-1)\11{\{\mw+\kw-1 \ge \pw\}}\\
&\le \cw2\kw \11{\{\mw+2\kw\ge \pw\}}
\defeq I_W^{(2)}(\statex).
\end{align*}
${\Delta V}_{W,1}^{(S,i-)}$ depends on whether $i$ is a rare or non-rare piece. When $i=r$ is some rare-piece, ${\Delta V}_{W,1}^{(S,r-)}$ further depends on whether swarm-$W$'s chunk-distribution is uniform or non-uniform, i.e., whether $\statex\in\sw1$ or $\statex\in\sw2$. If $\statex\in\sw1$, then $\nmbr{i}{W}\ne \nmbrn{W}$, therefore the departure reduces both the total mismatch and the highest chunk-count by 1. This gives,
\begin{align*}
&{\Delta V}_{W,1}^{(S,r-)}\\
&=
\left(\left(\totalmismatch{W}-\eta\nmbrn{W}-(1-\eta) \right)^+\right)^2-
\left(\left(\totalmismatch{W}-\eta\nmbrn{W}   \right)^+\right)^2\\
&\le\begin{cases}
(1-\eta)^2-2(1-\eta) \left(\totalmismatch{W}-\eta\nmbrn{W} \right) &\begin{array}{@{}l@{}} \text{if $\totalmismatch{W}\ge$} \\ (1-\eta)+\eta\nmbrn{W}   \end{array}\\
0 &\text{otherwise}.
\end{cases}
\\
&\le\begin{cases}
2(1-\eta)^2-2(1-\eta) \left(\totalmismatch{W}-\eta\nmbrn{W} \right) &\begin{array}{@{}l@{}} \text{if $\totalmismatch{W}\ge$} \\ 2(1-\eta)+\eta\nmbrn{W}   \end{array}\\
0 &\text{otherwise}.
\end{cases}
\end{align*}
If $\statex\in\sw{2}$, the download of piece $r$ makes it the only non-rare piece in the next state with its chunk-count equal to $\nmbrn{W}$. The chunk-count of every other piece decreases to $\nmbrn{W}-1$. The total mismatch $\totalmismatch{W}$ changes from 0 to $\kw-1$. Consequently,
\begin{align*}
{\Delta V}_{W,1}^{(S,r-)} &=
\left(\left(\kw-1-\eta\nmbrn{W}\right)^+\right)^2-
\left(\left(-\eta\nmbrn{W} \right)^+\right)^2\\
&\le\kw^2\le 4\kw^2.
\end{align*}
Combining the two cases for $i=r$, we have
\begin{align*}
&{\Delta V}_{W,1}^{(S,r-)} \\
&\le \begin{cases}
2(1-\eta)^2-2(1-\eta) \left(\totalmismatch{W}-\eta\nmbrn{W} \right) \le 0 
&\begin{array}{@{}l@{}}
\text{if $\totalmismatch{W}\ge$}\\
2(1-\eta) + \eta \nmbrn{W},
\end{array}\\
4\kw^2 &\text{otherwise}.
\end{cases}\\
&\defeq \udl{\psi}_W(\statex).
\end{align*}
When $i=n$ is some non-rare piece\footnote{Non-rare pieces in swarm-$W$ exist if and only if $\statex\in\sw1$.}, then the highest chunk-count $\nmbrn{W}=\nmbr{i}{W}$ stays the same and the total mismatch $\totalmismatch{W}$ increases by $\kw-1$. Overall, this causes an increment of $\kw-1\le\kw$ in $\totalmismatch{W}-\eta\nmbrn{W}$. Therefore, 
\begin{align*}
&{\Delta V}_{W,1}^{(S,n-)} \\
&=
\left(\left(\totalmismatch{W}-\eta\nmbrn{W}+\kw \right)^+\right)^2-
\left(\left(\totalmismatch{W}-\eta\nmbrn{W}   \right)^+\right)^2\\
&\le \begin{cases}
\kw^2 + 2\kw\left(\totalmismatch{W}-\eta\nmbrn{W} \right) &\text{if $\totalmismatch{W} -2(1-\eta) \ge \eta\nmbrn{W}$},\\
(\kw+2)^2 &\text{otherwise}.
\end{cases}\\
&\le 
\begin{cases}
\kw^2 +2\kw\left(
\totalmismatch{W}-\eta\nmbrn{W}
\right) &\text{if $\totalmismatch{W}-2(1-\eta) \ge \eta\nmbrn{W}$},\\
4\kw^2 &\text{otherwise}\\
\end{cases}\\
&\defeq \ov{\psi}_W(\statex).
\end{align*}
To utilize the negative drift from the download of swarm-$W$'s rare pieces, we partition the state-space into regions $\mcl{R}_W^{1}=\{\statex:\totalmismatch{W}-2(1-\eta)\ge\eta\nmbrn{W}\}$ and  $\mcl{R}_W^2=\mcl{S}\setminus\mcl{R}_W^1$. (Note that $\mcl{R}_W^{1} \subseteq \sw{1} $ and $\mcl{R}_W^{2} \supseteq \sw{2}$). We can now write
\begin{align*}\label{eq:stab:vwsiminus}
&{\Delta V}_W^{(S,i-)}\\
&\le \ov{\psi}_W \11{\{i\in\rwc\}} + \udl{\psi}_W \11{\{i\in \rw\}} -\cw1 + I_W^{(2)}\\
&=\11\{\statex\in\regrw{W}{2}\} 4\kw^2+\dots\\
&\hspace{10pt}+\11\{\statex\in\regrw{W}{1}\}\left(\ov{\psi}_W\11\{i\in\rwc\} +\udl{\psi}_W\11\{i\in\rw\} \right)-\dots\\
&\hspace{10pt} -\cw1+I_W^{(2)}\\
&\defeq \Psi_W^{(i-)}(\statex).\numberthis
\end{align*}

\subsubsection{Download of a Single Piece $i\in W$ without Peer Departure}
Here, a $(W,S)$-type peer with $W\setminus S\supsetneq \{i\}$ downloads piece $i$ and remains in the system. The total number of swarm-$W$ peers remains the same and the chunk-count of piece $i$ (and thus $\pw$) gets incremented by 1. The resulting potential change, denoted by $\Delta V_W^{(S,i+)}$ is component-wise upper-bounded as
\begin{align*}
{\Delta V}_{W,2}^{(S,i+)} &=-\cw1,\\
{\Delta V}_{W,3}^{(S,i+)} &=\cw2\left[\left(\mw-\pw-1\right)^+-\left(\mw-\pw\right)^+\right]\\
&\le
-\cw2\11{\{\mw-1 \ge \pw\}}\\ 
&\le
-\cw2\11{\{\mw-2 \ge \pw\}}
\defeq -I_W^{(1)}(\statex).
\end{align*}
Like ${\Delta V}_{W,1}^{(S,i-)}$, one can show that ${\Delta V}_{W,1}^{(S,i+)}$ is upper-bounded by
$\ov{\psi}_W\11{\{i\in\rwc\}}+\udl{\psi}_W\11{\{i\in \rw\}}$. Therefore,
\begin{align*}\label{eq:stab:vwsiplus}
&{\Delta V}_W^{(S,i+)}\\
&\le \ov{\psi}_W\11{\{i\in\rwc\}}+\udl{\psi}_W\11{\{i\in \rw\}}-\cw1-I_W^{(1)}\\
&=\11\{\statex\in\regrw{W}{2}\} 4\kw^2+\dots\\
&\hspace{10pt} +\11\{\statex\in\regrw{W}{1}\}\left(\ov{\psi}_W\11\{i\in\rwc\}+\udl{\psi}_W\11\{i\in\rw\} \right)\\
&\hspace{10pt} -\cw1 -I_W^{(1)}\\
&\defeq \Psi_W^{(i+)}(\statex).\numberthis
\end{align*}
Looking at \eqref{eq:stab:vwsiplus} and \eqref{eq:stab:vwsiminus}, we note that by choosing $\cw1>4\kw^2$, say $\cw1\ge 8\kw^2$, we ensure that the term $\11\{\statex\in\regrw{W}{2}\}4\kw^2-\cw1$ is non-positive. Intuitively, this means that over the region $\regrw{W}{2}$, as concerns the potential-change, our upper-bounds treat the non-rare and rare pieces in the same way.

\subsubsection{Download of a Single Extra Piece}
It can be observed that for any swarm-$W$, no potential change is induced in $V_W$ by the download of its extra pieces.

\subsubsection{Two-Sided Piece Exchanges}
Now, we consider the state-transitions that involve two mutually-ally peers, each transferring a piece to the other, as a result of some tit-for-tat contact initiated by one of them. In 4), we noted that no potential change is induced by the download of extra pieces. Therefore, as concerns the potential changes, all those transitions that involve the transfer of two extra pieces can be ignored. Similarly, those transitions that involve the transfer of an extra-piece coupled with some rare/non-rare piece can be viewed as (single) download of the accompanying rare/non-rare piece, which we have covered in 1)-3). We therefore focus only on those transitions that involve the transfer of two non-extra pieces. For this, consider two mutually-ally peers, say a $\ws$-peer missing piece $i\in W$ (and holding piece $j\in V$) and a $\vt$-peer missing piece $j$ (and holding piece $i$), who download the respective missing pieces as a result of a tit-for-tat contact initiated by one of them. We denote this transition by $\{(S,i\pm)_W,(T,j\pm)_V\}$ where $``-''$ indicates departure of the peer and $``+''$ indicates otherwise. From 1)-3), we note that the potential change induced by this piece exchange will depend on
\begin{itemize}
\item The type of piece $i$ ($i=r\in \rw$ or $i=n\in\rwc$),
\item The type of piece $j$ ($j=r'\in R_{V}$ or $j=n'\in {R}_V^c$),
\item Whether $\ws$-peer stays in or leaves the system, ($W\setminus S\supsetneq\{i\}$ or $W\setminus S=\{i\}$), and
\item Whether $\vt$-peer stays in or leaves the system ($V\setminus T\supsetneq\{j\}$ or $V\setminus T=\{j\}$).
\end{itemize}
Table \ref{tab:upperbounds_potchange_pieceexchanges} summarizes the upper-bounds on the potential change associated with each of those cases. Importantly, in each case, the upper-bound has been decomposed as the sum of two upper-bounds for the corresponding single piece download events. The bounds are obvious for the case $W\ne V$ and with some work, are fairly easy to establish when $W=V$. For completeness, they are derived in Appendix \ref{sec:appendix:pieceexchanges}.
\begin{table}[!b]
	\caption{Upper-Bounds on Potential Changes associated with Simultaneous Download of pieces $i$ and $j$ by mutually ally-peers $(W,S)$ and $(V,T)$ respectively}
	\centering
	\begin{tabular}{|c|l|c|}
			\hline
			Category &Transition &Potential Change\\
			\hline
			\rule{0pt}{10pt}
			1 &$(i,+)_W,(j,+)_V$ &$\Psi_W^{(i+)} + \Psi_V^{(j+)}$\\
			\hline
			\multirow{2}{*}{2} &$(i,+)_W,(j,-)_V$  &$\Psi_W^{(i+)} + \Psi_V^{(j-)}$\\
			\cline{2-3}
			&$(i,-)_W,(j,+)_V$  &$\Psi_W^{(i-)} + \Psi_V^{(j+)}$\\
			\hline
			3 &$(i,-)_W,(j,-)_V$  &$\Psi_W^{(i-)} + \Psi_V^{(j-)}$\\
			\hline
		\end{tabular}\label{tab:upperbounds_potchange_pieceexchanges}%
\end{table}
\subsection{Upper Bounds on Unit-Transition Drifts}
Having established upper-bound estimates of the potential changes, we can now proceed to evaluate the unit-transition drift ${QV}(\statex)$. We have established that for any swarm $W$, the upper-bound on potential-change induced by the download of piece $i\in W$ depends only on its type ($i\in\rw$ or $i\in\rwc$), whether or not the download is accompanied with the departure of the peer ($i+$ or $i-$), and the current chunk-distribution in swarm-$W$ ($\Psiwr{W}{r+}$ and $\Psiwr{W}{r-}$ depend on whether $ \statex\in\regrw{W}{1} $ or $ \statex\in\regrw{W}{2} $). Thus, for any state $\statex$, we can write
\begin{align*}\label{eq:stab:qvdef}
{{QV}}\left(\statex\right) &= \sum_{W\in\mcl{W}} {{QV}}_W^{(\emptyset,+)} + \sum_{W\in\mcl{W}} \underbrace{{{QV}}_{W}^{(\rw+)}+{{QV}}_{W}^{(\rwc+)}}_{\defeq {{QV}}_W^{(W+)}}\\
&\hspace{30pt}+\underbrace{{{QV}}_{W}^{(\rw-)}+{{QV}}_{W}^{(\rwc-)}}_{\defeq {{QV}}_W^{(W-)}},\numberthis
\end{align*}
where
\begin{align}\label{eq:stab:qvws}
\begin{split}
{QV}_{W}^{(\emptyset,+)}(\statex)&\defeq q_W^{(\emptyset,+)}{\Delta V}_W^{(\emptyset,+)},\\
{QV}_{W}^{(\rw +)}(\statex)&\defeq \sum_{\substack{r\in R_W,\\ W\setminus S\supsetneq \{r\}}} q_W^{(S,r+)}{\Delta V}_W^{(S,r+)},
\\
{QV}_{W}^{(\rw -)}(\statex) &\defeq \sum_{\substack{r\in R_W,\\ W\setminus S=\{r\}}} q_W^{(S,r-)}{\Delta V}_W^{(S,r-)},\\
{QV}_{W}^{(\rwc +)}(\statex)&\defeq \sum_{\substack{n\in \rwc,\\ W\setminus S\supsetneq \{n\}}} q_W^{(S,n+)}{\Delta V}_W^{(S,n+)},
\\
{QV}_{W}^{(\rwc -)}(\statex) &\defeq \sum_{\substack{n\in\rwc,\\ W\setminus S=\{n\}}} q_W^{(S,n-)}{\Delta V}_W^{(S,n-)}.
\end{split}
\end{align}

\subsubsection{Empty Peer Arrivals}
Let $\slmdas \defeq \sum_{W \in \mcl{W}}\lambda_W$ and $\c1 \defeq \max_{W \in \mcl{W}} \kw\cw1$. Using \eqref{eq:qwemptysetplus} and \eqref{eq:stab:vwemptysetplus},
\begin{align*}\label{eq:stab:qvarrivals}
\sum_{W\in\mcl{W}} {QV}_W^{(\emptyset,+)} \le \slmdas \c1. \numberthis
\end{align*}

\subsubsection{Downloads of Primary Pieces without Peer Departure}
The potential-change from the download of rare-pieces without accompanying peer-departure via the rarest-first (RF) mechanism is non-positive. ($\Psi_W^{(r+)}$ is always non-positive). In Lemma \ref{lem:rfrn} in Appendix \ref{sec:appendix:rfrn}, we upper-bound the corresponding drift by what we would have obtained if we were to replace RF by random-novel (RN), i.e.,
\begin{align*}\label{eq:stab:qvwrplus}
{QV}_W^{(\rw+)} &\le \sum_{\substack{r \in \rw, \\ W \setminus S \supsetneq \{r\}}} \frac{ \xws }{\x} \left[ \frac{ LU }{ |\rs| } \right.\\
&\hspace{-10pt}\left. + \Delta_p\xi_1 \sumvtr \frac{\xvt}{|\trs|} \right] \Psiwr{W}{r+}.\numberthis
\end{align*}
\emph{Remark}: Since \eqref{eq:stab:qvwrplus} is also an upper-bound for the download of rare-pieces using the random-novel (RN) mechanism, the stability proof of RFwPMS also works for ``RNwPMS'' -- a piece-selection policy same as RFwPMS except that rare-pieces are downloaded randomly.

Using \eqref{eq:qwsnplus}, \eqref{eq:qwsnplus2}, \eqref{eq:stab:vwsiplus}, and the fact that $\11\{\statex\in\regrw{W}{2}\}4\kw^2-\cw1 - I_W^{(1)} \le 0$, we get
\begin{align*}\label{eq:stab:qvwnplus}
QV^{(\rwc+)}&\le  \sum_{\substack{n\in \rwc,\\ W\setminus S\supsetneq\{n\}}}  \udl{q}_W^{(S,n+)}\left[
\11\{\statex\in\regrw{W}{2}\}4\kw^2-\cw1 \right.\\
&\hspace{-40pt} -\left.I_W^{(1)} \right] + \sum_{\substack{n\in \rwc,\\W\setminus S=\{n\}}} \frac{\xws}{\x} \udl{\Gamma}_W^{(n)}\xi_2\ov{\psi}_W\11{\{\statex\in \regrw{W}{1}\}}\zeta_W^{(n)}.\numberthis
\end{align*}
Lemma \ref{lem:getrid} in Appendix \ref{sec:appendix:getrid} combines \eqref{eq:stab:qvwrplus} and \eqref{eq:stab:qvwnplus} to give
\begin{align*}\label{eq:stab:qvwwplus}
{QV}_W^{(W+)} & \le\frac{\xw}{\x}
\sum_{i\in W} \frac{\udl{\Gamma}_W^{(i)}a_W^{(i)}}{\kw}\left(1-\freq{i}{W}-\gamma_W^{(i)}\right) \\
&\hspace{-30pt} \times \left[
\11{\{\statex\in\regrw{W}{1} \}}\left(\kw\xi_2\ov{\psi}_W\11{\{i\in\rwc\}}+\udl{\psi}_W \11{\{i\in \rw\}}\right)\right.\\
&\left. +\11{\{\statex\in\regrw{W}{2}\}}4\kw^2-\cw1-I_W^{(1)}
\right],\numberthis
\end{align*}
where, we have introduced $\gamma_{W}^{(i)}$ as a short-hand for $\frac{\sum_{S:W\setminus S=\{i\}}\xws}{\x_W}$, i.e., the fraction of swarm-$W$ peers missing only piece $i$ of file-$W$.

\subsubsection{Downloads of Primary Pieces with Peer Departure}
Using \eqref{eq:qwsiminus}, \eqref{eq:stab:vwsiminus}, and the fact that $\11\{\statex\in\regrw{W}{2}\}4\kw^2-\cw1+\11\{\statex\in\regrw{W}{1}\} \udl{\psi}_W \11\{i\in\rw\}\le 0 $, we have
\begin{align*}\label{eq:stab:qvwwminus}
{QV}_W^{(W-)} &\le \frac{\xw}{\x} \sum_{i\in W} \Gammalower{W}{i} a_W^{(i)} \gamma_W^{(i)} \\
&\hspace{-30pt} \left[
\11{\{\statex\in\regrw{W}{1}\}}\left( \xi_2\ov{\psi}_W\11{\{i\in\rwc\}}+
\udl{\psi}_W\11{\{i\in \rw\}}\right) \right.\\
&\hspace{-10pt}\left.+ 
\11{\{\statex\in\regrw{W}{2}\}}4\kw^2-\cw1+\xi_2 I_W^{(2)}
\right],\numberthis
\end{align*}

\subsubsection{Combining Everything}
Let us define
\begin{align*}\label{eq:stab:psiwi}
\Psi_W^{(i)}(\statex)&\defeq \11\{\statex\in\regrw{W}{2}\}4\kw^2\\ 
&\hspace{-30pt} +\11\{\statex\in\regrw{W}{1} \}\left(\kw\xi_2\ov{\psi}_W\11\{i\in\rwc\}+\udl{\psi}_W\11\{i\in\rw\}   \right)\\
&\hspace{-30pt} -\cw1.\numberthis
\end{align*}
Then, using \eqref{eq:stab:qvarrivals}, \eqref{eq:stab:qvwwplus}, and \eqref{eq:stab:qvwwminus}, in \eqref{eq:stab:qvdef}, we get
\begin{align*}\label{eq:stab:qv1}
QV(\statex)&\le \sum_{W\in\mcl{W}} \frac{\xw}{\x} \widetilde{QV}_W,\numberthis
\end{align*}
where
\begin{align*}\label{eq:stab:qvwtilde}
\widetilde{QV}_W(\statex) &\defeq \slmdas\c1 + \sum_{i\in W} \frac{\Gammalower{W}{i}a_W^{(i)} }{\kw} \left[  
\left(1-\freq{i}{W}\right)\Psi_W^{(i)}\right.\\
&\hspace{-10pt} \left. -\left(1-\freq{i}{W}-\gamma_{W}^{(i)} \right)I_W^{(1)}+\gamma_W^{(i)}\kw\xi_2I_W^{(2)}
\right].\numberthis
\end{align*}
With this initial setup in place, the rest of the proof is provided as a succession of lemmas in Appendix \ref{sec:appendix:stability}.
\section{Discussion}\label{sec:discussion}
\subsection{Sharing vs Suppression Trade-off for Non-Rare Pieces}
Note that with $\aw\in (0,1]$ sufficiently small, if we take the limit $\bw \rightarrow \infty$, swarm-based RFwPMS converges to a swarm-based version of RF (which is unstable in high arrival-rate regime), whereas if we take the limit $\bw \rightarrow 0$, swarm-based RFwPMS converges to a swarm-based version of MS~\cite{ms}. Thus, our expectation is that by choosing $\aw$ sufficiently small, and $\bw$ appropriately, an appropriate trade-off between the sharing and suppression of the non-rare pieces can be found. Via numerical simulations, we found that the expected sojourn time reduces with decreasing $\aw$, and choosing $\bw$ close to 1.5 appears to minimize it.

\subsection{Inter-Swarm Collaboration}
Our multi-swarm model in section \ref{sec:model} is general in terms of inter-swarm behavior of peers and their secondary download preferences. Here, we discuss three different behaviors that are covered:
\begin{enumerate}
\item[a)]
\textit{Altruistic Swarms}: In most wired P2P environments (e.g., the Internet), peers are generally insensitive to the consumption of their download and upload bandwidths, and they may download extra pieces in order to help other swarms. Such behavior can be captured in our model by setting $ \mcl{W}_{W} = \mcl{W} $ and $ \mcl{F}_{W} = \mcl{F} $ for every $W \in \mcl{W}$. From \cite{ssus}, a network in which all swarms are altruistic is called a \textit{universal swarm network}.
\item[b)]
\textit{Opportunistic Swarms}:
A different type of altruism is when peers do not download any extra pieces but share their pieces with those of other swarms who need them. We obtain this by setting $\mcl{W}_{W} = \mcl{W}$ and $\mcl{F}_{W} = W$ for every $W \in \mcl{W}$.
\item[c)]
\textit{Selfish Swarms}: In wireless P2P networks, peers are generally sensitive to the consumption of their download and upload bandwidths. This holds by setting $\mcl{W}_{W}=\{W\}$ and $\mcl{F}_{W}=W$. Thus, the peers do not download any extra pieces nor do they upload any piece to other swarms.
\end{enumerate}

\subsection{Piece-Selection Policies for Excess-Cache}
In our original version of RFwPMS, random-selection was assumed for the download of extra pieces, but the stability result in Theorem \ref{thm:stability} extends to any piece-selection policy one may use for the excess-cache. Without loss of generality, consider a piece-selection policy for the excess cache, whose transferable set is denoted by $E(\statex,\wh{T}, W, S)$, i.e., $E$ satisfies
$$  E(\statex, \wh{T}, W, S) \subseteq \left( \wh{T} \cap \mcl{F}_{W} \right) \setminus \left( S \cup W \right). $$
Recall that the Lyapunov function given by \eqref{eq:stab:v} and \eqref{eq:stab:vwdef} is not affected by a piece download from the set $E$. Consequently, all the bounds on $QV(\statex)$ including Lemma \ref{lem:qvupperbound} still hold. The final step of combining Lemma \ref{lem:qvupperbound} and Proposition \ref{prop:foster_kingman} has to be modified, however, as under some policies, the Markov process may no longer be irreducible with the current definitions of the state-space (\ref{eq:statespace}) and state (\ref{eq:state}). For instance, take a single-swarm network in which $E=\emptyset$, and the incoming swarm denoted by swarm-$W$ satisfies $\mcl{F}_{W}=\mcl{F} \supsetneq W.$ Then the set of all states in which some peer holds a piece from the set $\mcl{F} \setminus W$ is not reachable from any state. In such cases, the set of all states that are reachable from the empty state is a closed irreducible set of states and should be defined as the state-space of the network. Stability then follows from combining Lemma \ref{lem:qvupperbound} and Proposition \ref{prop:foster_kingman}. We summarize this discussion in the following proposition.
\begin{prop}
For the multi-swarm network model with non-persistent peers as described in Section \ref{sec:model}, swarm-based RFwPMS with any piece-selection policy for the excess-cache, is stable over the parameter region 
\begin{align*}
\left\{U > 0, \bsl{\lambda} > \mathbf{0}, L\ge 1, \Delta_p>0, \text{ and } \kw \ge 2 \ \forall\ W \in \mcl{W}\right\}.   
\end{align*}
\end{prop}

\subsection{Autonomous Swarms}
One more case to consider is when all the swarms in the network operate in isolation from each other. Specifically, a peer belonging to swarm-$W$ contacts and exchanges pieces with peers in the same swarm. The fixed seed, on the other hand, divides its uploading capacity across swarms, providing a static non-zero fraction of its total capacity to each swarm; optimal partition of the seed capacity is for future work.
Such swarms are called \textit{autonomous swarms} in \cite{ssus}. The stability of swarm-based RFwPMS holds for such swarms as well.
\begin{cor}
Consider a multi-swarm network where each swarm $W \in \mcl{W}$ behaves autonomously and the seed has allocated a static non-zero fraction of its total capacity for each swarm-$W$ (say $U_{W} > 0$). Then, the  network is stable under swarm-based RFwPMS over the parameter region 
\begin{align*}
\left\{U > 0, \bsl{\lambda} > \mathbf{0}, L\ge 1, \Delta_p>0, \text{ and } \kw \ge 2 \ \forall\ W \in \mcl{W}\right\}.   
\end{align*}
\end{cor}
\begin{IEEEproof}
Each swarm-$W$ can be considered as an isolated single-swarm network with fixed seed of capacity $U_{W} > 0$. Stability then follows by applying Theorem \ref{thm:stability} to each swarm.
\end{IEEEproof}

\subsection{Optimistic-Unchoking and Tit-for-Tat}
In our description of the Peer-Contact Policy in Section \ref{subsec:contactpolicy}, peer-$(1)$ makes a tit-for-tat contact with peer-$(2)$, it is assumed that the probability of peer-$(k)$ committing to transfer a piece to peer-$(-k)$, in the event that it does not benefit from peer-$(-k)$ is some fixed value, $p \in [0,1]$. However, from the stability analysis in Appendix \ref{sec:appendix:stability}, one may note that this assumption can be relaxed in two ways. i) If the network enforces optimistic-unchoking on the incoming peers, i.e., $Y_{opt}=1$, then the system is inherently stable regardless of what $p$ value is chosen by each peer in any of its tit-for-tat contacts -- indeed it may depend on the history of peer-$(k)$'s interactions with peer-$(-k)$. ($\Delta_p$ remains positive implying $\xi_2$ in \eqref{eq:Gammaratios} remains finite). ii) In the case that $Y_{opt}=0$, the stability of the network still holds with variable $p$ values as long as peers are forced to choose $p$ values larger than or equal to some positive threshold value $p_0 \in (0,1]$. ($\Delta_p$ remains positive implying $\xi_2$ in \eqref{eq:Gammaratios} remains finite). In particular, this means that our stability result holds for a history based tit-for-tat mechanism where each peer keeps track of its interactions with other peers. From i) and ii), we get the following proposition.
\begin{prop}
Consider the multi-swarm network model with non-persistent peers (as described in Section \ref{sec:model}) with the change that now each peer, in any tit-for-tat contact, can choose a variable $p$ value in $[p_0,1]$. Then, swarm-based RFwPMS is stable over the parameter region
\begin{align*}
&\left\{\mu > 0, U > 0, \bsl{\lambda} > \mathbf{0}, L\ge 1, (p_0>0 \text{ or } Y_{opt}=1), \text{ and } \right.\\
&\hspace{40pt} \left. \kw \ge 2 \ \forall\ W \in \mcl{W}\right\}.
\end{align*}
\end{prop}

\subsection{Comparison with Work of \cite{ms,ms:arxiv,ms:journal}}
As mentioned in Section \ref{sec:relatedwork}, \cite{ms:journal} introduced a threshold based version of mode-suppression, where the non-rare pieces are (completely) suppressed only when the largest-mismatch crosses a fixed constant threshold. Compared with this, swarm-based RFwPMS inhibits the replication of non-rare pieces by smoothly decreasing the non-rares sharing factor $\zeta_W^{(n)}$ in proportion with the largest-mismatch value. In Section \ref{subsec:sim_flashcrowd}, we note that a smoother-suppression like this should be preferred compared to a strictly threshold-based suppression. 

Another notable observation is that \cite{ms,ms:arxiv,ms:journal} perform random-selection on the set of available rare pieces. We assert that knowing the chunk distribution should allow for more advantage than just random-selection, and that a load-balancing scheme like rarest-first should do a better job in reducing the duration of transient phases. This gets manifested in Section \ref{subsec:sim_flashcrowd} and is also in line with the practical conclusions made in \cite{rfenough} (albeit without a theoretical analysis).

\subsection{Effect of Policy Parameters $\{\aw\}$:}
An interesting question is whether or not the stability result for swarm-based RFwPMS can be extended to the case when $\alpha_W=0$ for at least one of the swarms. From Lemma \ref{lem:upperboundnnonrares}, we see that the positive component of $QV_W^{(\rwc)}$ can be upper-bounded by a fixed constant, namely 
\begin{align}\label{eq:disc:dw}
\begin{split}
D_W &= D_W^{(1)} + D_W^{(2)}\11\{\mcl{W}_{W}\setminus \{W\}\ne\emptyset\}, \text{ where } \\
D_W^{(1)} &= 12e^{-2}\eta^{-2}\xi_2(LU+\Delta_p)\bw^2\kw^7\\
D_W^{(2)} &= 3e^{-2}\eta^{-1}\xi_2(LU+\Delta_p)\bw^{1+\aw^{-1}}\kw^{5+\aw^{-1}}.
\end{split}
\end{align}
If $W$ does not have any ally-swarm besides $W$, i.e., $\mcl{W}_W\setminus \{W\} = \emptyset$, then $D_W$ is independent of $\aw$. This immediately gives the below proposition.
\begin{prop}
For the multi-swarm model with non-persistent peers as described in Section \ref{sec:model}, swarm-based RFwPMS is stable over the parameter region
\begin{align*}
\left\{U > 0, \bsl{\lambda} > \mathbf{0}, L\ge 1, \Delta_p>0, \text{ and } \kw \ge 2 \ \forall\  W \in \mcl{W}\right\}.    
\end{align*}
where $\aw$ can be chosen to be zero for any swarm which does not have any other ally-swarms (besides itself).
\end{prop}

\begin{figure}[!ht]
	\centering
	\includegraphics[width=\linewidth]{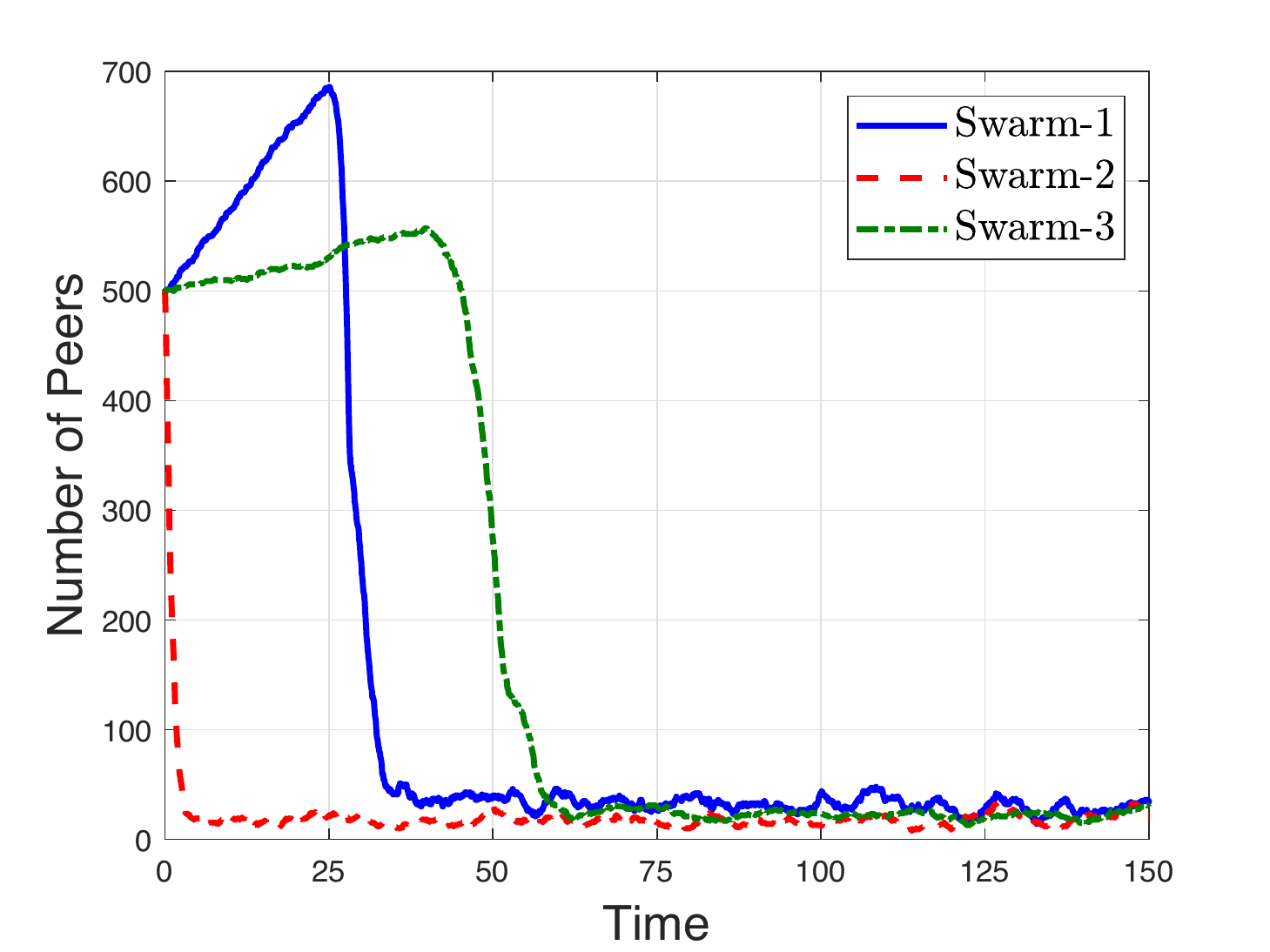}
	\caption{Stability-check of a Three-Swarm Network when $\alpha_W = 0$. The file-configurations were $W_1=[10], W_2=[6, 15]$, $W_3=[16, 30]$ and all swarms interacted altruistically. Initial state of each swarm was one-club -- all peers in the swarm missing one specific piece of the file. Assignment of other model parameters was $\bsl{\lambda} = (8, 4, 2), U=\frac{1}{3}, \mu=1, \wh{\mu}=\frac{1}{3}, L=3, p=0.5, Y_{opt}=1, \beta_W=1.5$.} \label{fig:stab_alpha_0}
\end{figure}

When swarm-$W$ has other ally-swarms, the constant $D_W \ra \infty$ as $\aw\ra 0$. Based on this, one may speculate that the stability of swarm-based RFwPMS is tightly coupled with the choice of $\{\aw\}$'s in the sense that the state of the system can possibly spiral out into ever increasing loads if $\aw=0$ for swarms $W$ that have other ally-swarms also. However, this does not seem to be the case in the numerical simulations we have performed. \figurename \ref{fig:stab_alpha_0} shows one such numerical snapshot for illustration. 
This corroborates the assertion that the parameter $\aw$ is a technical artifact in our setting, put in place to circumvent the issue of the positive component of the drift from the download of non-rare pieces in the case of ever increasingly large states where $\dnmbr{n}{W}$ is large. For the aforementioned reasons, we believe that swarm-based RFwPMS is stable in the setting when $\aw$'s can assume zero values. We state this as a conjecture below.

\begin{conjecture}\label{conjctr:alpha0}
For the multi-swarm model with non-persistent peers as described in Section \ref{sec:model}, swarm-based RFwPMS is stable over the parameter region
\begin{align*}
\left\{U > 0, \bsl{\lambda} > \mathbf{0}, L\ge 1, \Delta_p>0, \text{ and } \kw \ge 2 \ \forall\  W \in \mcl{W}\right\},    
\end{align*}
where $\aw$ can be chosen to be zero for any $W\in\mcl{W}$.
\end{conjecture}

\subsection{Scalability of swarm-based RFwPMS}
Scalability is also necessary for a P2P network, i.e., the system-wide throughput should scale with the number of peers. For the multi-swarm model presented in Section \ref{sec:model}, this means that the expected steady-state sojourn time in the network should be upper-bounded by a constant independent of $\bsl{\lambda}$. Unfortunately, analytically showing the scalability of stochastic models similar to the one we considered in this work is a hard endeavor and we leave this as an open problem. However, using an approach similar to that used in \cite{ms:arxiv,ms:journal}, we are able to show (under a mild assumption) the scalability of RFwPMS in the single-swarm case. This result is formally stated below.
\begin{thm}\label{thm:scalability}
For the single-swarm case of the multi-swarm model described in Section \ref{sec:model} (i.e., $\mcl{W}=\{W\}, \mcl{F}=W, \xw=\x$), provided that the empty-peers are always push-contacted in all tit-for-tat contacts, swarm based RFwPMS is scalable over its stability region. That is, the steady-state expected sojourn-time, denoted by $\ov{T}_W$, is upper-bounded by a constant independent of the arrival-rate of empty-peers. In particular, 
\begin{align*}
\ov{T}_W = 
\begin{cases}
O\left(\kw^3\right) &\text{if } 2\kw^2\lambda_W\le LU,\\
O\left(\kw^6\right) &\text{if } 2\kw^2\lambda_W> LU.
\end{cases}
\end{align*}
\end{thm}

The proof is established using Kingman's moment bound \cite{hajekbook} which uses the well-known technique of designing a suitable non-negative function and setting the steady-state expectation of its unit-transition drift equal to zero. Provided that the Markov chain is positive-recurrent, one usually hopes to obtain some convenient bound on the queue-length. For reference, please see Proposition \ref{prop:foster_kingman} in Appendix \ref{sec:appendix:fosterlyapunov}. 

In the next section, we provide the proof of Theorem \ref{thm:scalability}.

\section{Scalability of swarm based RFwPMS in the Single-Swarm}\label{sec:scalability}
Positive-recurrence of $\{\statex(t):t\ge 0\}$ is guaranteed by Theorem \ref{thm:stability}. Thus, to establish scalability, the key novelty is to first partition the stability region into two distinct load-regimes based on the size of total arrival-rate $\lambda_W$ in comparison to the seed's aggregate upload capacity $LU$ and then design a suitable non-negative function for each. The regimes are
\begin{enumerate}
\item[i)] 
$\mcl{L}_1=\left\{ 
\lambda_W < 0.5\kw^{-2}LU \right\}$; and
\item[ii)] 
$\mcl{L}_2=\left\{ 
\lambda_W \ge 0.5\kw^{-2} LU \right\}$.
\end{enumerate}
Let us denote the stationary distribution of $\{\statex(t):t\ge 0\}$ by $\{\rho(\statex):\statex\in\mcl{S}\}$, so that the expected number of peers in steady-state is given by $\mb{E}_{\rho}[\x]=\mb{E}_{\rho}[\xw]$, and by Little's Law $\ov{T}_W = \lambda_W^{-1}\mb{E}_{\rho}[\xw]$.

\subsection{Low Load Regime $\mcl{L}_1$}
For regime $\mcl{L}_1$, we have the following Lemma.
\begin{lem}\label{lem:scalability_lambdasmall}
For the single-swarm case of the multi-swarm model described in Section \ref{sec:model}, with swarm-based RFwPMS as the piece-selection policy, if the network parameters belong to $\mcl{L}_1$, then the steady-state expected sojourn time $\ov{T}_W$ is independent of $\lambda_W$. In particular, $\ov{T}_W=O(\kw^3)$. 
\end{lem}
\begin{IEEEproof}
The intuition behind this upper-bound is as follows: In $\mcl{L}_1$, the total arrival-rate $\lambda_W$ is sufficiently small compared to the seed's aggregate upload-rate $LU$; therefore, we expect the seed to quickly flush the system without any contribution from the normal peers. Keeping this in mind, the $\mu,\hat{\mu}$-independent bound, $\frac{\kw^2(\kw+1)}{LU}$, can be established by using Kingman's moment bound on a suitable non-negative function. Let
\begin{align*}
V_1(\statex)= \left(\kw \xw-\pw \right)^2 + \kw \xw-\pw.
\end{align*}
It can be observed that $V$ heavily penalizes the states with large number of missing pieces.\footnote{$\kw\xw -\pw=\sum_{i\in W}\xw-\nmbr{i}{W}$ is the number of pieces of $W$ missing in swarm-$W$.} The arrival of an empty-peer in swarm $W$ causes a unit-increase in $\xw$. So, the potential change associated with the arrival of an empty-peer is given by
\begin{align*}\label{eq:scal:l1:vemptysetplus}
{\Delta V}_1^{(\emptyset,+)}\le \Phi^{(\emptyset,+)} \defeq \kw^2+2\kw\left(\kw\xw-\pw\right)+\kw.\numberthis
\end{align*}
The download of a piece $i\in W$ by any $\ws$ peer causes a unit-decrease in the difference $\kw\xw-\pw$. Therefore, 
\begin{align*}\label{eq:scal:l1:vsiplusminus}
{\Delta V}_1^{(S,i\pm)}\le \udl{\Phi}\defeq -2\left(\kw\xw-\pw\right).\numberthis
\end{align*}
Upper-bounding ${QV}(\statex)$ by considering peer-arrivals and downloads of the rarest-pieces (from $\setrares{\mcl{F}}$) by the fixed seed only,
\begin{align*}\label{eq:scal:l1:qvupperbound}
&{QV_1}(\statex) \le \lambda_W \Phi^{(\emptyset,+)}+ \sum_{\udl{r}\in \setrares{\mcl{F}}}\left(1-\freqr{W}\right)\frac{LU}{|\setrares{\mcl{F}}|}\udl{\Phi}\\
&\labelrel{=}{eqr:scal:l1:a} \lambda_W \Phi^{(\emptyset,+)} +  LU\kw^{-1}\udl{\Phi}\\
&\labelrel{=}{eqr:scal:l1:b}\lambda_W \left(\kw^2+2\kw\left(\kw\xw-\pw\right)+\kw\right) \\
&\hspace{60pt} - 2LU\kw^{-1} \left(\kw\xw-\pw \right)\\
&=\lambda_W (\kw^2+\kw)\\
&\hspace{20pt} -2\left(LU\kw^{-1}-\lambda_W  \kw \right)\left(\kw\xw-\pw\right)\\
&\labelrel{\le}{eqr:scal:l1:c}\lambda_W (\kw^2+\kw)-\kw^{-1}LU\xw.\numberthis
\end{align*}
Here, \eqref{eqr:scal:l1:a} uses $1-\freqr{W}\ge \kw^{-1}$ (Lemma \ref{lem:chunkcounts}); \eqref{eqr:scal:l1:b} uses \eqref{eq:scal:l1:vemptysetplus} and \eqref{eq:scal:l1:vsiplusminus}; \eqref{eqr:scal:l1:c} follows from $\lambda_W<0.5\kw^{-2}LU$ and $\xw \le \kw\xw-P_W$.

Now, applying Kingman's moment-bound to \eqref{eq:scal:l1:qvupperbound} with $f(\statex) = \kw^{-1}LU\xw$ and $g(\statex) = \lambda (\kw+\kw^2)$ gives
\begin{align*}
\mb{E}_{\rho}[\xw] \le \frac{\lambda_W \kw^2(\kw+1) }{LU} \text{ and }\\
\ov{T}_W \le \frac{ \kw^2(\kw+1)}{LU}=O\left(\kw^3\right).
\end{align*}
\end{IEEEproof}

\subsection{High Load Regime $\mcl{L}_2$}
For regime $\mcl{L}_2$, we have the following Lemma.
\begin{lem}\label{lem:scalability_lambdabig}
For the single-swarm case of the multi-swarm model described in Section \ref{sec:model}, with swarm-based RFwPMS as the piece-selection policy, if the network parameters belong to $\mcl{L}_2$ and the empty-peers are always push-contacted in all tit-for-tat contacts, then the steady-state expected sojourn time $\ov{T}_W$ is independent of $\lambda_W$. In particular, $\ov{T}_W=O(\kw^6)$ for $\beta_W = O(\kw^{-3})$. 
\end{lem}
\begin{IEEEproof}
Let
\begin{align*}
V_2(\statex) &= V_{W,1}(\statex)+V_{W,2}(\statex) \\
&= \left(\left( 
\totalmismatch{W}-\eta\nmbrn{W}	
\right)^+\right)^2 +C_W^{(1)} \left(\xw -\nmbrn{W} \right).
\end{align*}
The above function uses the terms $V_{W,1}$ and $V_{W,2}$ of $V_W$ given in \eqref{eq:stab:vwdef}. Therefore, by ignoring the terms associated with $V_{W,3}(\statex)$ and noting that $\mcl{W}=\{W\}$, \eqref{eq:stab:qv1}, we get
\begin{align*}
QV_2(\statex) &\le \lambda_W\kw\cw1 + \sum_{i\in W} \frac{\Gammalower{W}{i}a_W^{(i)} }{\kw} \left[ \left(1-\freq{i}{W}\right)\Psi_{W}^{(i)} \right].
\end{align*}

\noindent \udl{Case 1 - $\statex\in \regrw{W}{2}$:} Here, 
\begin{align*}
&{QV}_2(\statex) \\
&\labelrel{\le}{eqr:scal:l2:20} \lambda_W\kw\cw1 +\sum_{i\in W} \frac{\Gammalower{W}{i} a_W^{(i)}}{\kw}\left[\left(1-\freq{i}{W}\right)\left(4\kw^2-\cw1\right) \right] \\
&\labelrel{\le}{eqr:scal:l2:21} \lambda_W\kw\cw1 + \frac{\Gammalower{W}{\udl{r} }}{\kw}\left[\left(1-\freqr{W}\right)\left(-0.5\cw1\right) \right] \\
&\labelrel{\le}{eqr:scal:l2:22} \lambda_W\kw\cw1+ \kw^{-2} \Gammalower{W}{\udl{r} } \left[-0.5\cw1\right]\\
&\labelrel{\le}{eqr:scal:l2:23} \lambda_W\kw\cw1 - 0.5\cw1\kw^{-2} (LU+\Delta_p \nmbrr{W})\\
&\labelrel{\le}{eqr:scal:l2:24} \lambda_W\kw\cw1 - 0.5\cw1\kw^{-2} (LU+\Delta_p(1-\eta)(\nmbrn{W}-2))\\
&\le \lambda_W\kw\cw1 - (1-\eta)\kw^{-2}\nmbrn{W} \left[0.5\cw1\Delta_p \right] \\ 
&\hspace{30pt} + \kw^{-2}\left( 2(1-\eta)\Delta_p - LU \right)^+ \left[ 0.5\cw1 \right]. \numberthis \label{eq:scal:l2:rw2}
\end{align*}
Here, \eqref{eqr:scal:l2:20} uses the definition of $\Psi_W^{(i)}$ (see \eqref{eq:stab:psiwi}); \eqref{eqr:scal:l2:21} uses $\cw1\ge 8\kw^2$ and then upper-bounds the summation by considering only the rarest-piece, that is denoted by $\udl{r}$; \eqref{eqr:scal:l2:22} uses $1-\freqr{W}\ge \kw^{-1}$ (Lemma \ref{lem:chunkcounts}); \eqref{eqr:scal:l2:23} uses $\Gammalower{W}{\udl{r}}\ge LU+\Delta_p\nmbrr{W}$ (see \eqref{eq:Gammalower2}); \eqref{eqr:scal:l2:24} uses $ \nmbrr{W} > (1-\eta)(\nmbrn{W}-2)$ because $\statex\in\regrw{W}{2}$.

\vspace{5pt}
\noindent \udl{Case 2 - $\statex\in \regrw{W}{1}$:} Here,
\begin{align*}
&{QV}_2(\statex) \\
&\labelrel{\le}{eqr:scal:l2:10} \lambda_W\kw\cw1 +\sum_{i \in W } \frac{ \Gammalower{W}{r}a_W^{(i)}}{\kw} \left[ \left( 1-\freq{i}{W} \right) \right.\\
&\hspace{10pt}\left. \times \left( \kw\xi_2\ov{\psi}_W\11\{i\in\rwc\}+\udl{\psi}_W\11\{i\in\rw\} -\cw1 \right) \right]\\
&\labelrel{\le}{eqr:scal:l2:11} \lambda_W\kw\cw1 +\sum_{n \in \rwc } \frac{ \Gammalower{W}{n}\zeta_W^{(n)}}{\kw} \left[ \left( 1-\freqn{W} \right) \kw\xi_2\ov{\psi}_W  \right] \\
&\hspace{10pt} +\sum_{r \in \rw } \frac{ \Gammalower{W}{r}}{\kw} \left[ \left( 1-\freq{r}{W} \right)\left( \udl{\psi}_W -\cw1 \right) \right]\\
&\labelrel{\le}{eqr:scal:l2:12} \lambda_W\kw\cw1 + D_W^{(1)} + \sum_{r \in \rw } \frac{ \Gammalower{W}{r}}{\kw} \left[ \left( 1-\freq{r}{W} \right) \udl{\psi}_W -\cw1 \right]\\
&\labelrel{\le}{eqr:scal:l2:13} \lambda_W\kw\cw1 + D_W^{(1)} \\
&\hspace{10pt} - 2(1-\eta)\kw^{-2} \left[ 2(1-\eta) \left( LU -2(1-\eta)\Delta_p \right) \right. \\
&\hspace{20pt} \left. + \left( (\kw-\eta)\nmbrn{W}-\sum_{i\ne\udl{r}}\nmbr{i}{W}  \right)  ( LU \wedge 2(1-\eta)\Delta_p ) \right]\\
&\labelrel{\le}{eqr:scal:l2:14} \lambda_W\kw\cw1 + D_W^{(1)} \\
&\hspace{10pt} - 2(1-\eta)\kw^{-2} \left[ 2(1-\eta) \left( LU -2(1-\eta)\Delta_p \right) \right. \\
&\hspace{80pt} \left. + (1-\eta)\nmbrn{W}( LU \wedge 2(1-\eta)\Delta_p ) \right]\\
&= \lambda_W\kw\cw1 + D_W^{(1)} \\
&\hspace{10pt} - (1-\eta)\kw^{-2} \nmbrn{W} \left[ 2(1-\eta) \left(LU \wedge 2(1-\eta)\Delta_p \right) \right] \\
&\hspace{20pt} + \kw^{-2}\left( 2(1-\eta)\Delta_p - LU\right)^+ \left[ 4(1-\eta)^2 \right].
\end{align*}
Here, \eqref{eqr:scal:l2:10} uses the definition of $\Psi_W^{(i)}$ (see \eqref{eq:stab:psiwi}); \eqref{eqr:scal:l2:11} uses $\cw1\ge 0$; \eqref{eqr:scal:l2:12} follows from Lemma \ref{lem:upperboundnnonrares}; \eqref{eqr:scal:l2:13} follows from Lemma \ref{lem:nw}; \eqref{eqr:scal:l2:14} uses $\nmbr{i}{W}\le\nmbrn{W}$.

Combining \eqref{eq:scal:l2:rw2} and \eqref{eq:scal:l2:rw2}, we get
\begin{align*}
&{QV}_2(\statex) \le \lambda_W\kw\cw1 + D_W^{(1)} \\
&\hspace{10pt} + \kw^{-2}\left(2(1-\eta)\Delta_p-LU \right)^+ \left(
0.5\cw1+4(1-\eta)^2\right)\\
&\hspace{20pt} - \kw^{-2}(1-\eta)\nmbrn{W} \min \left\{0.5\cw1\Delta_p, 2(1-\eta)LU, \right. \\
&\hspace{140pt} \left. 4(1-\eta)^2\Delta_p \right\}\\
&\labelrel{\le}{eqr:scal:l2:a} 8\lambda_W\kw^3 + D_W^{(1)} + 5\left(2(1-\eta)\Delta_p-LU \right)^+ \\
&\hspace{20pt} - 2(1-\eta)^2\kw^{-2}\nmbrn{W} \min \{ LU, 2(1-\eta)\Delta_p \}.\numberthis\label{eq:scal:l2:qvupperbound}
\end{align*}
Here, \eqref{eqr:scal:l2:a} follows from choosing $\cw1=8\kw^2$ and using $4(1-\eta)^2 \le \kw^2$. 

Applying Kingman's moment bound to \eqref{eq:scal:l2:qvupperbound} with $f(\statex)=2(1-\eta)^2\kw^{-2}\min\{LU,2(1-\eta)\Delta_p\} \nmbrn{W}$ and $g(\statex)=8\lambda_W\kw^3 +D_W^{(1)} + 5\left(2(1-\eta)\Delta_p-LU \right)^+$, we get
\begin{align*}
&\mb{E}_{\rho}[\xw - x_W^{\emptyset}] 
\labelrel{\le}{eqr:scal:l2:nonempty} \mb{E}_{\rho}[\pw] \le \kw \mb{E}_{\rho}[\nmbrn{W}] \\
&\le \frac{\left(
8\lambda_W\kw^3 + 5\left(2(1-\eta)\Delta_p-LU \right)^+
+ D_W^{(1)} \right)\kw^3
}{
2(1-\eta)^2\min\{LU, 2(1-\eta)\Delta_p \}
}\\
&=\frac{\kw^3\left(8\lambda_W\kw^3 + 5\left(2(1-\eta)\Delta_p-LU\right)^+\right) + D_W^{(1)}\kw^3}
{2(1-\eta)^2 \min\{LU, 2(1-\eta)\Delta_p \}}.
\end{align*}
Here, \eqref{eqr:scal:l2:nonempty} uses the fact that each non-empty peer must possess at least one piece of the file. 

Now, we divide the single-swarm system into two subsystems; one with empty-peers and the other with peers who hold some piece of $W$. The two subsystems are connected in tandem as shown in \figurename \ref{fig:scalability:1}.
\begin{figure}[!t]
\centering
\includegraphics[width=\linewidth]{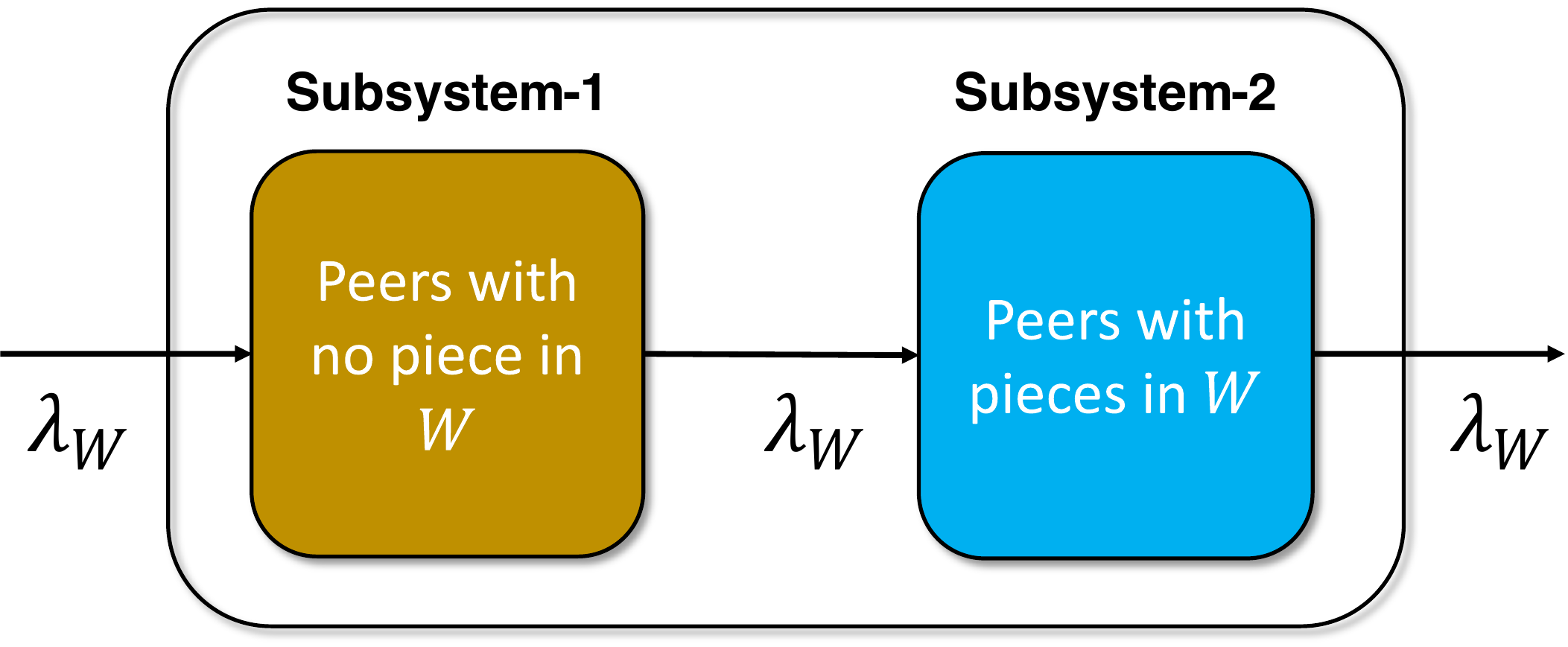}
\caption{Subsystems 1 and 2 in Tandem (Single-Swarm Version of the Multi-swarm Model described in Section \ref{sec:model}).}\label{fig:scalability:1}
\end{figure}
Only the empty peers arrive in subsystem-1, so the arrival rate in subsystem-1 is $\lambda_W$. Since both systems are stable in steady-state, the departure rate of subsystem-1 (which is also the arrival rate of subsystem 2) and the departure rate of subsystem-2 are $\lambda_W$. Let $\ov{T}_{W,1}$ and $\ov{T}_{W,2}$ denote the steady-state expected sojourn-times in subsystems 1 and 2 respectively, so that by Little's Law, 
$\mb{E}_{\rho}\left[\xw\right]=\lambda_W \left(\ov{T}_{W,1}+ \ov{T}_{W,2}\right)$. 
Under swarm based RFwPMS, consider two situations -- one in which a $\ws$-peer contacts a $(W,T)$ peer where $T\ne \emptyset$ and the other in which it contacts a $(W,\emptyset)$ peer. Then under the assumption, that empty-peers are always push-contacted, $(W,S)$ peer will push-contact $(W,\emptyset)$ peer with probability 1. This combined with the fact that $S\setminus T \subseteq S\setminus \emptyset$ ensures that the probability of an empty-peer downloading a piece in the next state transition is always lower-bounded by that of a peer from subsystem-2. Thus, $\ov{T}_{W,1}\le \ov{T}_{W,2}$, which gives 
\begin{align*}
&\mb{E}_{\rho}[\xw] \le 2\lambda_W \ov{T}_{W,2} = 2\mb{E}[\xw-x_W^{\emptyset}], \text{ and} \\
&\ov{T}_W \le 
\frac{\kw^3\left(8\kw^3 + 5\lambda_W^{-1}\left(2(1-\eta)\Delta_p-LU\right)^+\right)}
{(1-\eta)^2 \min\{LU, 2(1-\eta)\Delta_p \}}\\
&\hspace{40pt} + \frac{ \lambda_W^{-1} D_W^{(1)} \kw^3 }{(1-\eta)^2 \min\{LU, 2(1-\eta)\Delta_p \}}.
\end{align*}
Using $\lambda_W \ge 0.5\kw^{-2}LU$, the fact that $D_W^{(1)}$ is $O(\beta_W^2\kw^7)$ from \eqref{eq:disc:dw}, and the choice of $\bw=O(\kw^{-3})$, we get
\begin{align*}
\ov{T}_W = O\left( \kw^6 \right).
\end{align*}
This establishes scalability.
\end{IEEEproof}
\section{Simulation Based Performance Evaluation}\label{sec:simulations}
Next, we investigate the stability, scalability and sojourn time performance of RFwPMS via numerical simulations. Since our stability result holds for any number of swarms, we will evaluate performance in both single-swarm and multi-swarm settings. Unless otherwise noted, in all cases, we set $\alpha_W$ to $10^{-9}$ and $\beta_W$ to $1.5$. Furthermore, the tabulated steady-state sojourn times are based on samples taken after the simulated Markov-Chain hit stationarity (simulation run-times were chosen long enough to collect sufficiently large number of samples from the stationary distribution). 

\subsection{Stability Check}
We illustrate the stability of RFwPMS by simulating four instances of a two-swarm network with the setting of ``hard tit-for-tat ($p=0$) and opportunistic-unchoking ($Y_{opt}=1$)'' -- the four instances correspond to altruistic, opportunistic, selfish, and autonomous inter-swarm behaviors. In each instance, the network consists of a master-file $\mcl{F}$ of 25 pieces, i.e., $\mcl{F}=[25]$. The two swarms entering the network are denoted by $W_1$ and $W_2$, each having a peer-arrival rate of 20. Peers from swarm-$W_1$ wish to download file $W_1=[15]$ whereas those from swarm-$W_2$ are interested in file $W_2=[10,25]$. We initiate the network in a state where both swarms are in the one-club scenario: both have 500 peers with all in swarm-$W_1$ missing piece 1 and all in swarm-$W_2$ missing piece 6. For the autonomous case, the seed's upload capacity is divided evenly among the two swarms, i.e., $(U_{W_1},U_{W_2})=(0.5,0.5)$.

\begin{figure*}[!t]
\centering
\subfloat[Altruistic Swarms.]{\includegraphics[width=0.45\linewidth]{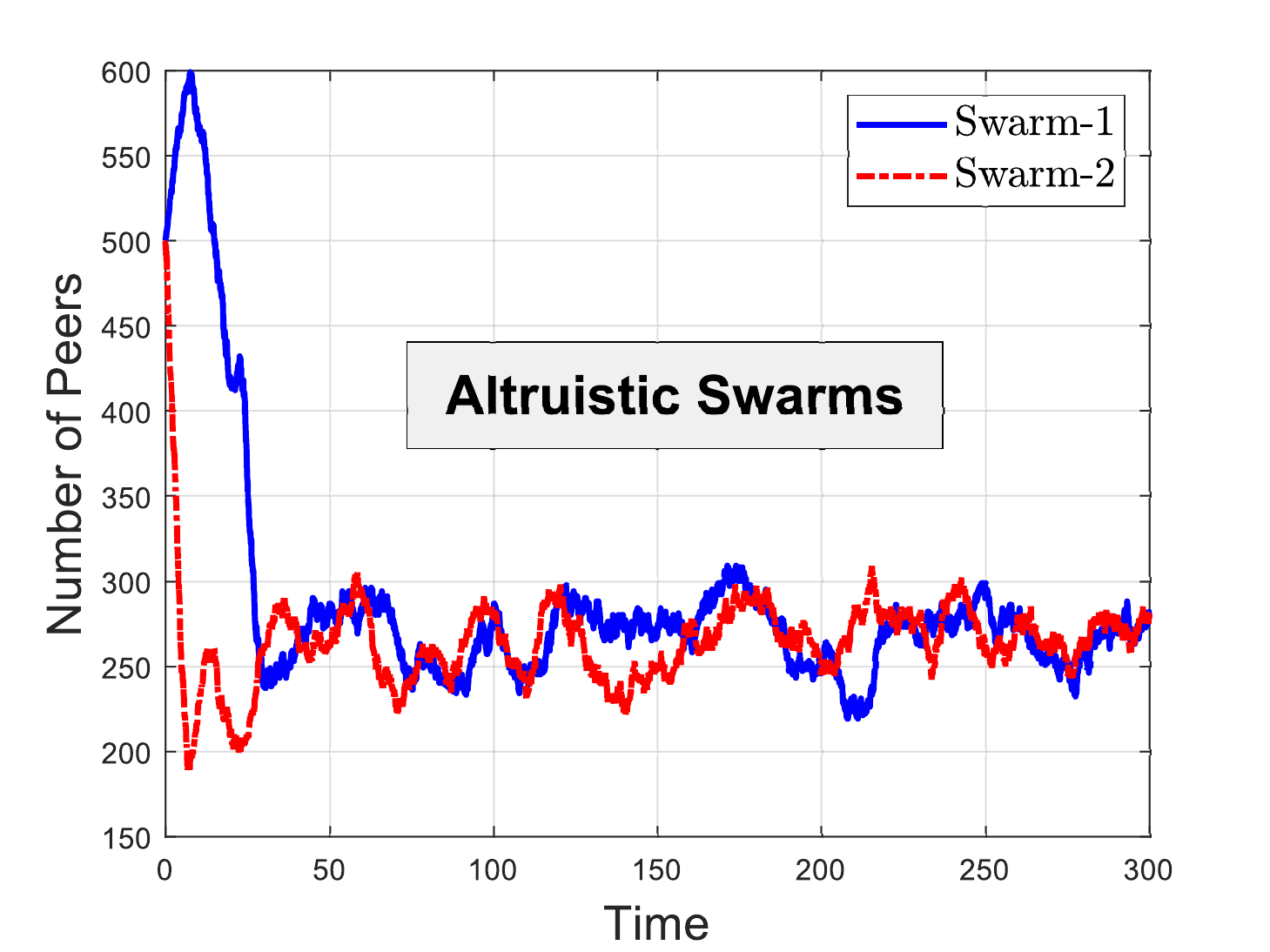}
\label{fig:stab:alt}}
\hfil
\subfloat[Opportunistic Swarms.]{\includegraphics[width=0.45\linewidth]{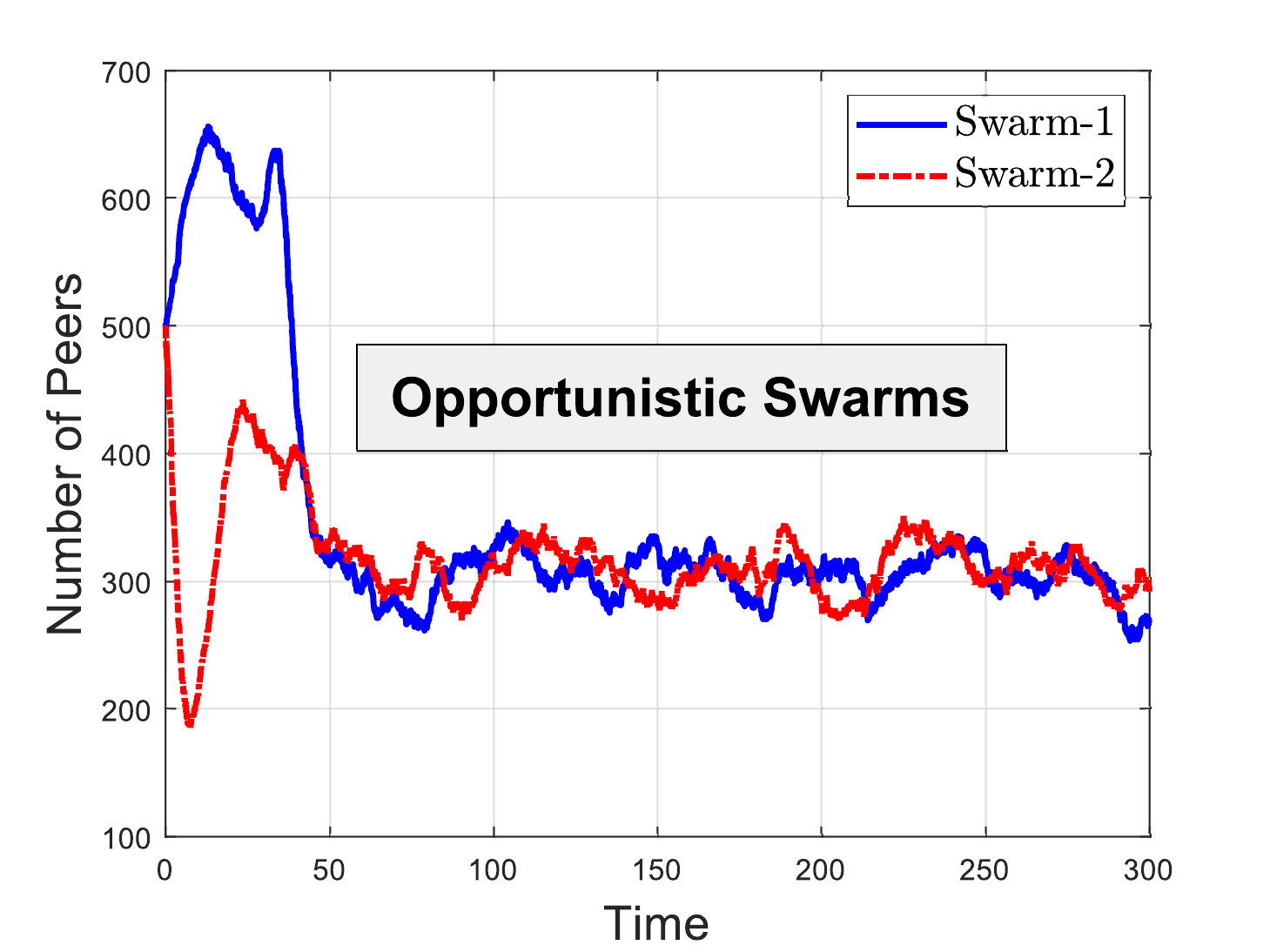}
\label{fig:stab:opp}}
\hfil
\subfloat[Selfish Swarms.]{\includegraphics[width=0.45\linewidth]{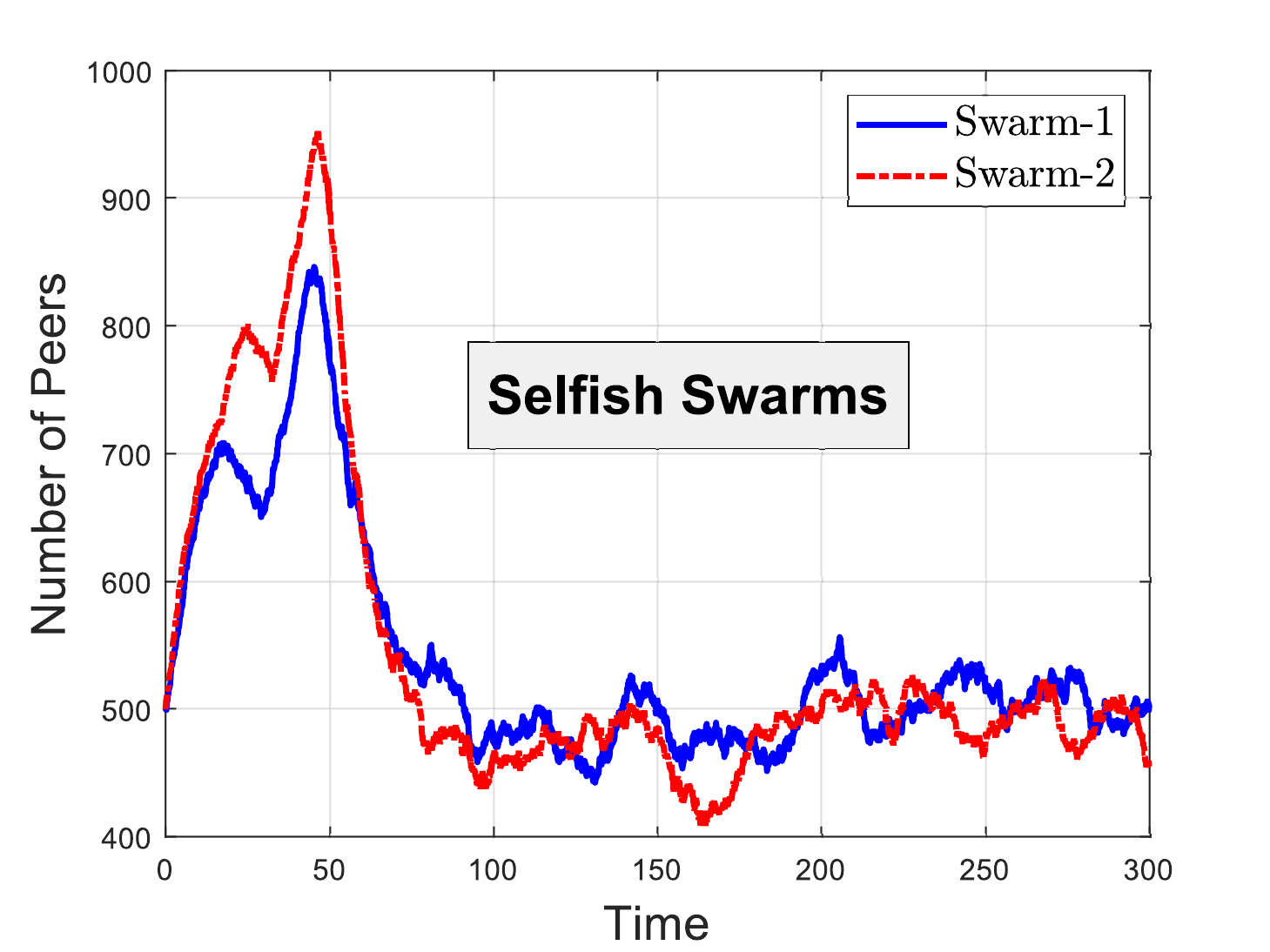}
\label{fig:stab:selfish}}
\hfil
\subfloat[Autonomous Swarms]{\includegraphics[width=0.45\linewidth]{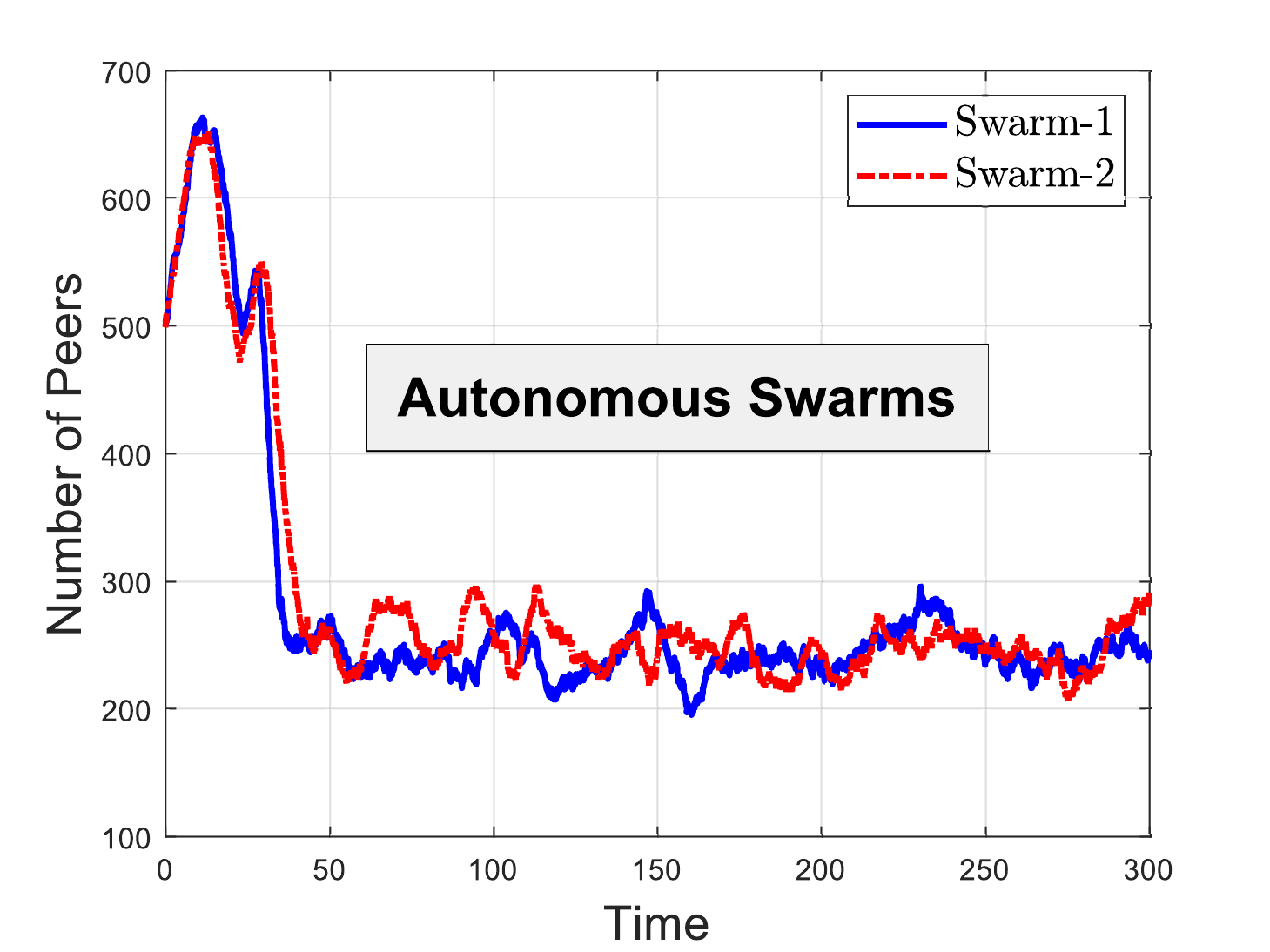}
\label{fig:stab:auto}}
\caption{Stability-check of RFwPMS in a Two-Swarm network for different inter-swarm behaviors. The primary files were $W_1=[15]$ and $W_2=[10, 25]$ and the master-file was set to their union. Assignment of remaining model parameters was $U=\mu=1$, $\wh{\mu}=\frac{1}{3}$, $L=3$, $p=0$, $Y_{opt}=1$. In the case of autonomous-swarms, the seed's upload capacity was divided equally among the swarms.}
\label{fig:stability}
\end{figure*}

\figurename \ref{fig:stability} shows the evolution of the swarm populations for the four inter-swarm behaviors. It is observed that each swarm is able to escape the one-club in finite time, after which a stable regime persists with minimal fluctuations. In the case of altruistic and opportunistic swarms, the population of swarm-$W_2$ suddenly drops in the beginning. This is because piece 6, which is missing in swarm-$W_2$, is widely available in swarm-$W_1$; due to altruism of swarm-$W_1$ peers, the one-club peers in swarm-$W_2$ quickly grab piece 6 and leave the network. In the opportunistic case, the population of swarm-$W_2$ increases almost linearly after the big initial drop. This is because once most of the one-club peers in swarm-$W_2$ have left, the network is dominated by swarm-$W_1$ peers. Therefore, most of the contacts of the new-comers in swarm-$W_2$ are with swarm-$W_1$ peers. Since the files are not identical, these new-comers cannot accumulate all pieces from swarm-$W_1$ peers and are forced to linger in the system -- till swarm-$W_1$'s population reduces enough and they get an opportunity for a useful contact.

\subsection{Scalability Check}\label{subsec:scalabilitycheck}
Here, we simulate twelve instances of a two-swarm network with the setting of ``soft tit-for-tat (choice of $p=0.5$) and no optimistic-unchoking ($Y_{opt}=0$)''. In each instance, the network's master-file comprises of 18 pieces ($\mcl{F}=[18]$) and the two swarms entering the network are interested in files $W_1=[10]$ and $W_2=[8,18]$. The twelve instances correspond to four inter-swarm behaviors (altruistic, opportunistic, selfish, and autonomous) and three arrival-rate configurations ($\bsl{\lambda}=(4,2)$, $4\bsl{\lambda}$ and $16 \bsl{\lambda}$). For the autonomous case, the seed's upload capacity is again split equally between the two swarms.

Table \ref{tab:expsorjtime} lists the (steady-state) expected sojourn times for the twelve model instances. We note that the expected sojourn time practically stays constant in all the four inter-swarm behaviors and improves as swarms interact more altruistically. Another observation is that the autonomous setting has lower expected sojourn times than the altruistic setting. This is not unexpected though; in autonomous setting, every peer makes contacts within their own swarm (no inter-swarm interactions), thus the likelihood that the next contact is useless is lower.

\begin{table}[!b]
	\caption{Expected Steady-State Sojourn Times \protect\linebreak for Different Arrival Rate Vectors. \protect\linebreak ($U=\mu=1, \wh{\mu}=\frac{1}{3}, L=3, p=0.5, Y_{opt}=0$).}
	\label{tab:expsorjtime}
	\centering
	\begin{tabular}{|c|c|c|c|}
			\hline
			\textbf{Swarms'} & \textbf{Arrival} & \multicolumn{2}{c|}{\textbf{$\mb{E}[\textbf{Sojourn Time}]$}} \\
			\cline{3-4} 
			\textbf{Behavior}&\textbf{Rates$^{\mathrm{a}}$}& \textit{$W_1 = [10]$} & \textit{$W_2 = [8,18]$} \\
			\hline
			\multirow{2}{*}{Altruistic} & $\bsl{\lambda}$ & $2.927 
			$  & $4.400 
			$ \\\cline{2-4}
			& $4\bsl{\lambda}$ & $3.088 
			$  & $3.990 
			$ \\\cline{2-4}
			& $16\bsl{\lambda}$ & $3.134 
			$  & $3.971 
			$ \\
			\hline
			\multirow{2}{*}{Opportunistic} & $\bsl{\lambda}$ & $3.704 
			$  & $5.042 
			$ \\\cline{2-4}
			& $4\bsl{\lambda}$ & $3.832 
			$  & $5.341 
			$ \\\cline{2-4}
			& $16\bsl{\lambda}$ & $3.956 
			$  & $5.570 
			$ \\
			\hline
			\multirow{2}{*}{Selfish} & $\bsl{\lambda}$ & $4.378 
			$  & $6.394 
			$ \\\cline{2-4}
			& $4\bsl{\lambda}$ & $4.590 
			$  & $6.482 
			$ \\\cline{2-4}
			& $16\bsl{\lambda}$ & $4.667 
			$  & $6.604 
			$ \\
			\hline
			\multirow{2}{*}{Autonomous} & $\bsl{\lambda}$ & $2.791 
			$  & $3.769 
			$ \\\cline{2-4}
			& $4\bsl{\lambda}$ & $2.712 
			$  & $2.667 
			$ \\\cline{2-4}
			& $16\bsl{\lambda}$ & $2.788 
			$  & $2.740 
			$ \\
			\hline
			\multicolumn{4}{l}{$^{\mathrm{a}} \bsl{\lambda}=(4,2)$. Simulation End-time: 1000 units}
		\end{tabular}%
\end{table}

\subsection{Sojourn Time Performance in the presence of Multiple Swarms}
\begin{figure*}[!t]
	\centering
	\subfloat[Expected Sojourn Times of Swarm-1.]{\includegraphics[width=0.4\linewidth]{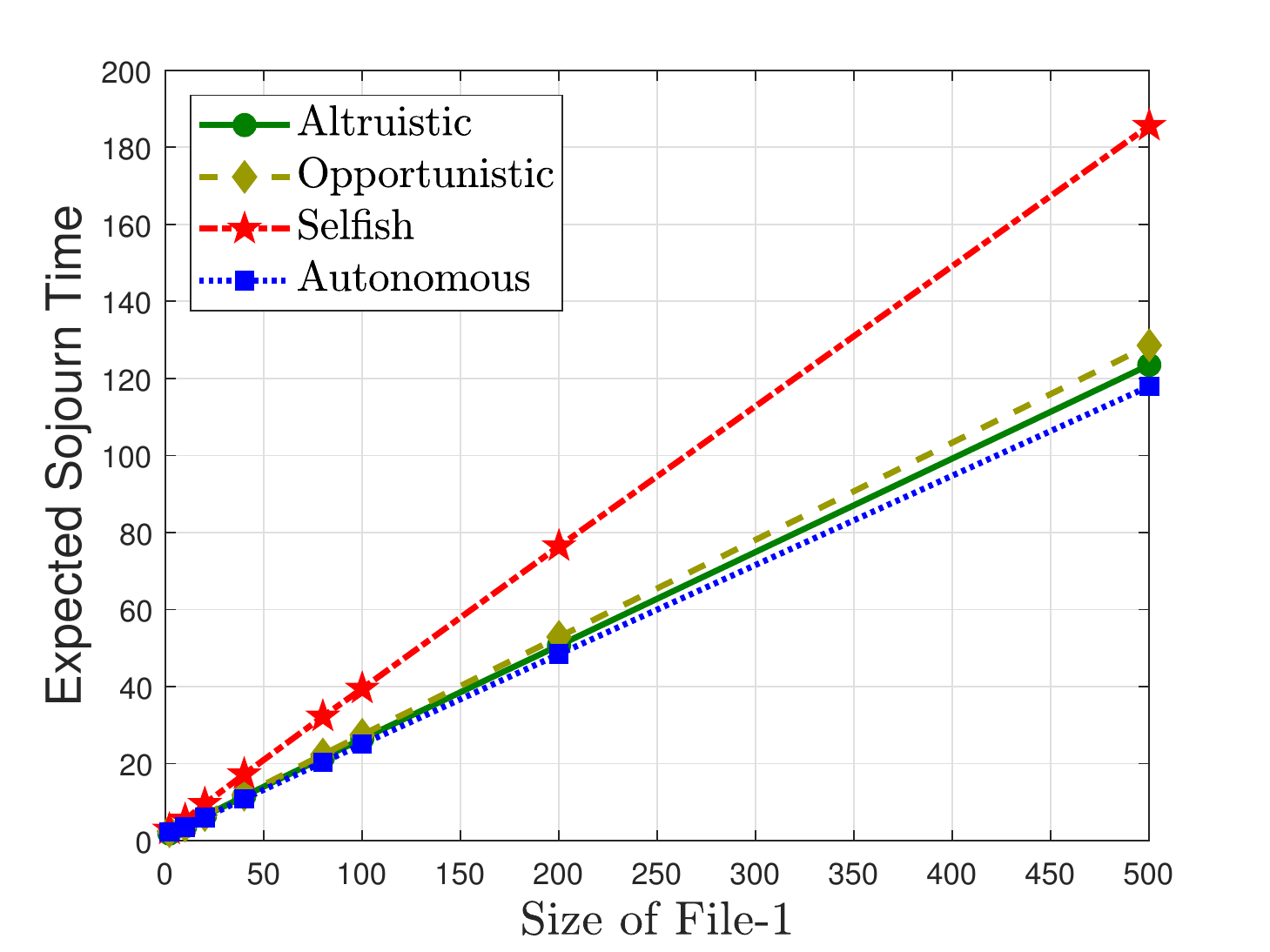}
		\label{fig:sjn_1}}
	\hfil
	\subfloat[Expected Sojourn Times of Swarm-2.]{\includegraphics[width=0.4\linewidth]{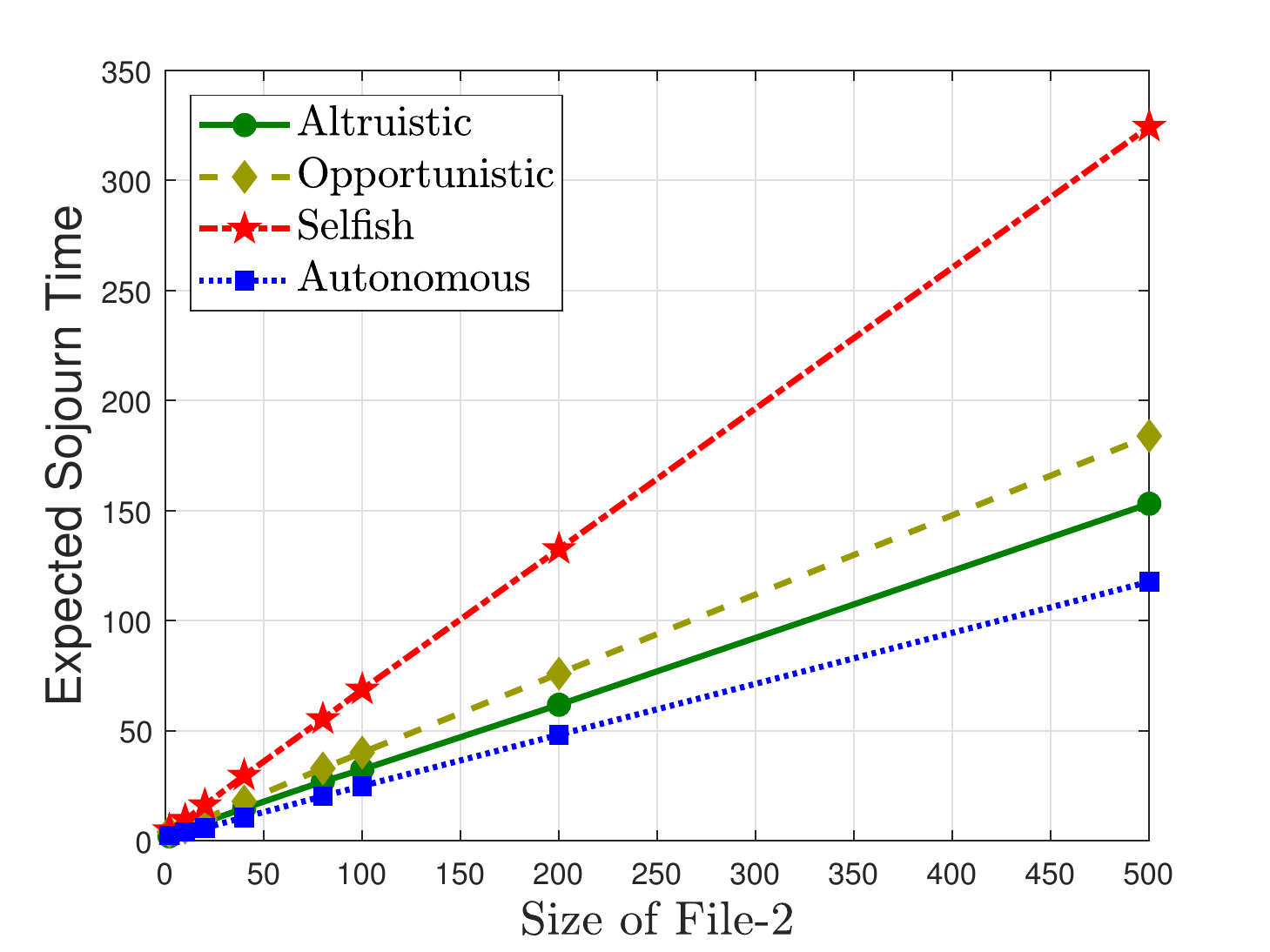}
		\label{fig:sjn_2}}
	\caption{Expected steady-state sojourn times vs file-sizes. \figurename \ref{fig:sjn_1} shows the expected sojourn times of swarm-1 and \figurename \ref{fig:sjn_2} shows those of swarm-2. Assignment of remaining model parameters was $\bsl{\lambda}=(6,2)$, $U = \mu = 1$, $\wh{\mu} = \frac{1}{3}$, $L = 3$, $p=0.5$, $Y_{opt} = 1$.}
	\label{fig:sjn}
\end{figure*}

How does the expected steady-state sojourn time scale with the file-size is another important performance measure. For this, we obtained an upper-bound estimate of $O(\kw^6)$ in Section \ref{sec:scalability}. Here, in \figurename \ref{fig:sjn} we verify this for a two-swarm network in a soft tit-for-tat setting (choice of $p=0.5$) with optimistic-unchoke ($Y_{opt}=1$) -- once again, under the four different inter-swarm behaviors.\footnote{The error-bars for confidence intervals are not shown due to their negligible magnitude (even with a choice of 99.9 percent confidence level).} 

Fixing the arrival-rate vector to $ \bsl{\lambda}=(6,2)$, increasing configurations of the two file-sizes were simulated such that for every choice of $W_1=[K_{W_1}]$, we set $W_2$ to $[\frac{1}{2}K_{W_1}, \frac{3}{2}K_{W_1}]$. (Again the seed's upload capacity was split evenly in the autonomous case). The sojourn-times for each swarm are observed to scale linearly with the file-size. Additionally, we note that for each swarm, the altruistic and autonomous cases are \textit{Pareto} better than opportunistic and selfish cases.

\subsection{RFwPMS and Mode Suppression}
Via simulations MS~\cite{ms} has been shown to outperform other previously proposed piece-selection policies. Here, we do a base-line comparison of RFwPMS and MS in a single-swarm network.

\subsubsection{Expected Steady-State Sojourn Times}
Table \ref{tab:expsorjdif} compares the expected steady-state sojourn times of RFwPMS, MS, and TMS\footnote{Threshold Mode-Suppression with suppression-threshold set to $2\kw$.} for different values of file-size $K_W$ with the arrival-rate fixed at $\lambda_W=4$. It is noted that using RFwPMS (with $\bw=1.5$), the expected sojourn time is indeed reduced compated to MS. This reduction is not expected to be significant when $K$ is large (roughly $100$ or more) as the increased chunk diversity in steady-state reduces the number of contacts in which only the non-rare pieces are offered. The steady-state sojourn-times of RFwPMS and TMS, on the other hand, are comparable for all values of $K$. 
\begin{table}[!tb]
	\caption{Expected Steady-State Sojourn Times for RFwPMS and MS. \protect\linebreak ($\lambda_W=4, U=\wh{\mu}=L=Y_{opt}=1, \beta_W=1.7$).}
	\label{tab:expsorjdif}
	\centering
	\begin{tabular}{|c|c|c|c|c|}
			\hline
			\multirow{2}{*}{$K_W$} & \multicolumn{3}{c|}{$\mb{E}[\textbf{Sojourn Time}]$} & \textbf{Perc. Improv.} \\
			\cline{2-4}
			&\textit{MS} 
			&\textit{TMS}
			&\textit{RFwPMS} & \textbf{over MS} \\
			\hline
			$2$ & $6.246$ &$5.022$ &$5.178$ & $17.103$ \\
			\hline
			$10$ & $18.250$ &$12.546$ &$12.525$ & $31.367$ \\
			\hline
			$20$ & $31.741$ &$23.020$ &$23.058$ & $27.356$ \\
			\hline
			$40$ & $55.648$ &$43.775$ &$43.750$ & $21.382$ \\
			\hline
			$80$ & $100.300$ &$84.374$ &$84.421$ & $15.831$ \\
			\hline
			$100$ & $121.804$ &$104.849$ &$104.610$ & $14.116$ \\
			\hline
			$200$ & $226.998$ &$205.300$ &$205.176$ & $9.613$ \\
			\hline
			$500$ & $533.737$ &$506.480$ &$506.351$ & $5.131$ \\
			\hline
			\multicolumn{4}{l}{Simulation End-time: 5000 units}
		\end{tabular}%
\end{table}

\subsubsection{Flash-Crowd Responses}\label{subsec:sim_flashcrowd}
\begin{figure*}[!b]
\centering
\subfloat[Number of Peers vs. Time.]{\includegraphics[width=0.45\linewidth]{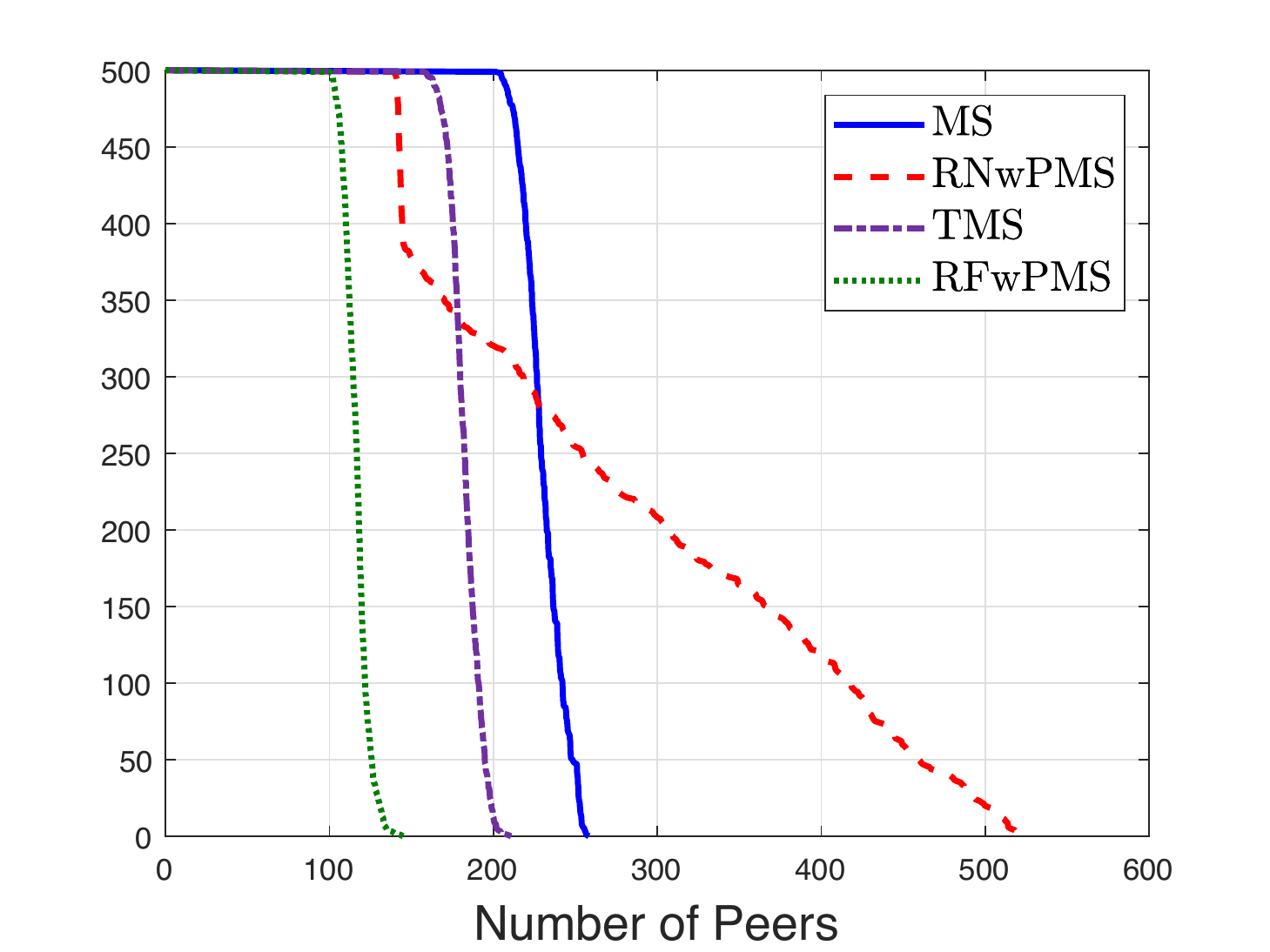}
\label{fig:fc_1}}
\hfil
\subfloat[Chunk-counts vs Time.]{\includegraphics[width=0.45\linewidth]{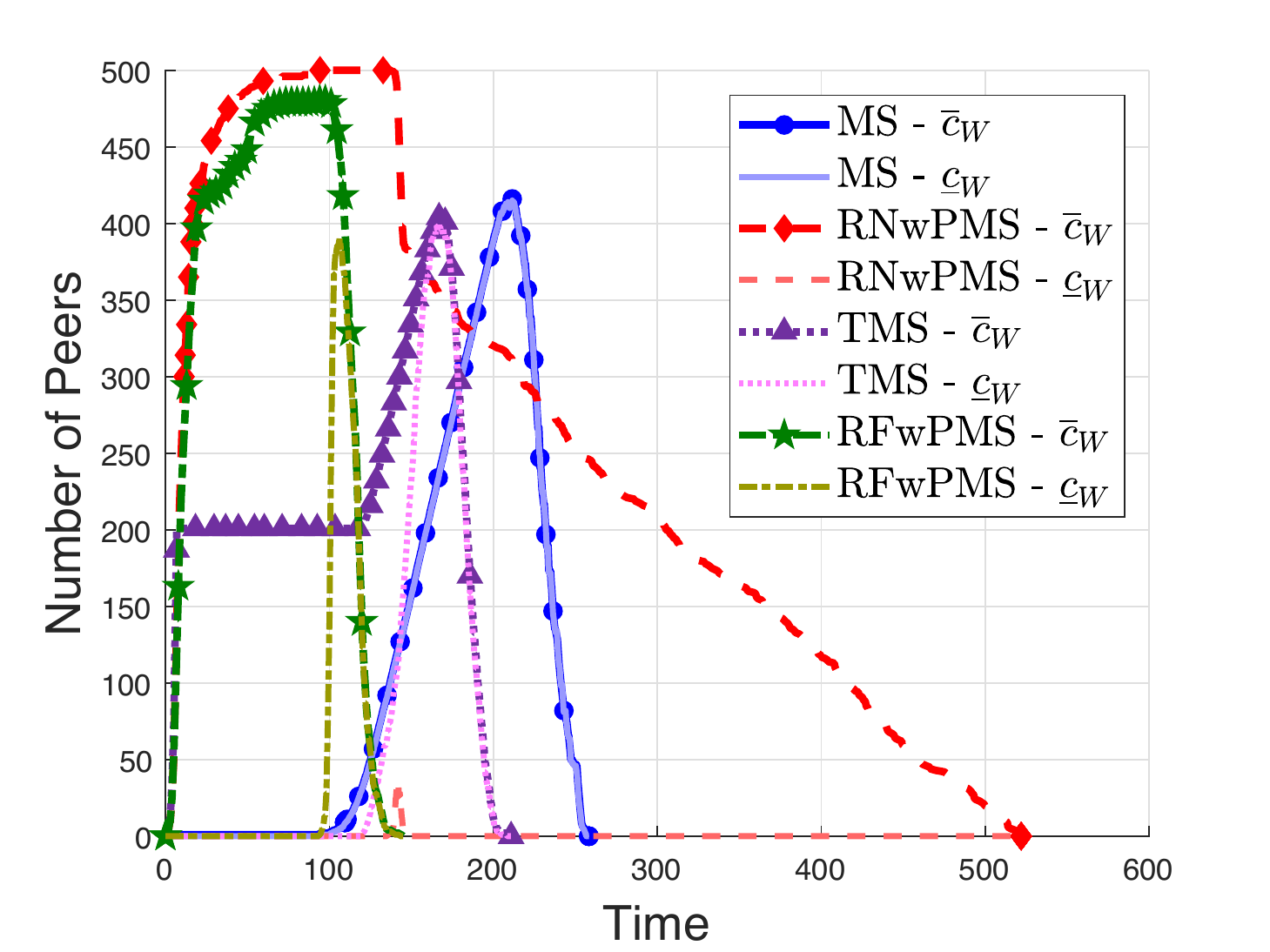}
\label{fig:fc_2}}
\caption{Flash-crowd responses of MS, RNwPMS, TMS, and RFwPMS when flash-crowd size is much larger than file-size. Initial population and file-size were set to 500 and 100 respectively. Assignment of other model parameters was $\lambda_W=0$, $U=\wh{\mu}=1$ $L = Y_{opt} = 1$).}
\label{fig:fc}
\end{figure*}
Real torrents seldom experience steady-state behavior as the arrival-rates vary over time with intermittent bursts of empty peers. Fig. \ref{fig:fc} compares the flash-crowd response of MS, TMS, RFwPMS, and RNwPMS (replacing RF by RN in our policy) for $\kw=100$ (``large") and an initial population of 500 empty peers ($\xw(0) = x_{W}^{(\phi)}\gg\kw$); RFwPMS flushes out the system in the least amount of time (about half the time as MS) whereas RNwPMS takes the most amount of time (see \figurename \ref{fig:fc_1}). The reason behind such responses can be understood from \figurename \ref{fig:fc_2}. Initially there are no pieces in the system and therefore the time till all $\kw$ pieces are introduced into the system depends solely on the seed's uploading capacity which is set to 1. Thus, for every policy, if $\kw$ is sufficiently large, it will take on average $\kw$ time units till every piece has been introduced into the system (see Fig. \ref{fig:fc_2}). MS forbids the download of non-rare pieces. Thus, there is exactly one copy of each piece when the seed has introduced the last piece into the system. 
After that, the number of copies of the non-rare piece and the rarest-piece go hand-in-hand with at most a difference of 1. TMS, on the other hand, allows the download of non-rare pieces till their chunk-count equals the threshold (twice the size of the file). After that, no peer is allowed to download those pieces. Therefore, by the time, the last piece is introduced into the system, there are many peers who have yet to download a large number of pieces to depart. 

Comparing with MS and TMS, both RFwPMS and RNwPMS allow the (probabilistic) download of non-rare pieces and since $\kw$ is sufficiently large, by the time the last $\kw^{\mathrm{th}}$ piece is introduced into the system, the largest mismatch $\udl{m}_W$ has increased considerably coupled with proportional decrease in $\zeta_W$. At this point, the largest mismatch is very large, but owing to the rarest-first mechanism, RFwPMS quickly stabilizes the largest mismatch by giving preference to \textit{the rarest of all available rare pieces} in all peer encounters. On the other hand, RNwPMS does not consider the frequency differences of the rare pieces. Thus, if $\kw$ is sufficiently large, the system is likely to get trapped in a one-club type state where almost all peers have all the pieces except the rarest one. Once this happens, only the seed can flush such peers from the system, which leads to a larger file-delivery time than that of MS.

\begin{figure*}[!t]
	\centering
	\subfloat[RFwPMS.]{\includegraphics[width=0.45\linewidth]{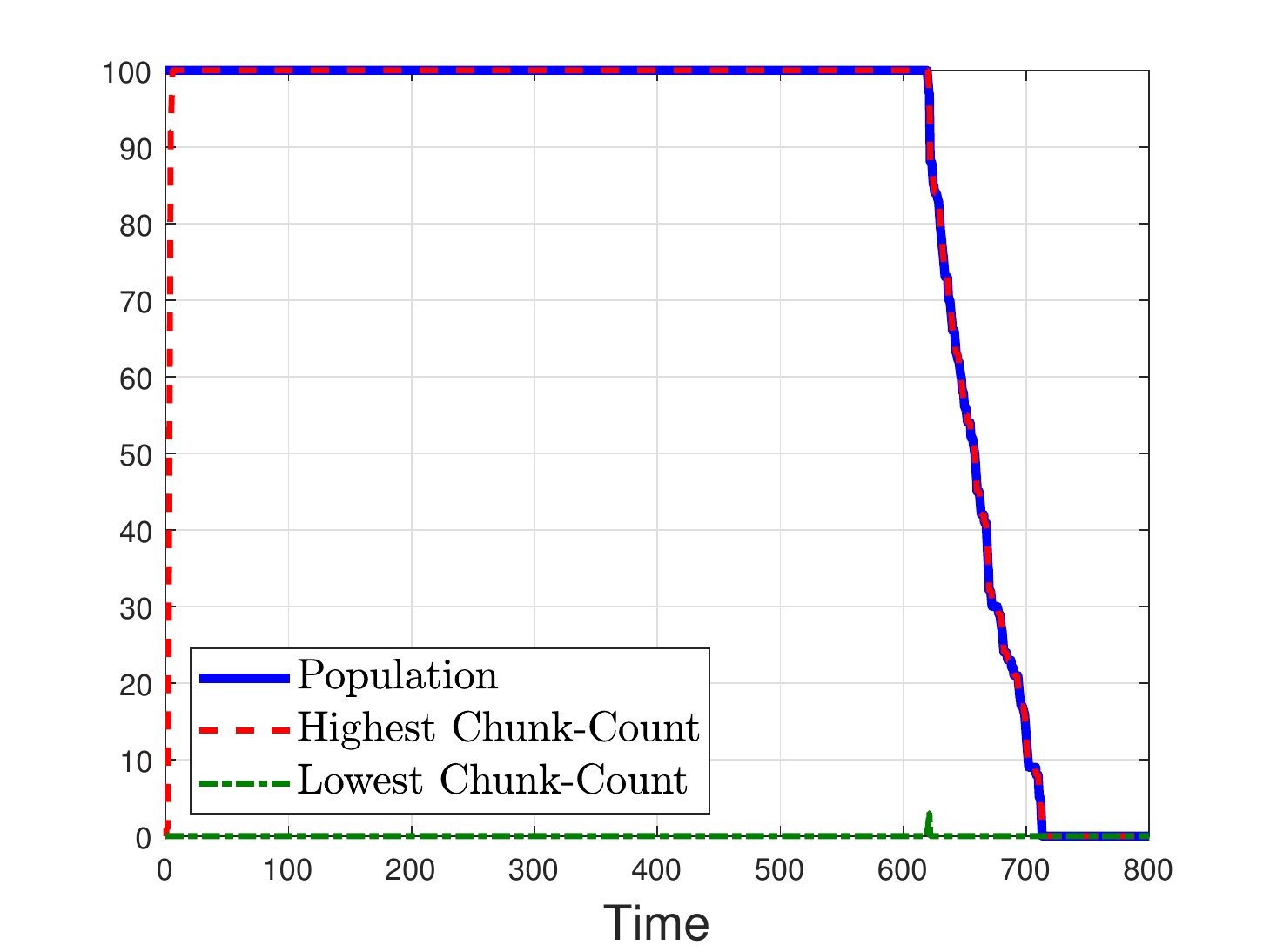}
		\label{fig:fc_3_a}}
	\hfil
	\subfloat[MS.]{\includegraphics[width=0.45\linewidth]{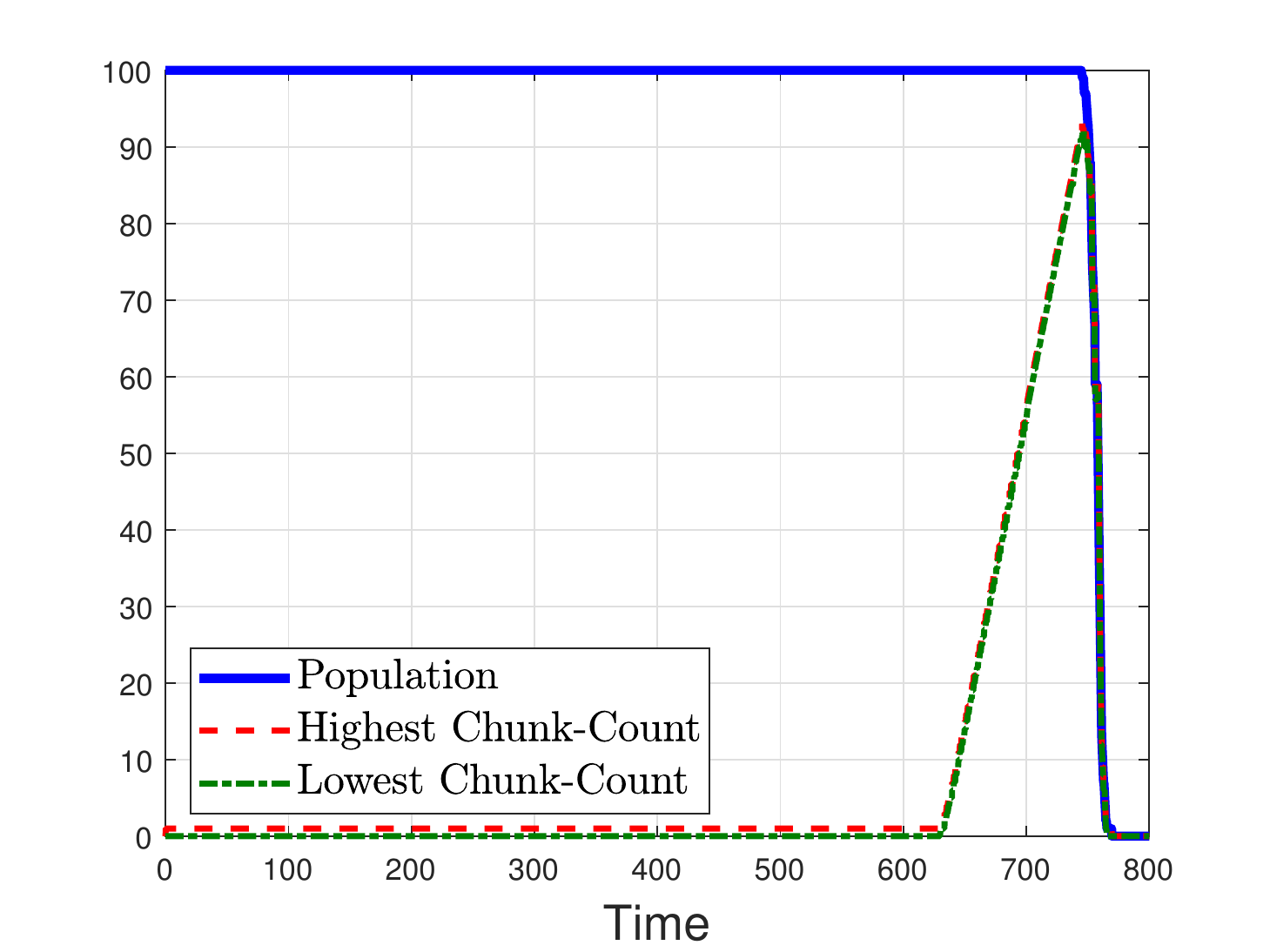}
		\label{fig:fc_3_b}}
	\caption{Flash-crowd responses of RFwPMS and MS when file-size is much larger than flash-crowd. Initial population and file-size were set to 100 and 600 respectively. Assignment of other model-parameters was $\lambda_W=0$, $U=\frac{1}{3}$, $\mu=\wh{\mu}=1$, $L=3$, $p=0.5$, $Y_{opt}=1$.}
	\label{fig:fc_3}
\end{figure*}
\begin{figure*}[!t]
	\centering
	\subfloat[$\beta_W=0.5$.]{\includegraphics[width=0.3\linewidth]{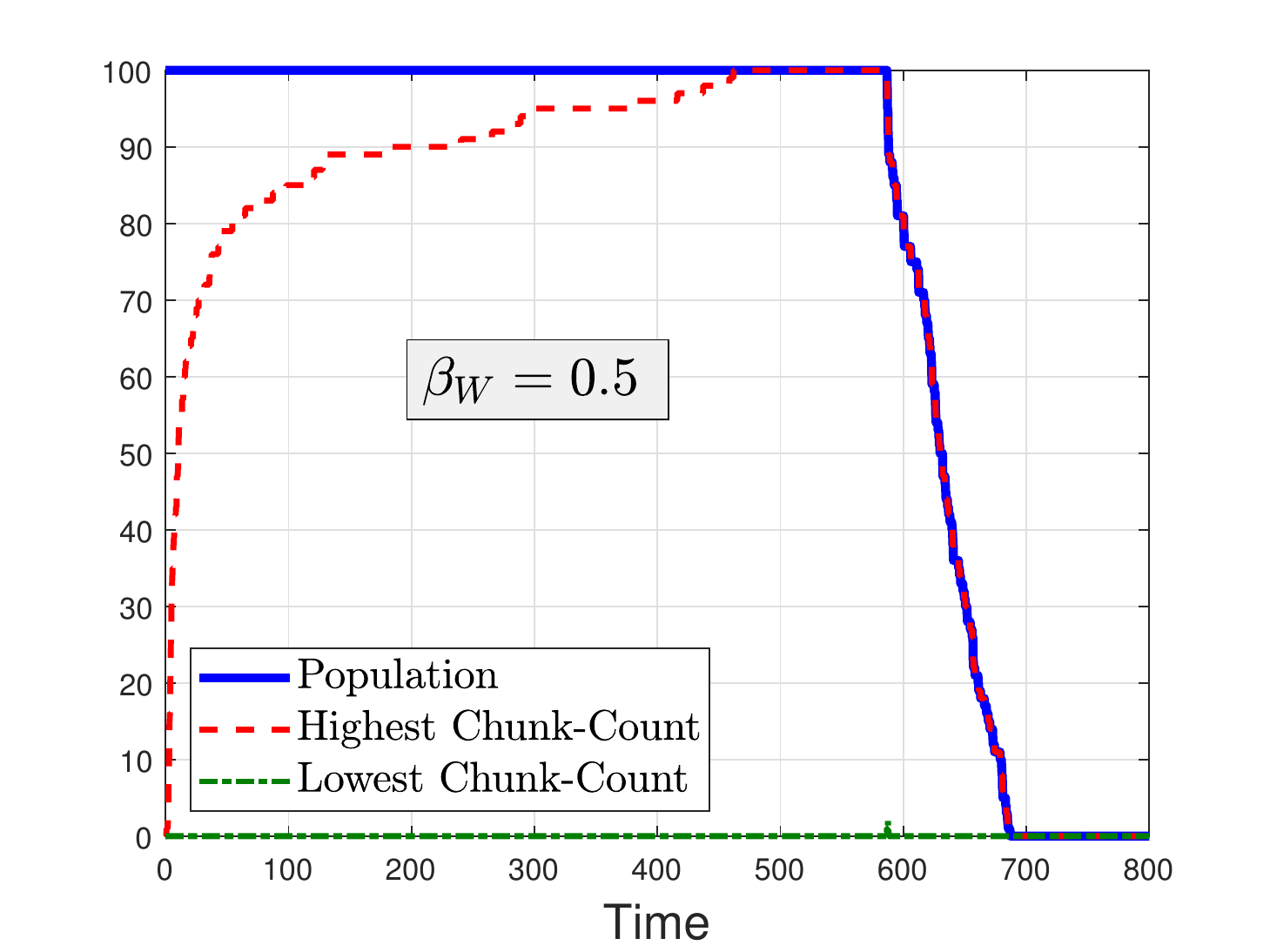}
		\label{fig:fc_4_a}}
	\hfil
	\subfloat[$\beta_W=0.4$.]{\includegraphics[width=0.3\linewidth]{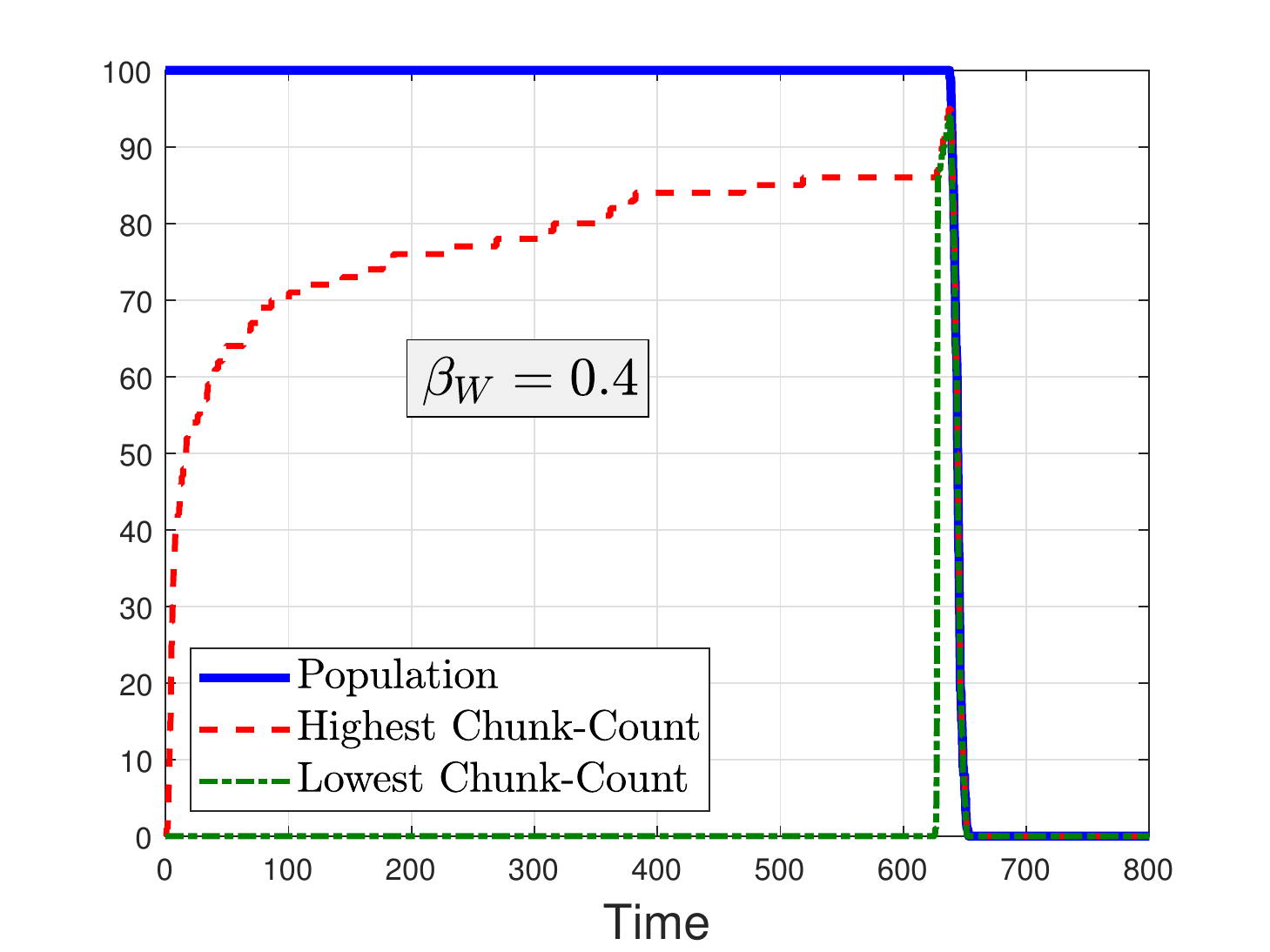}
		\label{fig:fc_4_b}}
	\hfil
	\subfloat[$\beta_W=0.2$.]{\includegraphics[width=0.3\linewidth]{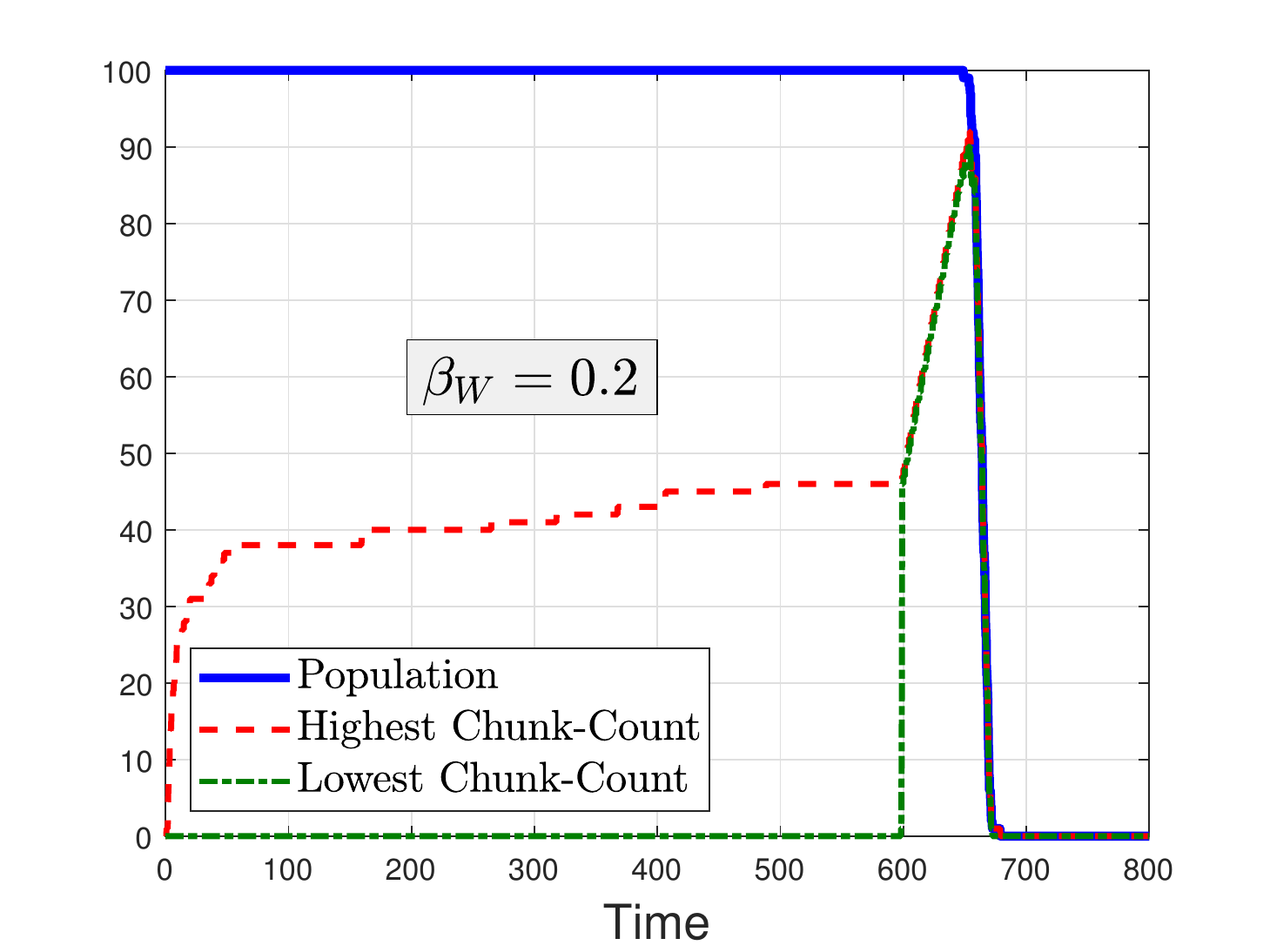}
		\label{fig:fc_4_c}}
	\caption{Flash-crowd response of RFwPMS when file-size is much larger than flash-crowd. The non-rares' sharing factor given by \eqref{eq:zetawnew} was used with three different choices of $\beta_W$. Mode parameters were the same as given in the caption of \figurename \ref{fig:fc_3}.}
	\label{fig:fc_4}
\end{figure*}
It is worth noting that RFwPMS can also get trapped in such one-club like states. This is specially true when the file-size $\kw$ is much larger than the size of the flash-crowd ($\kw \gg |\statex_W|(0)=x_{W}^{(\phi)}(0)$). As an example, \figurename \ref{fig:fc_3_a} shows such a flash-crowd situation in which RFwPMS gets locked into the one-club like state, whereas MS, like before, avoids this (\figurename \ref{fig:fc_3_b}). (Note that the flush-out time of MS is still larger than RFwPMS). The locking-in of RFwPMS in this one-club like state is because the non-rares' sharing factor $\zeta_W$ defined in \eqref{eq:zetaw} suppresses non-rare pieces via the ratio of the (largest) mismatch to the file-size. Since the mismatch is upper-bounded by the network-population (which in this case is much smaller than $\kw$), the suppression is simply not small enough to avoid the download of non-rare pieces.\footnote{This locking-in of RFwPMS would still occur if we suppress each piece $i$ through the ratio $\frac{\nmbr{i}{W} - \nmbrr{W}}{\kw}$.} This observation while simple is worth noting as it is reflective of what should happen in practice. Usually files shared over such networks would have chunk-sizes much larger than the size of the flash-crowd. In such transient situations, to obtain a good trade-off between suppression and sharing, it would make sense to have a non-rares' sharing factor that takes into account the flash-crowd size (besides network's parameters and the file-size). One way we could choose a $\zeta_W$ that can work for both types of flash-crowds would be to replace it by
\begin{align*}\label{eq:zetawnew}
\exp\left(-\frac{\Delta_1}{\beta_W LU} \frac{\mismatchr{W} + \dnmbr{i}{W} }{\min\{\kw, |\statex|\}}\right).\numberthis
\end{align*}
To this end, \figurename \ref{fig:fc_4} shows the flash-crowd response of RFwPMS for the aforementioned choice of $\zeta_W$ with different values of $\beta$ namely $0.2$, $0.4$, and $0.5$. When $\beta_W$ is set to 0.4, it seems to replicate the response of RFwPMS in \figurename \ref{fig:fc_2} and attains the least flush-out time.

\section{Conclusions}\label{sec:conclusion}
In this work, we proposed and studied a piece-selection policy, (swarm-based) RFwPMS, for a BitTorrent-like P2P network with multiple swarms of non-persistent users. RFwPMS combines rarest-first (for rare pieces) with an adjustable sharing-versus-suppression choice for non-rare pieces. Using Lyapunov drift analysis, we proved the stability of RFwPMS in a multi-swarm setting, independent of how peers use their excess cache, and whether or not swarms collaborate with each other. By using the Kingman's moment bound technique, we established the scalability of RFwPMS in the single-swarm case. Our numerical simulations also demonstrated evidence that RFwPMS can reduce the steady-state expected sojourn times as well as file-delivery times during a flash-crowd. 
From our numerical experiments, we note the following:
\emph{In the setting of non-persistent peers, the question of complete vs partial suppression makes sense for both steady-state and flash-crowd performance. In high arrival rate regime, the difference between the two in terms of average file-download times diminishes due to wide variety of chunk-profiles. On the other hand, in low arrival-rate regime, a fine choice of partial suppression that keeps into account the network population would lead to better file-download times.}

Lastly, since RFwPMS uses rarest-first with a minor modification (use of the non-rares' sharing factor), it is amenable to be incorporated into BitTorrent-like protocols. 

In regards to future-work, proving (or disproving) Conjecture \ref{conjctr:alpha0} and establishing the scalability of RFwPMS in the multi-swarm case are interesting directions to explore.

\appendices


\renewcommand{\theequation}{\thesection.\arabic{equation}}

\section{Two-sided Piece Exchanges in the Same Swarm}\label{sec:appendix:pieceexchanges}
Here, we consider the event that both $(W,S)$ and $(W,T)$ peers download pieces $i$ and $j$ respectively. 

\subsection{Change in $V_{W,1}$} 

\udl{Case 1: Both $i$ and $j$ are swarm-$W$'s non-rare pieces}: If both peers remain in the system, $\nmbrn{W}, \nmbr{i}{W}, \nmbr{j}{W} $ increase by 1; $\nmbr{i'}{W}$ stays the same for all $i'\in W\setminus \{i,j\}$. If both peers depart, $\nmbrn{W}$, $\nmbr{i}{W}$, $\nmbr{j}{W}$ decrease by 1; $\nmbr{i'}{W}$ decreases by 2 for all $i'\in W\setminus \{i,j\}$. If one peer departs and the other stays, $\nmbrn{W}, \nmbr{i}{W}, \nmbr{j}{W}$ stay the same; $\nmbr{i'}{W}$ decreases by 1 for all $i'\in W\setminus\{i,j\}$. Overall, in all cases, the total mismatch $\totalmismatch{W}$ increases by $\kw-2$ and the highest chunk-count reduces by at most 1. Overall, this causes an increment of at most $\kw-2+\eta_W\le \kw$ in $\totalmismatch{W}-\eta\nmbrn{W}$. Therefore,
\begin{align*}
&{\Delta V}_{W,1}^{\{(S,i\pm),(T,j\pm)\}}\\
&\le \left(\left( \totalmismatch{W}-\eta\nmbrn{W}+\kw \right)^+\right)^2-\left(\left( \totalmismatch{W}-\eta\nmbrn{W} \right)^+\right)^2\\
&\le \ov{\psi}_W \le 2\ov{\psi}_W.
\end{align*}

\udl{Case 2A - $i$ and $j$ are swarm-$W$'s rare-pieces and $\statex\in\mcl{S}_W^1$}: If both peers remain in the system, $\nmbrn{W}$ stays the same; $\nmbr{i}{W}$, $\nmbr{j}{W}$ increase by 1; $\nmbr{i'}{W}$ stays the same for all $i'\in W\setminus\{i,j\}$. If both peers depart the system, $\nmbrn{W}$ decreases by 2; $\nmbr{i}{W}$ and $\nmbr{j}{W}$ each decrease by 1; $\nmbr{i'}{W}$ also decreases by 2 for all $i'\in W\setminus\{i,j\}$. If one peer departs and the other stays, $\nmbrn{W}$ and $\nmbr{i'}{W}$ ($i'\in W\setminus \{i,j\}$) decrease by 1; $\nmbr{i}{W}$ and $\nmbr{j}{W}$ remain the same. Overall, in all cases, the total mismatch $\totalmismatch{W}$ reduces by 2 and $\nmbrn{W}$ decreases by 2 at most. Therefore,
\begin{align*}
&{\Delta V}_{W,1}^{\{(S,i\pm),(T,j\pm)\}}\\
&\le
\left(\left( \totalmismatch{W}-\eta\nmbrn{W}-2(1-\eta) \right)^+\right)^2 -
\left(\left( \totalmismatch{W}-\eta\nmbrn{W}\right)^+\right)^2\\
&=\begin{cases}
2\left[2(1-\eta)^2-2(1-\eta)\left(\totalmismatch{W}-\eta\nmbrn{W}\right)\right] 
&\begin{array}{@{}l@{}}
\text{if $\totalmismatch{W}\ge$}\\
2(1-\eta)+\eta\nmbrn{W}     
\end{array}\\
0 &\text{otherwise}.
\end{cases}\\
&\le 2\udl{\psi}_W.
\end{align*}

\udl{Case 2B - $i$ and $j$ are swarm-$W$'s rare-pieces and $\statex\in\mcl{S}_W^2$}: If both peers remain in the system, $\nmbrn{W}, \nmbr{i}{W}, \nmbr{j}{W}$ increase by 1; $\nmbr{i'}{W}$ remains the same for all $i'\in W\setminus \{i,j\}$. If both peers depart the system, $\nmbrn{W}, \nmbr{i}{W}, \nmbr{j}{W}$ decrease by 1; $\nmbr{i'}{W}$ decreases by 2 for all $i'\in W\setminus \{i,j\}$. If one peer departs and the other stays, $\nmbrn{W}, \nmbr{i}{W}, \nmbr{j}{W} $ stay the same; $\nmbr{i'}{W}$ decreases by 1 for all $i'\in W\setminus \{i,j\}$. Overall, in all cases, the total mismatch $\totalmismatch{W}$ increases from 0 to $\kw-2$ and $\nmbrn{W}$ decreases by at most 1. Therefore,
\begin{align*}
&{\Delta V}_{W,1}^{\{(S,i\pm),(T,j\pm)\}}\\
&\le\left(\left( \kw-2+\eta-\eta\nmbrn{W}\right)^+\right)^2-\left(\left( -\eta\nmbrn{W}\right)^+\right)^2\\
&\le (\kw+2)^2\le2(\kw+2)^2\le 2\udl{\psi}_W.
\end{align*}

\udl{Case 3 - $i$ is a rare-piece and $j$ a non-rare piece}: If both peers stay in the system, $\nmbrn{W}, \nmbr{i}{W}, \nmbr{j}{W}$ increase by 1; $\nmbr{i'}{W}$ stays the same for all $i'\in W \setminus\{i,j\}$. If both peers depart, $\nmbrn{W}, \nmbr{i}{W},\nmbr{j}{W} $ decrease by 1; $\nmbr{i'}{W}$ decreases by 2 for all $i'\in W\setminus \{i,j\}$.  If one peer departs and the other stays, $\nmbrn{W},\nmbr{i}{W}, \nmbr{j}{W}$ stay the same; $\nmbr{i'}{W}$ decreases by 1 for all $i'\in W\setminus\{i,j\}$. Overall, in all cases, the total mismatch $\totalmismatch{W}$ increases by $\kw-2$ and $\nmbrn{W}$ decreases by at most 1. Therefore,
\begin{align*}
&{\Delta V}_{W,1}^{\{(S,i\pm),(T,j\pm)\}}\\
&=\left(\left(\totalmismatch{W}-\eta\nmbrn{W}+\kw-2+\eta \right)^+ \right)^2-\left(\left(\totalmismatch{W}-\eta\nmbrn{W}\right)^+ \right)^2\\
&
\labelrel{\le}{eqr:twosided:31} \left(\left(\totalmismatch{W}-\eta\nmbrn{W}+\kw-1 \right)^+ \right)^2-\left(\left(\totalmismatch{W}-\eta\nmbrn{W} \right)^+ \right)^2\\
&
\labelrel{\le}{eqr:twosided:32} \left(\left(\totalmismatch{W}-\eta\nmbrn{W}+\kw \right)^+ \right)^2-\left(\left(\totalmismatch{W}-\eta\nmbrn{W} \right)^+ \right)^2\\
&+\left(\left(\totalmismatch{W}-\eta\nmbrn{W}-1 \right)^+ \right)^2-\left(\left(\totalmismatch{W}-\eta\nmbrn{W} \right)^+ \right)^2\\
&
\labelrel{\le}{eqr:twosided:33}
\left(\left(\totalmismatch{W}-\eta\nmbrn{W}+\kw \right)^+ \right)^2-\left(\left(\totalmismatch{W}-\eta\nmbrn{W} \right)^+ \right)^2\\
&+\left(\left(\totalmismatch{W}-\eta\nmbrn{W}-(1-\eta) \right)^+ \right)^2-\left(\left(\totalmismatch{W}-\eta\nmbrn{W} \right)^+ \right)^2\\
&
\le\ov{\psi}_W+\udl{\psi}_W.
\end{align*}
Here, step \eqref{eqr:twosided:31} uses $\eta<1$; \eqref{eqr:twosided:33} uses $\eta>0$; step \eqref{eqr:twosided:32} uses the inequality
\begin{align*}
&\left((x+a)^+\right)^2-\left((x)^+\right)^2 \le \left((x+1+a)^+\right)^2-\left((x)^+\right)^2 \\
&+ \left((x-1)^+\right)^2-\left((x)^+\right)^2 \text{ for all $x\in\mb{R}, a\ge 0$}.
\end{align*}
The case when $i$ is a non-rare piece and $j$ is a rare piece is similar.

The above inequalities ensure that we can write
\begin{align*}\label{eq:twosided1}
{\Delta V}_{W,1}^{\{(S,i\pm),(T,j\pm)\}}&\le\sum_{i'=i,j} \ov{\psi}_W \11{\{i'\in\rwc\}}+\udl{\psi}_W \11{\{i'\in R_W\}}.\numberthis
\end{align*}

\subsection{Change in $V_{W,2}$} 
For $V_{W,2}$, note that $\kw\xw-\pw=\sum_{i\in W}\xw-\nmbr{i}{W}$. If both peers stay in the system, $\xw$ stays the same; $\nmbr{i}{W}, \nmbr{j}{W}$ increase by 1; $\nmbr{i'}{W}$ stays the same for all $i'\in W\setminus\{i,j\}$. If both peers depart, $\xw$ decreases by 2; $\nmbr{i}{W},\nmbr{j}{W}$ decrease by 1; $\nmbr{i'}{W}$ decrease by 2 for all $i'\in W\setminus\{i,j\}$. If one peer leaves and the other departs; $\xw$ decreases by 1; $\nmbr{i}{W},\nmbr{j}{W}$ stay the same; $\nmbr{i'}{W}$ decreases by 1 for all $i'\in W \setminus\{i,j\}$. In all cases, the deficit $\xw-\nmbr{i'}{W}$ stays the same for all $i'\in W\setminus\{i,j\}$ and decreases by 1 for $i'\in\{i,j\}$. Thus, we can write
\begin{align*}\label{eq:twosided2}
{\Delta V}_{W,2}^{\{(S,i\pm),(T,j\pm)\}}&= -2\cw2.\numberthis
\end{align*}

\subsection{Change in $V_{W,3}$} 
The change in $V_{W,3}$ depends on whether peers depart or not, and does not depend on the type of $i$ and $j$. There are three cases. If both peers stay in the system, $\pw$ increases by 2. Therefore, 
\begin{align*}\label{eq:twosided3}
{\Delta V}_{W,3}^{\{(S,i+),(T,j+)\}}&\le
-2\cw2\11{\{\mw-2\ge \pw \}} = -2I_W^{(1)}.\numberthis
\end{align*}
If only one of the two peers depart, then $\pw$ decreases by $\kw-2$. Therefore,
\begin{align*}\label{eq:twosided4}
&{\Delta V}_{W,3}^{\{(S,i+/-),(T,j-/+)\}}\\
&\le
\cw2(\kw-2) \11{\{\mw+\kw-2\ge\pw\}}\\
&\le \cw2\kw\11\{  \mw+2\kw\ge\pw  \}-\cw2\11\{\mw-2\ge\pw\}\\
&=I_W^{(2)}-I_W^{(1)}. \numberthis
\end{align*}
If both peers depart the system, $\pw$ decreases by $2(\kw-1)$. Therefore,
\begin{align*}\label{eq:twosided5}
&{\Delta V}_{W,3}^{\{(S,i-),(T,j-)\}}\\
&\le \cw2(2(\kw-1) ) \11{\{ \mw+2(\kw-1) \ge \pw\}}\\
&\le 2\cw2\kw\11\{ \mw+2\kw \ge \pw \} 
= 2I_W^{(2)}.\numberthis
\end{align*}
Equations \eqref{eq:twosided1} to \eqref{eq:twosided5} establish the bounds in Table \ref{tab:upperbounds_potchange_pieceexchanges}.

\section{Bounding Rarest-First For Rare Pieces Using Random-Novel For Rare Pieces}\label{sec:appendix:rfrn}
\begin{lem}\label{lem:rfrn}
For all $\statex$,
\begin{align*}
&{QV}_W^{(\rw+)} \le \sum_{\substack{r \in \rw, \\ S: W \setminus S \supsetneq \{r\}}} \frac{ \xws }{\x} \left[ \frac{ LU }{ |\rs| } \right.\\
&\left. + \Delta_p\xi_1 \sumvtr \frac{\xvt}{|\trs|} \right] \Psiwr{W}{+}.
\end{align*}
\end{lem}
\begin{IEEEproof}
Using \eqref{eq:qwsrplus}, \eqref{eq:stab:vwsiplus} and the fact that $\Psi_W^{(r+)}\le 0 $ for every $r\in\rw$,
\begin{align*}
&{QV}_W^{(\rw+)} \le \sum_{\substack{r \in \rw, \\ S: W \setminus S \supsetneq \{r\}}} \frac{ \xws }{ \x } \left[ \frac{ LU \indrares{\mcl{F}} }{ |\setrares{\mcl{F}}| } \right.\\
&\hspace{10pt} \left. + \Delta_p\xi_1 \sumvtr \frac{ \xvt \indrares{T} }{|\setrares{T}| } \right]\Psiwr{W}{r+} \\
&\labelrel{=}{eqr:rfrn:11} \sum_{\substack{r \in \rw, \\ S: W \setminus S \supsetneq \{r\}}} \frac{ \xws }{ \x } \left[ \frac{ LU \indrares{\mcl{F}} }{ |\setrares{\mcl{F}}| } \right.\\
&\hspace{10pt} \left. + \Delta_p\xi_1 \sumvtr \frac{ \xvt \indrares{T} }{|\setrares{T}|}  \right] \Psiwr{W}{+}.\numberthis\label{eq:rfrn0}
\end{align*}
Here, \eqref{eqr:rfrn:11} uses the fact that $\Psi_W^{(r+)}$ is the same for all $r\in\rw$, which we have denoted by $\Psi_W^{(+)}$. Then, we can upper bound the first term as follows.
\begin{align*}
&\frac{\text{First Term }}{LU} = \sum_{\substack{r \in \rw, \\ S: W \setminus S \supsetneq \{r\}}} \frac{ \xws }{ \x }  \frac{ \indrares{\mcl{F}}  }{ | \setrares{\mcl{F}} | } \Psiwr{W}{+} \\
&\labelrel{=}{eqr:stepa1} \sumskrnew \frac{ \xws }{ \x } \frac{ \indrares{\mcl{F}}  }{ | \setrares{\mcl{F}} | } \Psiwr{W}{+}\\
&= \sumskrarestnew \frac{ \xws }{ \x } \frac{ 1 }{ |\setrares{\mcl{F}}| } \Psiwr{W}{+}\\
&= \sum \limits_{ S:|S \cap W|<\kw-1} \frac{ \xws }{ \x } \Psiwr{W}{+}\\
&= \sum \limits_{ S:|S \cap W|<\kw-1} \frac{ \xws }{ \x } \frac{|\rs|}{|\rs|} \Psiwr{W}{+}\\
&=\sumskrnew \frac{ \xws }{ \x } \frac{1}{ | \rs | } \Psiwr{W}{+}\\
&\labelrel{=}{eqr:stepa3} \sum_{\substack{r \in \rw, \\ S: W \setminus S \supsetneq \{r\}}} \frac{ \xws }{ \x } \frac{1}{ | \rs | } \Psiwr{W}{+}.\numberthis\label{eq:rfrn:firstterm}
\end{align*}
In steps \eqref{eqr:stepa1} and \eqref{eqr:stepa3}, we change the order of summation. Similarly, we can upper bound the second term as follows.
\begin{align*}
&\frac{\text{Second Term }}{\Delta_p\xi_1} = \hspace{-15pt}  \sumrsvt \frac{ \xws }{ \x } \frac{ \xvt \indrares{T} }{ |\setrares{T}| } \Psiwr{W}{+} \\
& \labelrel{=}{eqr:stepb1} \sum \limits_{ \substack{ S:|S \cap W|<\kw-1,\\ V: W\in\mcl{W}_V, T: \trs \ne \emptyset,\\ r:r\in\trs  } } \frac{ \xws }{ \x }
\frac{\xvt \indrares{T} }{ |\setrares{T}| } \Psiwr{W}{+}\\
&= \sum \limits_{ \substack{ S:|S \cap W|<\kw-1,\\ V: W\in\mcl{W}_V, T: \trs \ne \emptyset,\\r:r\in\setrares{T}  } } \frac{ \xws }{ \x }
\frac{\xvt }{ |\setrares{T}| } \Psiwr{W}{+}\\
&=\sum \limits_{ \substack{ S:|S \cap W|<\kw-1,\\ V: W\in\mcl{W}_V, T: \trs \ne \emptyset  } } \frac{ \xws }{ \x }
\xvt \Psiwr{W}{+}\\
&=\sum \limits_{ \substack{ S:|S \cap W|<\kw-1,\\ V: W\in\mcl{W}_V, T: \trs \ne \emptyset  } } \frac{ \xws }{ \x }
\frac{ | \trs | \xvt }{ |\trs| } \Psiwr{W}{+}\\
&= \sum \limits_{ \substack{ S:|S \cap W|<\kw-1,\\ V: W\in\mcl{W}_V, T: \trs \ne \emptyset,\\r:r\in\trs  } } \frac{ \xws }{ \x }
\frac{ \xvt }{ |\trs| } \Psiwr{W}{+}\\
&\labelrel{=}{eqr:stepb3}  \sumrsvt \frac{ \xws }{ \x }
\frac{ \xvt }{ |\trs| } \Psiwr{W}{+}.\numberthis\label{eq:rfrn:secondterm}
\end{align*}
In steps \eqref{eqr:stepb1} and \eqref{eqr:stepb3}, we change the order of summation. 

Note that both \eqref{eq:rfrn:firstterm} and \eqref{eq:rfrn:secondterm} are upper-bounds for the RNwPMS (swapping out RF by RN). Substituting the two in \eqref{eq:rfrn0}, we get
\begin{align*}
&{QV}_W^{(\rw+)} \le \sum_{\substack{r \in \rw, \\ S: W \setminus S \supsetneq \{r\}}} \frac{ \xws }{\x} \left[ \frac{ LU }{ |\rs| } \right.\\
&\hspace{10pt} \left. + \Delta_p\xi_1 \sumvtr \frac{\xvt}{|\trs|} \right] \Psiwr{W}{r+}.
\end{align*}
\end{IEEEproof}

\section{Combining ${QV}_W^{(\rw+)}$ and ${QV}_W^{(\rwc+)}$}\label{sec:appendix:getrid}
\begin{lem}\label{lem:getrid}
For all $\statex$,
\begin{align*}
QV_W^{(W+)}&\le \frac{\xw}{\x} \sum_{i\in W}\frac{\udl{\Gamma}_W^{(i)}a_W^{(i)}}{\kw}\left(1-\freq{i}{W}-\gamma_W^{(i)}\right) \\
&\hspace{-30pt} \times \left[ \11{\{\statex\in\regrw{W}{1}\}}\left(\kw\xi_2\ov{\psi}_W\11{\{i\in\rwc\}}+\udl{\psi}_W\11{\{i\in \rw\}}\right) \right.\\
&\hspace{-10pt}\left. +\11{\{\statex\in\regrw{W}{2}\}}4\kw^2-\cw1-I_W^{(1)}\right].
\end{align*}
\end{lem}
\begin{IEEEproof}
Let
\begin{align*}
g(i,\statex)&=\11\{\statex\in\regrw{W}{2}\}4\kw^2-\cw1-I_W^{(1)} \\
&\hspace{10pt} +\11\{\statex\in\regrw{W}{1},i\in\rw\}\udl{\psi}_W \le 0.
\end{align*}
Adding \eqref{eq:stab:qvwrplus} with the first term in \eqref{eq:stab:qvwnplus}, and using the definitions of $\udl{q}_W^{(S,n+)}$ (see \eqref{eq:qwsnplus2}) and $\Psi_W^{(i+)}$ (see \eqref{eq:stab:vwsiplus}), we get
\begin{align*}\label{eq:added0}
&\sum\limits_{\substack{i\in W,\\S:W\setminus S \supsetneq\{i\}}} \frac{\xws}{\x} \left[ LU\left(\frac{\zeta_W^{(i)} \11{\{i\in\rwc, \rw\setminus S=\emptyset\}}}{|W \setminus S|} \right.\right.\\
&\hspace{10mm} \left.\left. +\frac{\11{\{i\in \rw\}}}{|\rw\setminus S|}\right) + 
\Delta_p\xi_1\sum_{\substack{V:W\in\mcl{W}_V\\T:i\in T }} \xvt \right.\\
&\hspace{10mm}\left. \times \left(\frac{\zeta_W^{(i)}\11{\{i\in\rwc,(T\cap \rw)\setminus S=\emptyset\}}}{|(T\cap W)\setminus S|} \right.\right.\\
&\hspace{10mm}\left.\left. +\frac{\11{\{i\in \rw\}}}{|(T\cap \rw)\setminus S|}\right)\right]g(i,\statex).\numberthis
\end{align*}
Here,
\begin{align*}
	&\frac{\text{First Term}}{LU}\\ 
	&= \left[\sum \limits_{\substack{
			i\in\rwc,\\ W\setminus S\supsetneq \{i\},\\ \rw\setminus S=\emptyset}} \frac{\zeta_W^{(i)}}{|W \setminus S|} + \sum \limits_{\substack{
			i\in \rw\\ W\setminus S\supsetneq \{i\}}} \frac{1}{|\rw \setminus S|}\right]\frac{\xws}{\x}g(i,\statex)\\
	&
	\labelrel{=}{eqr:stepb10}\left[\sum \limits_{\substack{
			|S\cap W|<\kw-1,\\ \rw \setminus S=\emptyset, i\in W\setminus S}} \frac{\zeta_W^{(i)}}{| W \setminus S|} \right. \\
			&\hspace{20pt} \left. + \sum \limits_{\substack{
			|S\cap W|<\kw-1,\\ \rw\setminus S\ne\emptyset, i\in \rw \setminus S}} \frac{1}{|\rw \setminus S|}\right]\frac{\xws}{\x}g(i,\statex)\\
	&
	\labelrel{\le}{eqr:stepb11}\left[\sum \limits_{\substack{
			|S\cap W|<\kw-1,\\ \rw\setminus S=\emptyset, i\in W\setminus S}} \frac{\zeta_W^{(i)}}{|W \setminus S|} \right.\\
			&\hspace{20pt} \left. + \sum \limits_{\substack{
			|S\cap W|<\kw-1,\\ \rw \setminus S\ne\emptyset, i\in \rwc \setminus S}} \frac{\zeta_W^{(i)}}{| W\setminus S|} \right.\\
			&\hspace{20pt} \left. + \sum \limits_{\substack{
			|S\cap W|<\kw-1,\\ \rw \setminus S\ne\emptyset, i\in \rw \setminus S}} \frac{1}{|W \setminus S|} \right]\frac{\xws}{\x} g(i,\statex)\\
	&
	=\left[\sum \limits_{\substack{
			|S\cap W|<\kw-1,\\ i\in \rwc \setminus S}} \frac{\zeta_W^{(i)}}{|W \setminus S|} \right.\\
			&\hspace{20pt} \left. + \sum \limits_{\substack{
			|S\cap W|<\kw-1,\\ \rw \setminus S\ne\emptyset, i\in \rw \setminus S}} \frac{1}{|W \setminus S|} \right]\frac{\xws}{\x} g(i,\statex)\\
	&
	\labelrel{=}{eqr:stepb12}\left[\sum \limits_{\substack{
			i\in\rwc,\\
			W\setminus S\supsetneq\{i\}}} \frac{\zeta_W^{(i)}}{|W \setminus S|} + \sum \limits_{\substack{
			i\in \rw, \\
			W\setminus S \supsetneq \{i\}}} \frac{1}{|W \setminus S|} \right]\frac{\xws}{\x} g(i,\statex)\\
	&
	\labelrel{\le}{eqr:stepb13}\sum \limits_{\substack{
			i\in W,\\W\setminus S\supsetneq\{i\}}} \frac{\xws}{\x} \frac{a_W^{(i)}}{\kw} h(i,\statex).\label{eq:firstterm}\numberthis
\end{align*}
Steps \eqref{eqr:stepb10} and \eqref{eqr:stepb12} change the order of summation; step \eqref{eqr:stepb11} uses $\zeta_W^{(i)}\le 1$, $|\rw \setminus S|\le |W \setminus S|$, and $g(n,\statex)\le g(r,\statex)$ for all $n\in\rwc,r\in\rw$; \eqref{eqr:stepb13} uses $|W \setminus S|\le \kw$. Similarly,
\begin{align*}
	&\frac{\text{ Second Term }}{\Delta_p\xi_1}\\
	&=\sum_{V:W\in\mcl{W}_V} \left[\sum\limits_{\substack{i\in\rwc, W\setminus S \supsetneq\{i\},\\i\in T, (T\cap \rw)\setminus S=\emptyset}}\frac{\zeta_W^{(i)}}{|(T\cap W)\setminus S|} \right.\\
	&\left. + \sum\limits_{\substack{i\in \rw, W\setminus S\supsetneq\{i\},\\i\in T}}\frac{1}{|(T\cap \rw)\setminus S|} \right]\frac{\xws\xvt}{\x}g(i,\statex)\\
	&\hspace{0mm}
	\labelrel{=}{eqr:stepb20}
	\sum_{V:W\in\mcl{W}_V}
	\left[\sum\limits_{\substack{S:|S\cap W|<\kw-1,\\ (T\cap \rw)\setminus S=\emptyset, i\in (T\cap W)\setminus S}}\frac{\zeta_W^{(i)}}{|(T\cap W)\setminus S|} \right.\\
	&\left. + \sum\limits_{\substack{|S\cap W|<\kw-1,\\ (T\cap \rw)\setminus S\ne\emptyset,\\
	i\in(T\cap \rw)\setminus S}}\frac{1}{|(T\cap \rw)\setminus S|} \right]\frac{\xws\xvt}{\x}g(i,\statex)\\
	&\hspace{0mm}
	\labelrel{\le}{eqr:stepb21}
	\sum_{V:W\in\mcl{W}_V}
	\left[\sum\limits_{\substack{S:|S\cap W|<\kw-1,\\ (T\cap \rw)\setminus S=\emptyset, i\in (T\cap W)\setminus S}}\frac{\zeta_W^{(i)}}{|(T\cap W)\setminus S|} \right.\\
	&\left. +
	\sum\limits_{\substack{|S\cap W|<\kw-1,\\ (T\cap \rw)\setminus S\ne\emptyset, i\in (T\cap\rwc)\setminus S}}\frac{\zeta_W^{(i)}}{|(T\cap W)\setminus S|} \right.\\
	&\left. + \sum\limits_{\substack{|S\cap W|<\kw-1,\\ (T\cap \rw)\setminus S\ne\emptyset, i\in(T\cap \rw)\setminus S}}\frac{1}{|T\setminus S|} \right]\frac{\xws\xvt}{\x}g(i,\statex)\\
	&\hspace{0mm}
	\labelrel{=}{eqr:stepb22}
	\sum_{V:W\in\mcl{W}_V}
	\left[\sum\limits_{\substack{i\in\rwc, W\setminus S\supsetneq\{i\}\\T:i\in T}}\frac{\zeta_W^{(i)}}{|(T\cap W)\setminus S|} \right.\\
	&\left. +
	\sum\limits_{\substack{i\in \rw, W\setminus S\supsetneq\{i\},\\T:i\in T}}\frac{1}{|(T\cap W)\setminus S|} \right] \frac{\xws\xvt}{\x}g(i,\statex)\\
	&\hspace{0mm}
	\labelrel{\le}{eqr:stepb23}
	\sum\limits_{\substack{i\in W,\\ W\setminus S\supsetneq\{i\}}}  \frac{\xws}{\x}\frac{a_W^{(i)}}{\kw}\sum_{\substack{V:W\in\mcl{W}_V, \\ T:i\in T}}\xvt g(i,\statex).\label{eq:secondterm}\numberthis
\end{align*}
Steps \eqref{eqr:stepb20} and \eqref{eqr:stepb22} change the order of summation; step \eqref{eqr:stepb21} uses $\zeta_W^{(i)}\le 1$, $|(T\cap \rw)\setminus S|\le |(T\cap W)\setminus S|$, and $g(n,\statex)\le g(r,\statex)$ for all $n\in\rwc,r\in\rw$; step \eqref{eqr:stepb23} uses $|(T\cap W)\setminus S|\le \kw$. Adding \eqref{eq:firstterm} and \eqref{eq:secondterm} gives the below upper-bound on \eqref{eq:added0},
\begin{align*}\label{eq:added}
&\sum\limits_{\substack{i\in W,\\W\setminus S\supsetneq\{i\}}}  \frac{\xws}{\x}\frac{\Gammalower{W}{i} a_W^{(i)}}{\kw} \left[ \11\{\statex\in\regrw{W}{2}\}4\kw^2-\cw1-I_W^{(1)}\right.\\
&\left.+\11\{\statex\in\regrw{W}{1},i\in\rw\}\udl{\psi}_W 
\right].\numberthis
\end{align*}
Adding \eqref{eq:added} with the second term in \eqref{eq:stab:qvwnplus}, we get
\begin{align*}
&{QV}_W^{(W+)}\le 
\sum\limits_{\substack{i\in W,\\W\setminus S\supsetneq\{i\}}}  \frac{\xws}{\x}\frac{\Gammalower{W}{i} a_W^{(i)}}{\kw} \left[\11\{\statex\in\regrw{W}{1}\}\right.\\
&\hspace{10pt} \left. \times \left( \11\{i\in\rwc\}\kw\xi_2\ov{\psi}_W + \udl{\psi}_W\11\{i\in\rw\}\right) \right.\\
&\hspace{20pt} \left. +\11\{\statex\in\regrw{W}{2}\}4\kw^2-\cw1-I_W^{(1)}
\right]\\
&\hspace{0pt}=\frac{\xw}{\x}\sum\limits_{\substack{i\in W,\\W\setminus S\supsetneq\{i\}}}  \frac{\xws}{\xw}\frac{\Gammalower{W}{i} a_W^{(i)}}{\kw} \left[\11\{\statex\in\regrw{W}{1}\}\right.\\
&\hspace{10pt} \left. \times \left( \11\{i\in\rwc\}\kw\xi_2\ov{\psi}_W + \udl{\psi}_W\11\{i\in\rw\}\right) \right.\\
&\hspace{20pt} \left. +\11\{\statex\in\regrw{W}{2}\}4\kw^2-\cw1-I_W^{(1)}
\right]\\
&\hspace{0pt}=\frac{\xw}{\x} \sum_{i\in W} \frac{\udl{\Gamma}_W^{(i)}a_W^{(i)}}{\kw}\left(1-\freq{i}{W}-\gamma_W^{(i)}\right) \left[
\11{\{\statex\in\regrw{W}{1}\}} \right. \\
&\hspace{10pt}\left. \times \left(\kw\xi_2\ov{\psi}_W\11{\{i\in\rwc\}}+\udl{\psi}_W\11{\{i\in \rw\}}\right) \right.\\
&\hspace{20pt}\left. +\11{\{\statex\in\regrw{W}{2}\}}4\kw^2-\cw1-I_W^{(1)}
\right],
\end{align*}
where we recall that $\gamma_{W}^{(i)}$ is the fraction of swarm-$W$ peers who have all the pieces of $W$ except $i$.
\end{IEEEproof}

\section{Exponential Decay of Positive Drift from Non-Rare Pieces in $\regrw{W}{1}$ }\label{sec:appendix:upperboundnonrares}
\begin{lem}\label{lem:upperboundnnonrares}
For all $\statex\in\regrw{W}{1}$,
\begin{align*}
&\sum_{n\in\rwc}\frac{\Gammalower{W}{n}\zeta_W^{(n)} }{\kw}\left[ \left(1-\freqn{W} \right)\kw\xi_2\ov{\psi}_W \right] \le D_W.
\end{align*}
where
\begin{align*}
D_W&=D_W(\eta,\xi_2,L,U,\Delta_p,\kw)\\
&\defeq D_W^{(1)} + D_W^{(2)}\11\{\mcl{W}_{W}\setminus \{W\}\ne\emptyset\},\\
D_W^{(1)} &= 12e^{-2}\eta^{-2}\xi_2(LU+\Delta_p)\bw^2\kw^7,\\
D_W^{(2)} &= 3e^{-2}\eta^{-1}\xi_2(LU+\Delta_p)\bw^{1+\aw^{-1}}\kw^{5+\aw^{-1}}.
\end{align*}
\end{lem}
\begin{IEEEproof}
\begin{align*}
&\sum_{n\in\rwc}\frac{\Gammalower{W}{n}\zeta_W^{(n)} }{\kw}\left[ \left(1-\freqn{W} \right)\kw\xi_2\ov{\psi}_W \right]\\
&\hspace{10pt} \labelrel{\le}{eqr:nrub:11} 
\sum_{n\in\rwc} \left(LU+2\Delta_p \left(\nmbrn{W} + \dnmbr{n}{W}\right)  \right) \xi_2 \left(  3\kw^2\nmbrn{W} \right) \zeta_W^{(n)}\\
&\hspace{10pt}\labelrel{\le}{eqr:nrub:12}  3\kw^2\xi_2\left(LU+2\Delta_p\right) 
\sum_{n\in\rwc} \left(\nmbrn{W} + \dnmbr{n}{W}\right) \nmbrn{W} \zeta_W^{(n)}.\numberthis\label{eq:upperboundnonrares0}
\end{align*}
Here, \eqref{eqr:nrub:11} uses $\ov{\psi}_W\le 3\kw^2\nmbrn{W}$ in $\regrw{W}{1}$; \eqref{eqr:nrub:12} uses $1\le \nmbrn{W} + \dnmbr{n}{W}$. Now, in $\regrw{W}{1}$, $\totalmismatch{W}\ge\eta\nmbrn{W}$. This implies that
\begin{align*}
\ov{m}_W=\nmbrn{W}-\nmbrr{W}\ge
\frac{\eta\nmbrn{W}}{\kw}.
\end{align*}
Therefore, 
\begin{align*}
\zeta_W^{(n)}\le \exp\left( -\frac{\eta\nmbrn{W}\kw^{-1} + \left(\dnmbr{n}{W} \right)^{\aw} }{\beta_W \kw} \right).
\end{align*}
If $\beta_W=0$, then  $\zeta_W^{(n)}=0$ (non-rare pieces are completely suppressed) and the Lemma follows trivially. If $\beta_W>0$, then
\begin{align*}
\sum_{n\in\rwc} \nmbrn{W}^2 \zeta_W^{(n)} &\hspace{0pt}\le \kw \max\limits_{y\ge 0} y^2\exp\left(-\frac{\eta}{\beta_W\kw^2}y^2 \right)\\
&\hspace{0pt}=\kw \left(2\frac{\bw\kw^2}{\eta}\right)^2e^{-2}\\
&\hspace{0pt}= 4e^{-2} \eta^{-2} \beta_W^2 \kw^5,\numberthis\label{eq:upperboundnonrares:firstterm}
\end{align*}
and
\begin{align*}
&\sum_{n\in\rwc} \nmbrn{W}\dnmbr{n}{W} \zeta_W^{(n)} = 
\sum_{n\in\rwc} \nmbrn{W}\dnmbr{n}{W} \zeta_W^{(n)}\11\{\mcl{W}_W\setminus \{W\}\ne\emptyset\}\\ &\hspace{0pt}\le \kw \max\limits_{y\ge 0} y\exp\left(-\frac{\eta}{\bw\kw^2}y \right)\\
&\hspace{30pt} \times
\max\limits_{z\ge 0} z\exp\left( -\frac{z^{\aw}}{\bw\kw}\right)\11\{\mcl{W}_W\setminus \{W\}\ne\emptyset\}\\
&\hspace{0pt}=\kw \left(\frac{\bw\kw^2}{\eta}\right)e^{-1}\\
&\hspace{30pt} \times 
\left( \bw\kw \right)^{\frac{1}{\aw}} e^{-1}  \11\{\mcl{W}_W\setminus \{W\}\ne\emptyset\}
\\
&\hspace{0pt}= e^{-2} \eta^{-1} \beta_W^{1+\aw^{-1}} \kw^{3+\aw^{-1}} \11\{\mcl{W}_W\setminus \{W\}\ne\emptyset\}.\numberthis\label{eq:upperboundnonrares:secondterm}
\end{align*}
Using \eqref{eq:upperboundnonrares:firstterm} and \eqref{eq:upperboundnonrares:secondterm} in \eqref{eq:upperboundnonrares0}, the Lemma follows.
\end{IEEEproof}

\section{Proof of Theorem \ref{thm:stability}}\label{sec:appendix:stability}
For each $W \in \mcl{W}$, let $\delta_W \in (0,1)$ be sufficiently small number to be chosen appropriately, and recall that the two blocks $\regrw{W}{1}=\{\statex \in \mcl{S}: \totalmismatch{W}-2(1-\eta) \ge \eta\nmbrn{W}\}$ and $\regrw{W}{2}= \mcl{S}\setminus \regrw{W}{1}$ form a partition of the state-space $\mcl{S}$. For each $k \in [2]$, we further divide the block $\regrw{W}{k}$ into three regions, namely
\begin{align*}
\regrw{W}{k1} &=\regrw{W}{k}\cap \left\{ 
(\kw-\eta)\nmbrn{W}-\sum_{i\ne \udl{r}}\nmbr{i}{W} \ge \delta_W\xw \right\},\\
\regrw{W}{k2} &=\regrw{W}{k}\cap \left\{ \delta_W\xw > (\kw-\eta)\nmbrn{W}-\sum_{i\ne \udl{r}}\nmbr{i}{W} \ge \dots \right.\\ 
&\hspace{30pt} \left. \ge \frac{(\mw-2)(1-\eta)}{\kw} \right\},\\
\regrw{W}{k3} &=\regrw{W}{k}\cap \left\{
(\kw-\eta)\nmbrn{W}-\sum_{i\ne \udl{r}}\nmbr{i}{W} < \delta_W\xw \text{ and } \right.\\
&\hspace{30pt} \left. (\kw-\eta)\nmbrn{W}-\sum_{i\ne \udl{r}}\nmbr{i}{W} < \frac{(\mw-2)(1-\eta)}{\kw}
\right\}.
\end{align*}
Below we prove \ref{thm:stability} through a series of lemmas, together which establish that the unit-transition drift conditions are satisfied by the candidate Lyapunov function given by \eqref{eq:stab:v} and \eqref{eq:stab:vwdef}.

\begin{lem}\label{lem:pw}
For all $W \in \mcl{W}$, $ k \in [2] $, and $\statex \in \regrw{W}{k1}$, the highest frequency $\freqn{W}$ is lower-bounded by $\frac{\delta_W}{\kw-\eta}$. Therefore, the total chunk-count $\pw$ is lower-bounded by $\frac{\delta_W}{\kw-\eta} \xw $.
\end{lem}
\begin{IEEEproof}
We can lower-bound $\pw$ by $\nmbrn{W}$. Given $ (\kw-\eta)\nmbrn{W}-\sum_{i\ne \udl{r}}\nmbr{i}{W} \ge \delta_W\xw $ in $\regrw{W}{k1}$, the Lemma is obvious.
\end{IEEEproof}

\begin{lem}\label{lem:deltasmallenough}
For all $W \in \mcl{W}$, $k \in [2]$, and $ \statex \in \regrw{W}{k2} \cup \regrw{W}{k3} $, the fraction of peers missing  piece $i\in W$, i.e., $1-\freq{i}{W}$ is lower-bounded by $0.5$ provided  $\delta_W\le 0.5(1-\eta)$.
\end{lem}
\begin{IEEEproof}
In $\regrw{W}{k2} \cup \regrw{W}{k3}$, we have $(\kw-\eta)\nmbrn{W}-\sum_{i\ne \udl{r}}\nmbr{i}{W}<\delta_W\xw$. Since $\nmbr{i}{W}$ is at most $\nmbrn{W}$, this implies that $(1-\eta)\nmbrn{W}<\delta_W\xw \implies \freqn{W}<\frac{\delta_W}{1-\eta}$. Now, for any piece $i\in W$, we have $ 1 - \freq{i}{W} \ge 1 - \freqn{W} > 1 - \frac{\delta_W}{1-\eta}$. By choosing $\delta_W\le 0.5(1-\eta)$, we ensure that $1-\freq{i}{W} \ge 0.5$.
\end{IEEEproof}

\begin{lem}\label{lem:gamma}
For all $W \in \mcl{W}$, $ k \in [2] $, and $\statex \in \regrw{W}{k2} \cup \regrw{W}{k3}$, the fraction of peers who are missing only piece $i \in W$, i.e., $\gamma^{(i)}_{W}=\sum_{S:W\setminus S=\{i\}} \frac{\xws}{\xw} $ is upper bounded by $ \frac{\delta_W}{1-\eta} $.
\end{lem}
\begin{IEEEproof}
A swarm-$ W $ peer missing only piece $ i $ of file-$W$ has all the other pieces of the file, therefore,
\begin{align*}
\sum_{S:W\setminus S =\{i\}} \xws &= \gamma^{(i)}_{W} \xw \\
&\le \pi_W^{\hat{(i)}} \xw \text{ for all } \hat{i} \in W \text{ such that } \hat{i} \ne i\\
&\le \freqn{W} \xw.
\end{align*}
Given $\freqn{W} < \frac{\delta_W}{1-\eta}$ in $ \regrw{W}{k2} \cup \regrw{W}{k3}$, the inequality follows.
\end{IEEEproof}

\begin{lem}\label{lem:nw}
For all $W\in\mcl{W}$, $\theta_1 > 0 $, and $ \theta_2 \ge 2(1-\eta)^2 $, let $g$ be defined as
\begin{align*}
g(\theta_1, \theta_2) &= \sum_{r\in \rw} \frac{\Gammalower{W}{r}}{\theta_1}  \left[\left( 1 - \freq{r}{W} \right) \left( \udl{\psi}_W - \theta_2 \right) \right]
\end{align*}
Then for all $\statex\in\regrw{W}{1}$, $g$ is upper-bounded by 
\begin{align*}
&-\frac{2(1-\eta)}{\theta_1\kw} \left[ 2(1-\eta) \left( LU -2(1-\eta)\Delta_p \right) \right. \\
&\hspace{10pt} \left. + \left( (\kw-\eta)\nmbrn{W}-\sum_{i\ne\udl{r}}\nmbr{i}{W}  \right)  ( LU \wedge 2(1-\eta)\Delta_p ) \right].
\end{align*}
Consequently, for any $\theta$ and $\eps>0$, there exists $\nwj{1} = \nwj{1}(\theta, \theta_1, \theta_2, \epsilon) \in \mathbb{R}^+$ such that for all $\statex \in \regrw{W}{1}$,
\begin{align*}
&\theta + g(\theta_1, \theta_2) \le -\eps 
\end{align*}
if $ (\kw-\eta)\nmbrn{W}-\sum_{i\ne\udl{r}}\nmbr{i}{W} \ge \nwj{1} $.
\end{lem}
\begin{IEEEproof}
Since $\statex\in\regrw{W}{1}$, $\udl{\psi}_W \le 0$. Given $\theta_2>0$, each term in the summation is negative. Let us denote the rarest-piece by $\udl{r}$, i.e., $\nmbr{\udl{r}}{W}=\nmbrr{W}$. By Lemma \ref{lem:chunkcounts}, $1-\freq{\udl{r}}{W}$ is bounded from below by $\frac{1}{\kw}$. Upper-bounding the summation in $g$ by considering only the rarest-piece $\udl{r}$, we get
\begin{align*}
&g(\theta_1, \theta_2) \\
&\le \frac{\Gammalower{W}{\udl{r}}}{\theta_1\kw} \left[ 2(1-\eta)^2 - 2(1-\eta) \left( \totalmismatch{W}-\eta\nmbrn{W} \right) - \theta_2 \right]\\
&\hspace{0pt}\labelrel{=}{eqr:nw:1} - \frac{2(1-\eta)\Gammalower{W}{\udl{r}}}{\theta_1\kw} 
\left[ \left((\kw-\eta)\nmbrn{W}-\sum_{i\ne\udl{r}}\nmbr{i}{W}-\nmbrr{W} \right) \right]\\
&\hspace{0pt} \labelrel{\le}{eqr:nw:2} - \frac{2(1-\eta) }{\theta_1\kw}\left( LU + \Delta_p \nmbrr{W} \right) \\
&\hspace{30pt} \left((\kw-\eta)\nmbrn{W}-\sum_{i\ne\udl{r}}\nmbr{i}{W}-\nmbrr{W} \right)\\
&\hspace{0pt} \defeq \wt{g}(\nmbrr{W}; \theta_1),
\end{align*}
where, \eqref{eqr:nw:1} uses $\theta_2 \ge 2(1-\eta)^2 $ and the definition of $\totalmismatch{W}$ (see \eqref{eq:totalmismatch}); and \eqref{eqr:nw:2} uses $\Gammalower{W}{\udl{r}}\ge LU+\Delta_p \nmbrr{W} $ (see \eqref{eq:Gammalower2}). 

Since $\statex\in\regrw{W}{1}$, we have
$\totalmismatch{W}-2(1-\eta)\ge \eta\nmbrn{W}$, which implies that $\nmbrr{W}\le (\kw-\eta)\nmbrn{W}-\sum_{i\ne\udl{r}}\nmbr{i}{W}-2(1-\eta)$. It can be verified that $\wt{g}(\nmbrr{W}; \theta_1)$ is a quadratic and strictly convex function of $\nmbrr{W}$. Therefore, it has a unique minimum and attains its maximum at the boundary points 0 and $(\kw-\eta)\nmbrn{W}-\sum_{i\ne\udl{r}}\nmbr{i}{W}-2(1-\eta)$. This gives
\begin{align*}
&g(\theta_1, \theta_2) 
\le \wt{g}(0) \wedge \wt{g}\left( (\kw-\eta)\nmbrn{W}-\sum_{i\ne\udl{r}}\nmbr{i}{W}-2(1-\eta) \right) \\
&=-\frac{2(1-\eta)}{\theta_1\kw} \left[ 2(1-\eta) \left( LU -2(1-\eta)\Delta_p \right) \right. \\
&\hspace{10pt} \left. + \left( (\kw-\eta)\nmbrn{W}-\sum_{i\ne\udl{r}}\nmbr{i}{W}  \right)  ( LU \wedge 2(1-\eta)\Delta_p ) \right].
\end{align*}
\end{IEEEproof}

\begin{lem}\label{lem:qvwupperbound}
For any $ \epsilon > 0 $ and any $W \in \mcl{W}$, there exists a countable set $\setaw$ and a finite constant $B_W \ge 0$ such that
\begin{align*}
\widetilde{QV}_W &\le - \eps \11{ \{\statex \in \setawc \} } + B_W \11{ \{\statex \in \setaw\}  }, \text{ and } \\
N_W &= \max \limits_{\statex \in \setaw} \xw < \infty.
\end{align*}
\end{lem}
\begin{IEEEproof} We will show that $ \widetilde{{QV}}_W$ is upper bounded by $- \eps$ over the region $ \regrw{W}{k} $ for each $k \in [2]$, except possibly, for a finite and bounded population of swarm-$W$.

\vspace{5pt}
\noindent 
\udl{\textbf{Case 1 - } $\statex \in \regrw{W}{1}$:}	
From \eqref{eq:stab:qvwtilde} and \eqref{eq:stab:psiwi}, we have
\begin{align*}
&\widetilde{QV}_W = \slmdas\c1 + \sum_{i\in W} \frac{\Gammalower{W}{i}a_W^{(i)} }{\kw} \left[  
\left( 1-\freq{i}{W}\right)\Psi_W^{(i)} \right.\\
&\left. - \left(1-\freq{i}{W}-\gamma_{W}^{(i)} \right)I_W^{(1)}+\gamma_W^{(i)}\kw\xi_2I_W^{(2)}
\right],
\end{align*}
where
\begin{align*}
\Psi_W^{(i)}(\statex) &= \kw\xi_2\ov{\psi}_W\11\{i\in\rwc\}+\udl{\psi}_W\11\{i\in\rw\} -\cw1.
\end{align*}

Region $\regrw{W}{11}$: 
Here, $\pw\ge \nmbrn{W}\ge \frac{\delta_W}{\kw-\eta}\xw$ (see Lemma \ref{lem:pw}). Therefore, for all large $\xw$, $I_W^{(1)}, I_W^{(2)}$ are both zero. This gives
\begin{align*}
&\widetilde{QV_W} \le \slmdas\c1 +\sum_{i \in W } \frac{ \Gammalower{W}{r}a_W^{(i)}}{\kw} \left[ \left( 1-\freq{i}{W} \right) \right.\\
&\hspace{10pt}\left. \left( \kw\xi_2\ov{\psi}_W\11\{i\in\rwc\}+\udl{\psi}_W\11\{i\in\rw\} -\cw1 \right) \right]\\
&\le \slmdas\c1 +\sum_{n \in \rwc } \frac{ \Gammalower{W}{n}\zeta_W^{(n)}}{\kw} \left[ \left( 1-\freqn{W} \right) \kw\xi_2\ov{\psi}_W  \right] \\
&\hspace{10pt} +\sum_{r \in \rw } \frac{ \Gammalower{W}{r}}{\kw} \left[ \left( 1-\freq{r}{W} \right)\left( \udl{\psi}_W -\cw1 \right) \right]\\
&\le \slmdas\c1 + D_W + \sum_{r \in \rw } \frac{ \Gammalower{W}{r}}{\kw} \left[ \left( 1-\freq{r}{W} \right) \udl{\psi}_W  \right],
\end{align*}
where $D_W=D_W(\eta,\xi_2,L,U,\Delta_p,\beta_W,\kw)$ is given by Lemma \ref{lem:upperboundnnonrares}. From Lemma \ref{lem:nw}, it follows that $\widetilde{QV}_W\le-\eps$ if $(\kw-\eta)\nmbrn{W}-\sum_{i\ne\udl{r}}\nmbr{i}{W}  \ge \nwj{1}(\theta=\slmdas\c1+D_W, \theta_1=K_W,\theta_2= \cw1,\eps)$. Since $(\kw-\eta)\nmbrn{W}-\sum_{i\ne\udl{r}}\nmbr{i}{W} \ge\delta_W\xw$, this is true for all large $\xw$.

Region $\regrw{W}{12}$: 
Here, we use the bounds, $-I_W^{(1)}\le 0$, $I_W^{(2)}\le \kw\cw2$, and $\gamma_W^{(i)}< \frac{\delta_W}{1-\eta}$ for every $i\in W$ (Lemma \ref{lem:gamma}). Then, for all large $\xw$,
\begin{align*}
&\widetilde{QV}_W \le \slmdas\c1 + \sum_{i \in W } \frac{ \Gammalower{W}{i}a_W^{(i)} }{\kw} \left[ \left( 1-\freq{i}{W} \right) \right.\\
&\hspace{10pt} \left. \times \left( \kw\xi_2\ov{\psi}_W\11\{i\in\rwc\}+\udl{\psi}_W\11\{i\in\rw\} -\cw1 \right) \right. \\
&\hspace{10pt} \left. + \frac{\delta_W}{1-\eta} \kw^2 \xi_2 \cw2 \right]\\
&\labelrel{\le}{eqr:reg12:1} 
\slmdas\c1 +\sum_{i\in W} \frac{\Gammalower{W}{i}a_W^{(i)} }{\kw} \left[\left(1-\freq{i}{W}\right) \right.\\
&\hspace{10pt}\left. \left(
\kw\xi_2\ov{\psi}_W\11\{i\in\rwc\}+\udl{\psi}_W\11\{i\in\rw\} -\cw1 \right.\right.\\
&\hspace{30pt} \left.\left. + \frac{2\delta_W}{1-\eta} \kw^2\xi_2\cw2\right) \right]\\
&\labelrel{\le}{eqr:reg12:2} \slmdas\c1 +\sum_{i\in W} \frac{\Gammalower{W}{i}a_W^{(i)} }{\kw} \left[\left(1-\freq{i}{W}\right) \right.\\
&\hspace{10pt} \left. \times \left(
\kw\xi_2\ov{\psi}_W\11\{i\in\rwc\}+\udl{\psi}_W\11\{i\in\rw\} \right.\right.\\
&\hspace{10pt} \left.\left.-0.5\cw1 \right) \right]\\
&\le \slmdas\c1 +\sum_{n\in \rwc} \frac{\Gammalower{W}{n}\zeta_W^{(n)} }{\kw} \left[\left(1-\freqn{W}\right)
\kw\xi_2\ov{\psi}_W   \right] \\
&\hspace{10pt} + \sum_{r\in \rwc} \frac{\Gammalower{W}{r}}{\kw} \left[\left(1-\freq{r}{W}\right)\left(
\udl{\psi}_W -0.5\cw1 \right) \right]\\
&\labelrel{\le}{eqr:reg12:3} \slmdas\c1 +D_W +\sum_{r\in \rw} \frac{\Gammalower{W}{r}}{\kw}
\left[\left(1-\freq{r}{W}\right) \udl{\psi}_W \right].
\end{align*}
Here, \eqref{eqr:reg12:1} uses $1-\freq{i}{W}\ge 0.5 $ for every $i\in W$ (Lemma \ref{lem:deltasmallenough}); \eqref{eqr:reg12:2} follows from choosing $\delta_W$ small enough so that $2\frac{\delta_W}{1-\eta}\kw^2\xi_2\cw2\le 0.5\cw1 $; \eqref{eqr:reg12:3} follows from Lemma \ref{lem:upperboundnnonrares}. From Lemma \ref{lem:nw}, it follows that $\widetilde{QV}_W\le-\eps$ if  $ (\kw-\eta)\nmbrn{W}-\sum_{i\ne\udl{r}}\nmbr{i}{W} \ge \nwj{1}(\theta=\slmdas\c1+D_W,\allowbreak \theta_1=K_W,\theta_2= 0.5\cw1, \eps )$. Since $(\kw-\eta)\nmbrn{W}-\sum_{i\ne\udl{r}}\nmbr{i}{W}\ge\frac{(\mw-2)(1-\eta)}{\kw}$, choosing $\mw$ large enough so that $ \frac{(\mw-2)(1-\eta)}{\kw} \ge\nwj{12} $ ensures that $\widetilde{QV}_W\le-\eps$ for large enough $\xw$.

Region $\regrw{W}{13}$: 
Here, $(\kw-\eta)\nmbrn{W}-\sum_{i\ne \udl{r}}\nmbr{i}{W} < \frac{(\mw-2)(1-\eta)}{\kw} $. This implies that $\nmbrn{W} < \frac{\mw-2}{\kw} $. Therefore, the total chunk-count $\pw$ is upper-bounded by $\kw\nmbrn{W}\le\mw-2 $. Consequently, $I_W^{(1)}$ and $I_W^{(2)}$ are both non-zero. We use the bounds $\udl{\psi}_W\11\{i\in\rw\} -\cw1\le 0$ and $\gamma_W^{(i)}<\frac{\delta_W}{1-\eta}$ for every $i\in W$ (Lemma \ref{lem:gamma}). Then, for all large $\xw$,
\begin{align*}
&\widetilde{QV}_W \le \slmdas\c1 +
\sum_{i\in W} \frac{\Gammalower{W}{i}a_W^{(i)} }{\kw} \left[ \left(1-\freq{i}{W} \right) \right.\\
&\hspace{5pt}\left. \times \left(\kw\xi_2\ov{\psi}_W\11\{i\in\rwc\} \right) - \left(1-\freq{i}{W}-\frac{\delta_W}{1-\eta}\right)\cw2 \right. \\
&\hspace{10pt} \left. + \frac{\delta_W}{1-\eta} \kw^2\xi_2\cw2\right]\\
&\labelrel{\le}{eqr:reg13:1} \slmdas\c1  +\sum_{i\in W} \frac{\Gammalower{W}{i}a_W^{(i)} }{\kw} \left[ \left(1-\freq{i}{W} \right) \right.\\
&\hspace{10pt} \left. \left(\kw\xi_2\ov{\psi}_W\11\{i\in\rwc\} \right) -0.5\cw2 \right. \\
&\hspace{10pt} \left. + \frac{\delta_W}{1-\eta} \cw2 \left( 1+\kw^2\xi_2 \right)\right]\\
&\labelrel{\le}{eqr:reg13:2} \slmdas\c1  +\sum_{i\in W} \frac{\Gammalower{W}{i}a_W^{(i)} }{\kw} \left[ \left(1-\freq{i}{W} \right) \right.\\
&\hspace{10pt} \left.\times \left(\kw\xi_2\ov{\psi}_W\11\{i\in\rwc\} \right) -0.25\cw2 \right]\\
&\le \slmdas\c1 + 
\sum_{n\in\rwc} \frac{\Gammalower{W}{n}\zeta_W^{(n)} }{\kw}\left[
\left(1-\freqn{W} \right)\left(\kw\xi_2\ov{\psi}_W  \right) \right] \\
&\hspace{10pt} +\sum_{r\in \rw} \frac{\Gammalower{W}{r}}{\kw} \left[-0.25\cw2 \right]\\
&\labelrel{\le}{eqr:reg13:3} \slmdas\c1 +D_W +\sum_{r\in \rw} \frac{\Gammalower{W}{r}}{\kw} \left[-0.25\cw2 \right] \\
&\labelrel{\le}{eqr:reg13:4} \slmdas\c1 +D_W-0.25\kw^{-1}LU\cw2\\
&\labelrel{\le}{eqr:reg13:5} -\eps.
\end{align*}
Here, \eqref{eqr:reg13:1} uses $1-\freq{i}{W}\ge 0.5$ for $i\in W$ (Lemma \ref{lem:deltasmallenough}); \eqref{eqr:reg13:2} follows from choosing $\delta_W$ small enough so that $\frac{\delta_W}{1-\eta}\cw2(1+\kw^2\xi_2)\le 0.25\cw2$; \eqref{eqr:reg13:3} follows from Lemma \ref{lem:upperboundnnonrares}; \eqref{eqr:reg13:4} upper-bounds the summation by considering only the rarest piece and using $\udl{\Gamma}_W^{(i)}\ge LU $ (see \eqref{eq:Gammalower2}); \eqref{eqr:reg13:5} follows by choosing $\cw2$ large enough so that $0.25\kw^{-1}LU\cw2\ge\slmdas\c1+D_W+\eps $.

\vspace{5pt}
\noindent \udl{\textbf{Case 2 - } $\statex \in \regrw{W}{2}$:}	
From \eqref{eq:stab:qvwtilde} and \eqref{eq:stab:psiwi}, we can write
\begin{align*}
\widetilde{QV}_W &\le \slmdas\c1 + \sum_{i \in W } \frac{ \Gammalower{W}{i}a_W^{(i)} }{\kw} \left[ \left( 1-\freq{i}{W} \right)\left(4\kw^2-\cw1 \right) \right.\\
&\hspace{10pt} \left. - \left ( 1- \freq{i}{W} - \gamma_W^{(i)} \right) I_W^{(1)} + \gamma^{(i)}_{W} \kw \xi_2 I_W^{(2)} \right].
\end{align*}
Region $\regrw{W}{21}$: 
Like in $\regrw{W}{11}$, $I_W^{(1)},I_W^{(2)}$ are both zero for large $\xw$. Since $\statex\in\regrw{W}{2} $, $\totalmismatch{W}-2(1-\eta)<\eta\nmbrn{W}$. This implies that $\ov{m}_W<\eta\nmbrn{W}+2(1-\eta) \implies \nmbrr{W}>(1-\eta)\nmbrn{W}-2\ge \frac{\delta_W(1-\eta)}{\kw-\eta}\xw-2$. Then, for all large $\xw$,
\begin{align*}
&\widetilde{QV}_W \le \slmdas\c1 +\sum_{i\in W} \frac{\Gammalower{W}{i} a_W^{(i)}}{\kw}\left[\left(1-\freq{i}{W}\right)\left(4\kw^2-\cw1\right) \right] \\
&\labelrel{\le}{eqr:reg21:0} \slmdas\c1 + \frac{\Gammalower{W}{\udl{r} }}{\kw}\left[\left(1-\freqr{W}\right)\left(-0.5\cw1\right) \right] \\
&\labelrel{\le}{eqr:reg21:1} \slmdas\c1 + \kw^{-2} \Gammalower{W}{\udl{r} } \left[-0.5\cw1\right]\\
&\labelrel{\le}{eqr:reg21:2} \slmdas\c1 - 0.5\cw1\kw^{-2} \Delta_p \nmbrr{W} \\
&\labelrel{\le}{eqr:reg21:3} \slmdas\c1 - 0.5\cw1\kw^{-2} \Delta_p \left(\frac{\delta_W(1-\eta)}{\kw-\eta}\xw-2 \right) \\
&\labelrel{\le}{eqr:reg21:4}-\eps.
\end{align*}
Here, \eqref{eqr:reg21:0} uses $4\kw^2-\cw1\le -0.5\cw1$ and then upper-bounds the summation by considering only the rarest-piece, that is denoted by $\udl{r}$; \eqref{eqr:reg21:1} uses $1-\freqr{W}\ge \kw^{-1}$ (Lemma \ref{lem:chunkcounts}); \eqref{eqr:reg21:2} uses $\Gammalower{W}{\udl{r}}\ge \Delta_p\nmbrr{W}$ (see \eqref{eq:Gammalower2}); \eqref{eqr:reg21:3} uses $ \nmbrr{W}\ge \frac{\delta_W(1-\eta)}{\kw-\eta}\xw-2$; \eqref{eqr:reg21:4} uses the fact that $\xw$ is large enough.

Region $\regrw{W}{22}$: 
Like $\regrw{W}{12}$ we use the bounds, $-I_W^{(1)}\le 0$, $I_W^{(2)}\le K_W\cw2$, and $\gamma_W^{(i)}< \frac{\delta_W}{1-\eta}$ for every $i\in W$ (Lemma \ref{lem:gamma}). Then,
\begin{align*}
&\widetilde{QV}_W \le \slmdas\c1+
\sum_{i\in W} \frac{\Gammalower{W}{i} a_W^{(i)}}{\kw} \left[\left(1-\freq{i}{W}\right)\left(4\kw^2-\cw1\right) \right.\\
&\hspace{10pt} \left. + \frac{\delta_W}{1-\eta} \kw^2\xi_2\cw2\right] \\
&\labelrel{\le}{eqr:reg22:0} \slmdas\c1 +
\sum_{i\in W} \frac{\Gammalower{W}{i} a_W^{(i)}}{\kw} \left[\left(1- \freq{i}{W} \right)
\left(-0.5\cw1\right) \right.\\
&\hspace{10pt} \left. + \frac{\delta_W}{1-\eta} \kw^2\xi_2\cw2\right] \\
&\labelrel{\le}{eqr:reg22:1} \slmdas\c1+
\sum_{i\in W} \frac{\Gammalower{W}{i} a_W^{(i)}}{\kw} \left[
\left(1- \freq{i}{W} \right)\left(
-0.5\cw1 \right.\right.\\
&\hspace{10pt} \left.\left. + \frac{2\delta_W}{1-\eta} \kw^2\xi_2\cw2\right) \right] \\
&\labelrel{\le}{eqr:reg22:2} \slmdas\c1+
\sum_{i\in W} \frac{\Gammalower{W}{i} a_W^{(i)}}{\kw} \left[
\left(1- \freq{i}{W} \right)\left(
-0.25\cw1 \right) \right] \\
&\labelrel{\le}{eqr:reg22:3} \slmdas\c1+
\sum_{i\in W} \frac{\Gammalower{W}{i} a_W^{(i)}}{\kw} \left[-0.125\cw1
\right] \\
&\labelrel{\le}{eqr:reg22:4} \slmdas\c1-0.125\cw1
\kw^{-1} \Gammalower{W}{\udl{r}} \\
&\labelrel{\le}{eqr:reg22:5} \slmdas\c1-0.125\cw1
\kw^{-1} \Delta_p\nmbrr{W} \\
&\labelrel{\le}{eqr:reg22:6} \slmdas\c1-0.125\cw1
\kw^{-1} \Delta_p\left(\frac{(\mw-2)(1-\eta)^2}{\kw(\kw-\eta)} -2 \right) \\
&\labelrel{\le}{eqr:reg22:7}-\eps.
\end{align*}
Here, \eqref{eqr:reg22:0} uses $4\kw^2-\cw1\le -0.5\cw1$; \eqref{eqr:reg22:1} uses $1-\freq{i}{W}\ge 0.5 $ for every $i\in W$ (Lemma \ref{lem:deltasmallenough}); \eqref{eqr:reg22:2} follows from choosing $\delta_W$ small enough so that $\frac{2\delta_W}{1-\eta} \kw^2\xi_2\cw2\le 0.25 \cw1 $; \eqref{eqr:reg22:3} uses $1-\freq{i}{W}\ge 0.5 $ for every $i\in W$ (Lemma \ref{lem:deltasmallenough}); \eqref{eqr:reg22:4} upper-bounds the summation by considering only the rarest piece (denoted by $\udl{r}$); \eqref{eqr:reg22:5} uses $\Gammalower{W}{\udl{r}}\ge\Delta_p\nmbrr{W}$ (see \eqref{eq:Gammalower2}); \eqref{eqr:reg22:6} uses $\nmbrr{W}\ge (1-\eta)\nmbrn{W}-2$ and $\nmbrn{W} \ge \frac{(\mw-2)(1-\eta) }{\kw} $; \eqref{eqr:reg22:7} follows by choosing $\mw$ large enough so that $0.125\cw1 \kw^{-1}\Delta_p\left(\frac{(\mw-2)(1-\eta)^2}{\kw(\kw-\eta)}-2 \right) \ge \slmdas\c1 +\eps$.

Region $\regrw{W}{23}$: 
Like $\regrw{W}{13}$, $I_W^{(1)}$ and $I_W^{(2)}$ are both non-zero. We use the bounds, $I_W^{(2)}\le K_W\cw2$, $4\kw^2-\cw1\le 0$, $\gamma_W^{(i)}<\frac{\delta_W}{1-\eta}$ for every $i\in W$ (Lemma \ref{lem:gamma}). Then,
\begin{align*}
&\widetilde{QV}_W \le \slmdas\c1 + \sum_{i \in W } \frac{ \Gammalower{W}{i} a_W^{(i)} }{\kw} \left[  - \left ( 1- \freq{i}{W} - \frac{\delta_W}{1-\eta} \right) \cw2 \right.\\
&\hspace{10pt} \left. + \frac{\delta_W}{1-\eta} \kw^2 \xi_2 \cw2 \right]\\
&\labelrel{\le}{eqr:reg23:1} \slmdas\c1 + \sum_{i \in W } \frac{ \Gammalower{W}{i} a_W^{(i)} }{\kw} \left[ -0.5\cw2 \right.\\
&\hspace{10pt} \left. + \frac{\delta_W}{1-\eta} \cw2\left(1 +  \kw^2 \xi_2 \right) \right]\\
&\labelrel{\le}{eqr:reg23:2} \slmdas\c1 + \sum_{i \in W } \frac{ \Gammalower{W}{i} a_W^{(i)} }{\kw} \left[ -0.25\cw2 \right]\\
&\labelrel{\le}{eqr:reg23:3} \slmdas\c1 - 0.25\kw^{-1}LU\cw2\\
&\labelrel{\le}{eqr:reg23:4}-\eps.
\end{align*}
Here, \eqref{eqr:reg23:1} uses $1-\freq{i}{W}\ge 0.5$ for every $i\in W$ (Lemma \ref{lem:deltasmallenough}); \eqref{eqr:reg23:2} follows from choosing $\delta_W$ small enough so that $\frac{\delta_W}{1-\eta}\cw2(1+\kw^2\xi_2)\le 0.25\cw2$; \eqref{eqr:reg23:3} upper-bounds the summation by considering only the rarest piece and uses $\Gammalower{W}{i}\ge LU$ (see \eqref{eq:Gammalower2}); \eqref{eqr:reg23:4} follows by choosing $\cw2$ large enough so that $0.25\kw^{-1}LU\cw2\ge \slmdas\c1+\eps $.
\end{IEEEproof}

\begin{lem}\label{lem:qvupperbound}
For any $\eps' > 0$, there exists a finite set $\mcl{A}$ and a finite constant $B$ such that
\begin{align*}
{QV}(\statex) \le - \eps' \11{ \{ \statex \in \mcl{A}^c \} }  + B \11{ \{ \statex \in \mcl{A} \} }.
\end{align*}
\end{lem}
\begin{IEEEproof}
Fix $\eps' > 0$. For any $\eps > 0$ and any $W \in \mcl{W}$, it follows from Lemma \ref{lem:qvwupperbound} that $ \widetilde{QV}_W \le - \eps $, except possibly over $\setaw$ where its population and corresponding term $\widetilde{QV}_W$ are bounded from above by $N_W$  and $B_W \ge 0$ respectively. We can write the state-space as
$ \mcl{S} = \bigcup\limits_{ \mcl{H} : \mcl{H} \subseteq \mcl{W} } \mcl{S}_{\mcl{H}}, $
where 
\begin{align*}
\mcl{S}_{\mcl{H}} &\defeq \left\{ \xw \le \Nw \text{ for all } W \in \mcl{H}, \text{ and } \right.\\
&\hspace{30pt} \left. \x_U > \Nu \text{ for all } U \in \mcl{W} \setminus \mcl{H} \right\}.
\end{align*}
\noindent \udl{\textbf{Case 1 - } $\mcl{H} = \emptyset$}: Let $\statex \in \mcl{S}_{\emptyset}$. Since $\xw > \Nw $ for all $W \in \mcl{W}$, from Lemma \ref{lem:qvwupperbound}, it follows that $\widetilde{QV}_W \le - \eps$ for all $W \in \mcl{W}$. From \eqref{eq:stab:qv1}, this gives ${QV} \le - \eps$. Choosing $\eps \ge \eps'$ ensures ${QV} \le - \eps'$.

\vspace{5pt}
\noindent \udl{\textbf{Case 2 - } $\mcl{H} = \mcl{W}$}: The set $\mcl{S}_{\mcl{W} }$ is finite. Therefore, for any $\statex \in \mcl{S}_{\mcl{W} }$, we can write
$ {QV} \le B_{\mcl{W}} \defeq \max \limits_{\statex \in \mcl{S}_{\mcl{W}} } \left( {QV} \right)^{+} < \infty$.

\vspace{5pt}
\noindent \udl{\textbf{Case 3 - } $\emptyset \ne \mcl{H} \subsetneq \mcl{W}$}:	Let $ \statex \in \mcl{S}_{\mcl{H}} $. We can upper-bound ${QV}$ as follows.	
\begin{align*}
{QV} &= \sum_{W \in \mcl{H}} \frac{ \xw }{ \x }  \widetilde{QV}_W + \sum_{U \in \mcl{W} \setminus \mcl{H} } \frac{ \x_U }{ \x } \widetilde{QV}_{U} \\
&\le \frac{ \sum_{W \in \mcl{H}} \Nw B_{W}}{\sum_{ U \in \mcl{W} \setminus \mcl{H} } \x_U} + \frac{ \sum_{U \in \mcl{W} \setminus \mcl{H} } \x_U }{ \sum_{W \in \mcl{H} } \Nw + \sum_{U \in \mcl{W} \setminus \mcl{H} } \x_U  } (- \eps) .
\end{align*}
Note that $\sum_{U \in \mcl{W} \setminus \mcl{H} } \x_U \ra \infty$ over the set $ \mcl{S}_{\mcl{H}} $, which implies
\begin{align*}
\frac{ \sum_{W \in \mcl{H}} N_W B_W}{\sum_{ U \in \mcl{W} \setminus \mcl{H} } \x_U} \to 0 \text{ and } \frac{ \sum_{U \in \mcl{W} \setminus \mcl{H}} \x_U }{ \sum_{W \in \mcl{H}} N_W + \sum_{U \in \mcl{W} \setminus \mcl{H}} \x_U  } \to 1.
\end{align*}
Therefore, for any $\phi > 0$, there exists $N_{ \mcl{H} } = N_{ \mcl{H} } (\phi) \in \mb{R}_{+}$ such that ${QV} \le \phi + (1 - \phi)(- \eps) $ for any $\statex \in \mcl{S}_{\mcl{H}}$ with $\sum_{U \in \mcl{W}_{2}} \x_U \ge N_{ \mcl{H} } $. Choosing $\eps = 2 \epsilon'$ and $0 < \phi \le \frac{\eps/2}{1+ \eps}$ ensures ${QV} \le - \eps'$.

For all $\mcl{H}$ such that $\emptyset \ne \mcl{H} \subsetneq \mcl{W}$, let us define
\begin{align*}
\mcl{A}_{ \mcl{H} } &\defeq \mcl{S}_{\mcl{H}} \cap \left\{ \sum_{U \in \mcl{H} } \x_U < N_{\mcl{H} } \right\} &&\quad \text{(finite)},\\
\text{and then } \mcl{A} &\defeq \left( \bigcup\limits_{ \phi \ne \mcl{H} \subsetneq \mcl{W} } \mcl{A}_{\mcl{H}} \right) \cup \mcl{S}_{\mcl{W}} &&\quad \text{(finite)},\\
B &\defeq \max \limits_{\statex \in \mcl{A} } \left( {QV}\left(\statex\right) \right)^+ < \infty.
\end{align*}
Then, for any state $\statex $, we can write
$$ {QV}(\statex) \le - \eps' \11{ \{ \statex \in  \mcl{A}^c \} }  + B \11{ \{ \statex \in \mcl{A} \} }. $$ establishing the result.	
\end{IEEEproof}

Lemma \ref{lem:qvupperbound} is the final stage. Combining Lemma \ref{lem:qvupperbound} and Proposition \ref{prop:foster_kingman}, we conclude that Theorem \ref{thm:stability} holds. As a final step, we now illustrate how the constants $\{\cw1,\cw2,\mw,\delta_W\}_{W\in\mcl{W}}$ can be set consistently.

\vspace{5pt}
\udl{\textbf{Setting $\{\cw1,\cw2,\mw,\delta_W\}_{W\in\mcl{W}}$}}:
Set some $\eps'>0$, $\eps=2\eps'$, and $\eta\in(0,1)$.

\begin{itemize}
\item For all $W\in\mcl{W}$, individually set $\cw1=8\kw^2$.
\item Set $\c1=\max_{W\in \mcl{W}}\kw \cw1$.
\item For all $W\in\mcl{W}$, individually set $\cw2$ such that $0.25\kw^{-1}LU\cw2\ge\slmdas\c1+D_W+\eps $.
\item For all $W\in\mcl{W}$, individually set $\delta_W$ such that $\delta_W\le0.5(1-\eta)$, $\frac{2\delta_W}{1-\eta}\kw^2\xi_2\cw2 \le 0.125\cw1 $, and $\frac{2\delta_W}{1-\eta}\kw^2\xi_2\le 0.25\cw2$.
\item For all $W\in\mcl{W}$, set $\mw$ so that $\frac{(\mw-2)(1-\eta)}{\kw}\ge \nwj{1} \big(\slmdas+D_W,\kw,0.5\cw1,\eps \big) $ and
$ 0.125\cw1\kw^{-1}\Delta_p\allowbreak \left( \frac{(\mw-2)(1-\eta)^2}{\kw(\kw-\eta)}-2 \right) \ge \slmdas\cw1 +\eps$, where $\nwj1$ is specified in Lemma \ref{lem:nw}.
\end{itemize}

\section{Foster-Lyapunov Theorem 
}\label{sec:appendix:fosterlyapunov}
\begin{prop}\label{prop:foster_kingman}
Let $\{ \mathbf{x}(t): t \ge 0\}$ be a continuous-time, time-homogeneous and irreducible Markov chain with state space $\mcl{S}$ and generator matrix $Q$. Suppose there exist a non-negative function $V : \mcl{S} \ra \mb{R}_{+}$, an $\eps' > 0$, a finite set $\mcl{A}$, and a finite constant $B$ such that $\{\statex : V(\statex) \le C \}$ is finite for all $C \in \mb{R}_{+}$, and the unit-transition drift ${QV}(\statex) $ is upper bounded as 
\begin{align*}
{QV}(\statex) \le - \eps' \11{ \{\statex \in  \mcl{A}^c \} }(\statex) + B \11{ \{ \statex \in \mcl{A} \} }( \statex ).
\end{align*}
Then, the process $\{\statex(t):t\ge 0\}$ is positive recurrent. The unit-transition drift ${QV}(\statex)$ is given by
\begin{align*}
{QV}(\statex) \defeq \sum \limits_{\statey \in \mcl{S}, \statey \ne \statex} q(\statex, \statey) \left( V(\statey) - V(\statex) \right). \numberthis \label{eq:{QV}def}
\end{align*}
Now, suppose $V'$, $f$, and $g$ are non-negative functions on $\mcl{S}$, and suppose ${QV}'(\statex) \le  -f(\statex) + g(\statex)$ for all $\statex \in \mcl{S}$. In addition, suppose $\{\statex(t):t\ge 0\}$ is positive-recurrent, so that the means, $\ov{f} = \pi f$ and $\ov{g} = \pi g$ are well-defined. Then $\ov{f} \le \ov{g}$. (In particular, if $g$ is
bounded, then $\ov{g}$ is finite, and therefore $\ov{f}$ is finite).
\end{prop}

\section{List of Important Symbols}
\label{sec:appendix:notationtable}
\begin{itemize}
\setlength\itemsep{2pt}
\item
$\mcl{F}$: Master-file.
\item
$K$: $|\mcl{F}|$.
\item
$W$: Denotes a file or the corresponding swarm.
\item
$\kw$: $|W|$.
\item
$\mcl{F}_W$: Set of all pieces downloadable by swarm-$W$.
\item
$\mcl{W}$: Set of all swarms entering the network.
\item
$\lambda_W$: Arrival rate of an empty swarm-$W$ peer.
\item
$ \bsl{\lambda} $: $(\lambda_W:W\in\mcl{W})$.
\item
$\slmdas$: $\sum_{W\in\mcl{W}} \lambda_W$.
\item
$\mcl{W}_W$: Ally-set of swarm-$W$.
\item
$\xws$: Number of $\ws$-type peers.
\item
$\statex$: State of the network. See \eqref{eq:state}.
\item
$\xw$: Number of peers in swarm-$W$. See \eqref{eq:xw}.
\item
$\x$: Number of peers in the network. See \eqref{eq:x}.
\item
$L$: Number of contact links with each peer.
\item
$Y_{opt}$: Binary parameter for optimistic-unchoke.
\item
$\mu$: Ticking rate of tit-for-tat link.
\item
$\widehat{\mu}$: Ticking rate of optimistic-unchoke link.
\item
$U$: Fixed seed's upload rate.
\item
$p$: In a given tit-for-tat contact, $p$ is the lower-bound on the probability of a peer push-contacting the other peer.
\item RF: Rarest-First.
\item RN: Random-Novel.
\item MS: Mode-Suppression.
\item TMS: Threshold-Mode-Suppression.
\item RFwPMS: Rarest-First with Probabilistic-Mode-Suppression.
\item LPS: Last-Piece Syndrome.
\item FPS: First-Piece Syndrome.
\item RNwPMS: Random-Novel with Probabilistic-Mode-Suppression.
\item
$i$: Typically used to denotes a piece of some file.
\item
$\freq{i}{W}$ / $\nmbr{i}{W}$: Frequency / chunk-count of piece $i$. See \eqref{eq:freqwi}
\item
$\freqn{W}$ / $\freqr{W}$: Maximum / Minimum chunk-frequency amongst pieces of $W$ in swarm-$W$. See \eqref{eq:maxminfreqw}.
\item
$\nmbrn{W}$ / $\nmbrr{W}$: Maximum / Minimum chunk-count amongst pieces of $W$ in swarm-$W$. See \eqref{eq:maxminnmbrw}.
\item
$\pw$: Total chunk-count in swarm-$W$. See \eqref{eq:chunkcountw}.
\item
$m_W^{(i)}$: Mismatch of piece $i$ in swarm-$W$. See \eqref{eq:mwi}.
\item
$\ov{m}_W$: Largest-mismatch in swarm-$W$. See \eqref{eq:largestmwi}.
\item
$\totalmismatch{W}$: Total-mismatch in swarm-$W$. See \eqref{eq:totalmismatch}.
\item
$\dnmbr{i}{W}$: Swarm-$W$'s complementary chunk-count of piece-$i$. See \eqref{eq:diw}.
\item
$\rw$: Set of rare pieces in swarm-$W$. See \eqref{eq:rw}.
\item
$\rwc$: $W\setminus\rwc$, set of non-rare pieces in swarm-$W$.
\item
$r$: Index for a rare piece of some swarm.
\item
$n$: Index for a non-rare piece of some swarm.
\item
$\udl{r}$: Index for a piece with the smallest chunk-count.
\item
$\setrares{\wh{T}}$: Set of rarest-pieces transferable from revealed cache-profile $\wh{T}$ to $\ws$-type peer. See \eqref{eq:setrares}.
\item
$\zeta_W^{(n)}$: Non-rares sharing factor of swarm-$W$. See \eqref{eq:zetaw} and \eqref{eq:zetawnew}.
\item
$A(\statex, \wh{T}, W, S)$: Transferable set. See Algorithm \ref{alg:transferable_set}.
\item
$\mcl{S}$: State-space of $\{\statex(t):t\ge 0\}$. See \eqref{eq:statespace}.
\item
$\Delta_t$: $2(L-\11\{Y_{opt}=1\} )\mu t+\11\{Y_{opt=1}\}\widehat{\mu} $.
\item
$\Gammaexact{W}{i}$ / $\Gammalower{W}{i}$ / $\Gammaupper{W}{i}$: Exact / Lower / Upper estimate for aggregate rate at which some ally-peer of swarm-$W$ push-contacts someone. See \eqref{eq:Gammalower2}, \eqref{eq:Gammaupper2} and \eqref{eq:Gammabounds}.
\item
$\xi_2$: An upper-bound on the ratio $\Gammaupper{W}{i}/\Gammalower{W}{i}$ for all $W\in\mcl{W}$ and $i\in W$. See \eqref{eq:Gammaratios}.
\item
$q_W^{(\emptyset,+)}$: $\lambda_W$.
\item
$q_W^{(S,i+)}$: Aggregate rate at which a $\ws$-peer downloads piece $i$ without departing the system.
\item
$q_W^{(S,i-)}$: Aggregate rate at which a $\ws$-peer downloads piece $i$ and departs the system.
\item
$\sw1$: $\{\statex:\rw(\statex)\subsetneq W\}$.
\item
$\sw2$: $\{\statex:\rw(\statex)= W\}$.
\item
$\udl{q}_W^{(S,n+)}$: Lower estimate for rate of $\ws$-peer downloading non-rare piece $n\in\rwc$. See \eqref{eq:qwsnplus}.
\item
$\gamma_W^{(i)}$: $\sum_{S:W\setminus S=\{i\}}\frac{\xws}{\xw} $. That is, the fraction of swarm-$W$ peers who need only piece $i$ to depart the system.
\item
$a_W^{(i)}$: $\zeta_W^{(i)} \11\{i\in\rwc\} + \11\{i\in\rw\} $.
\item
$V$: $\sum_{W\in\mcl{W}} V_W (\statex)$. See \eqref{eq:stab:v}.
\item
$V_W$: See \eqref{eq:stab:vwdef}.
\item
$\regrw{W}{1}$: $\{\statex:\totalmismatch{W}-2(1-\eta)\ge\eta\nmbrn{W}\}$.
\item
$\regrw{W}{2}$: $ \mcl{S}\setminus \regrw{W}{1}$.
\item
$\ov{\psi}_W$: 
\begin{align*}
&\left( \kw^2 +2\kw \left(
\totalmismatch{W} - \eta\nmbrn{W}
\right) \right) \11[\statex\in\regrw{W}{1}] \\
&\hspace{20pt} + 4\kw^2\11[\statex\in\regrw{W}{2}].
\end{align*}
\item
$\udl{\psi}_W$: 
\begin{align*}
&\left( 2(1-\eta)^2 - 2(1-\eta)\left(\totalmismatch{W} - \eta\nmbrn{W} \right) \right) \11[\statex\in\regrw{W}{1}] \\
&\hspace{20pt} + 4\kw^2\11[\statex\in\regrw{W}{2}].  
\end{align*}
\item
$I_W^{(1)}$: $\cw2\11\{\mw-2\ge \pw\} $.
\item
$I_W^{(2)}$: $\cw2\kw\11\{\mw+2\kw\ge\pw\} $.
\item
$\Psi_W^{(i-)}$: See \eqref{eq:stab:vwsiminus}.
\item
$\Psi_W^{(i+)}$: See \eqref{eq:stab:vwsiplus}.
\item
$QV$: See \eqref{eq:stab:qvdef} and \eqref{eq:stab:qvws}.
\item
$\Psi_W^{(i)}$: See \eqref{eq:stab:psiwi}.
\item
$\widetilde{QV}_W$: Used in upper estimate of $QV$. See \eqref{eq:stab:qv1} and \eqref{eq:stab:qvwtilde}.
\item
$\delta_W$: Small number in $(0,1)$ used in Appendix \ref{sec:appendix:stability}.
\item
$N_W^{(1)}$: Used in Lemma \ref{lem:nw}. 
\item
$\mcl{A}_W$, $B_W$, $N_W$: Used in Lemma \ref{lem:qvwupperbound}.
\item
$\mcl{H}$, $\mcl{S}_{\mcl{H}}$, $\mcl{A}_{\mcl{H}}$, $\mcl{A}$, $B$, $B_{\mcl{W}}$: Used in Lemma \ref{lem:qvupperbound}.
\end{itemize}


%



\section*{Acknowledgment}
This work was funded in part, by NSF via grants ECCS2038416, EPCN1608361, EARS1516075, CNS1955777, and CCF2008130 for V. Subramanian, grants EARS1516075, CNS1955777, and CCF2008130 for N. Khan, grant CCF1934986 for M. Moharrami, and by the University of Michigan via Rackham Predoctoral Fellowship for M. Moharrami.

\ifCLASSOPTIONcaptionsoff
  \newpage
\fi



\bibliographystyle{IEEEtr}
\bibliography{references}
%



%

\begin{IEEEbiography}
[{\includegraphics[width=1in, height=1.25in, clip, keepaspectratio]{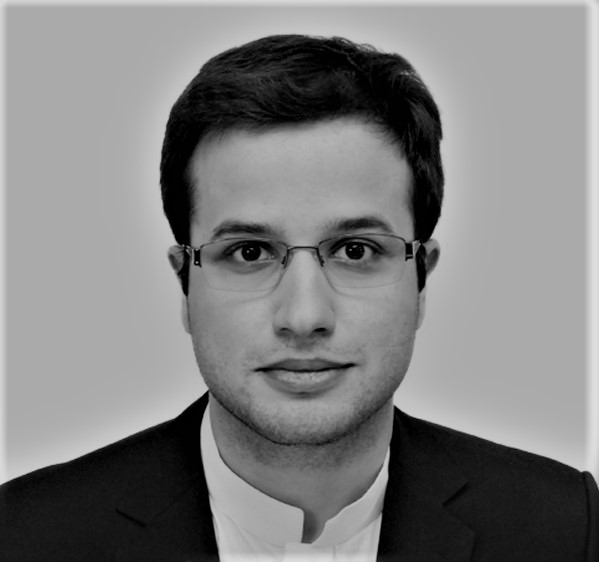}}]{Nouman Khan} (Member, IEEE) is a Ph.D candidate in the department of Electrical Engineering and Computer Science (EECS) at the University of Michigan, Ann Arbor, MI, USA. He received the B.S. degree in Electronic Engineering from the GIK Institute of Engineering Sciences and Technology, Topi, KPK, Pakistan, in 2014 and the M.S. degree in Electrical and Computer Engineering from the University of Michigan, Ann Arbor, MI, USA in 2019. His research interests include stochastic systems and their analysis and control.
\end{IEEEbiography}

\begin{IEEEbiography}[{\includegraphics[width=1in, height=1.25in, clip, keepaspectratio]{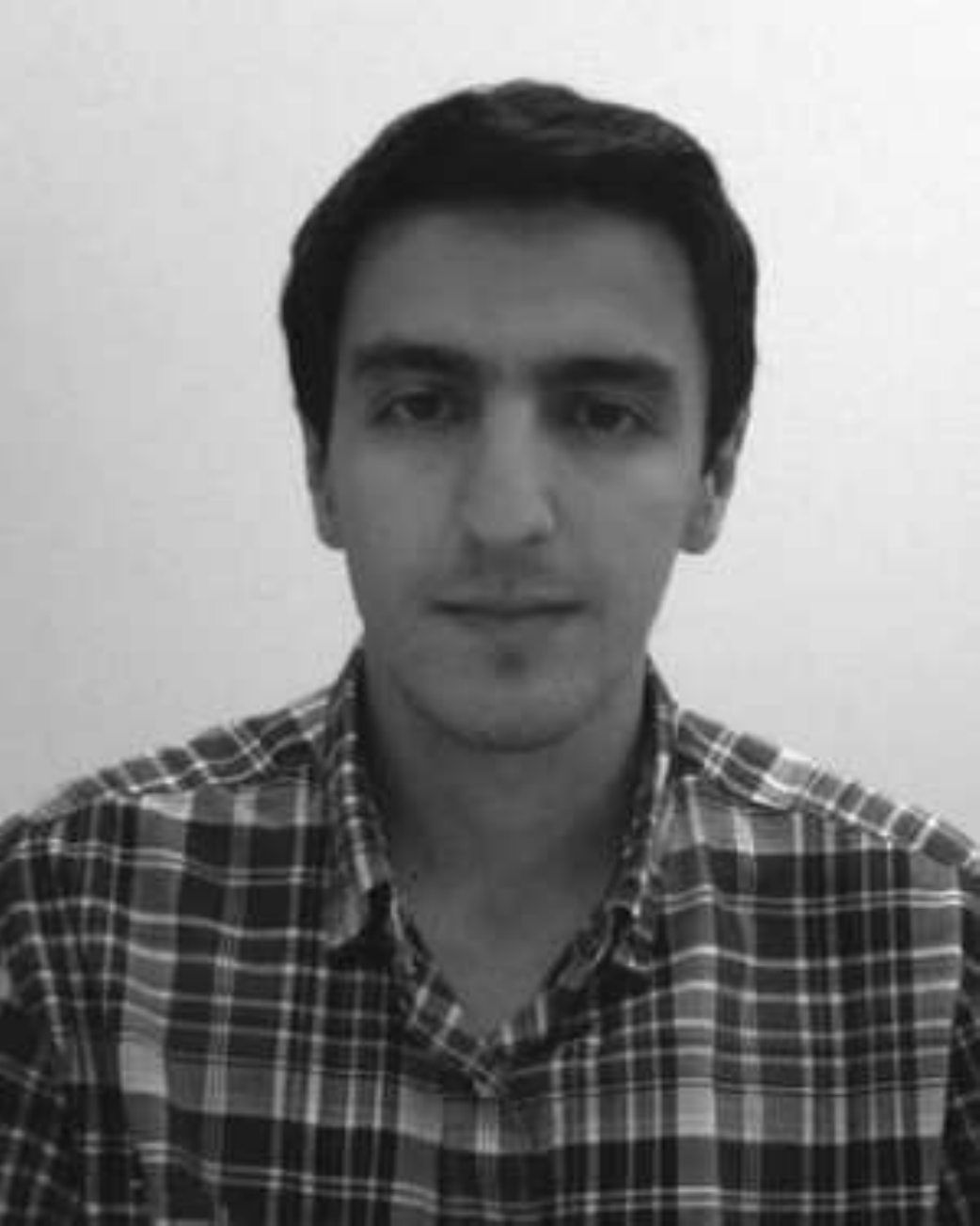}}]{Mehrdad Moharrami} is a TRIPODS Postdoctoral Research Fellow at the University of Illinois at Urbana Champaign. He received his B.Sc. from the Sharif University of Technology, Iran, in Mathematics, as well as Electrical Engineering. He holds an M.Sc. in Electrical Engineering and an M.Sc. in Mathematics from the University of Michigan. He received his Ph.D. in Electrical Engineering in 2020. His research interests include Markov decision processes, reinforcement learning, and random graph models for economics, learning, and computation.
\end{IEEEbiography}

\begin{IEEEbiography}[{\includegraphics[width=1in, height=1.25in, clip, keepaspectratio]{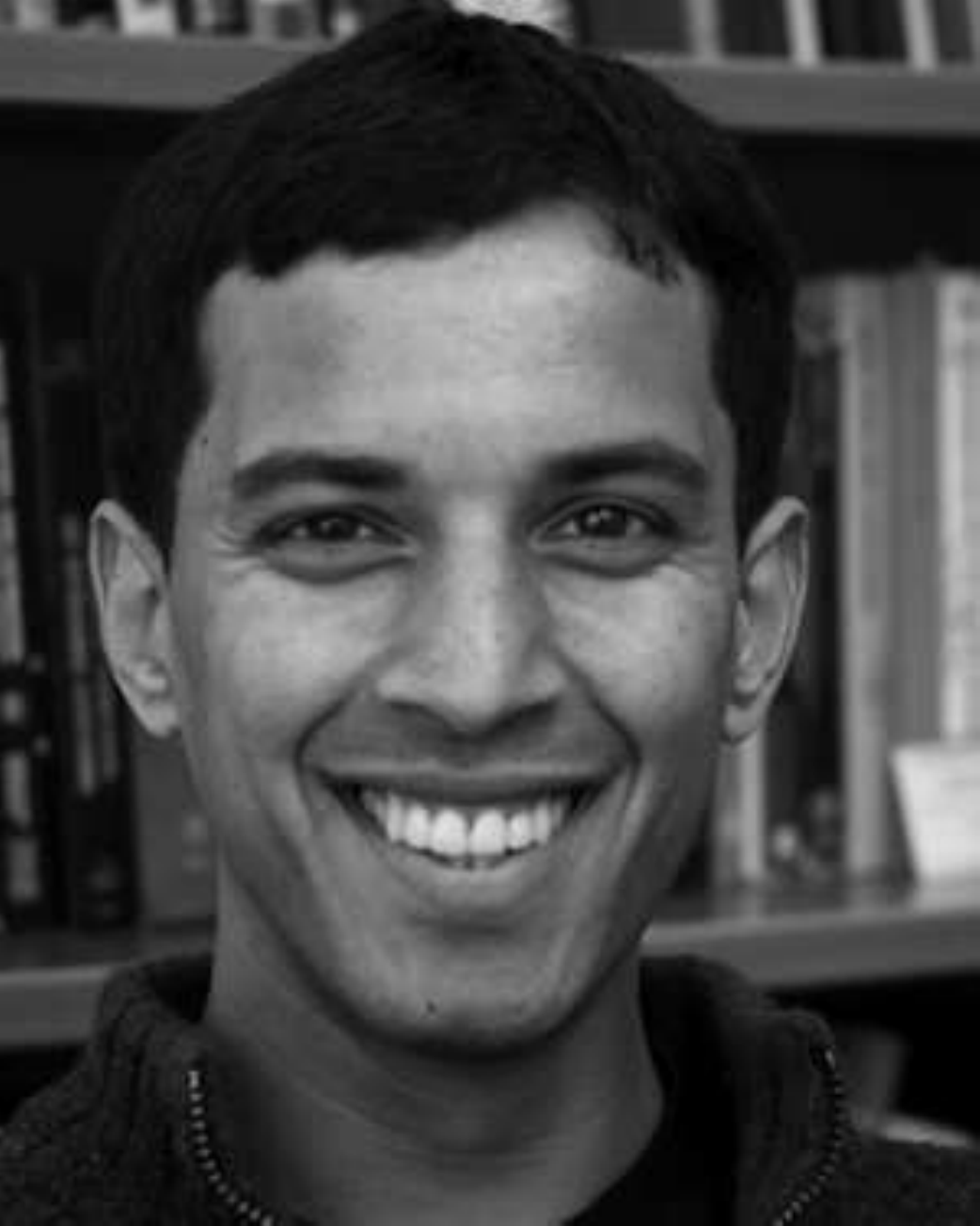}}]{Vijay Subramanian} (Senior Member, IEEE) received the Ph.D. degree in electrical engineering from the University of Illinois at Urbana-Champaign, Champaign, IL, USA, in 1999. He was a Researcher with Motorola Inc., and also with Hamilton Institute, Maynooth, Ireland, for a few years following which he was a Research Faculty with the Electrical Engineering and Computer Science (EECS) Department, Northwestern University, Evanston, IL, USA. In 2014, he joined the University of Michigan, Ann Arbor, MI, USA, where he is currently an Associate Professor with the EECS Department. His research interests are in stochastic analysis, random graphs, game theory, and mechanism design with applications to social, as well as economic and technological networks. 
\end{IEEEbiography}




\end{document}